\documentclass[twocolumn]{aastex63}

\usepackage{amsmath}
\usepackage{longtable}

\received{April 25, 2022}
\revised{June 22, 2022}
\accepted{June 27, 2022}
\published{August 12, 2022}

\shorttitle{TOI-1452\,$\rm b$}
\shortauthors{Cadieux et al.}

\graphicspath{{./}{figures/}}

\begin{document}

\title{TOI-1452\,b: SPIRou and TESS reveal a super-Earth in a temperate orbit transiting an M4 dwarf}

\correspondingauthor{Charles Cadieux}
\email{charles.cadieux.1@umontreal.ca}

\suppressAffiliations

\author[0000-0001-9291-5555]{Charles Cadieux} % CONFIRMED AS FIRST AUTHOR
\affiliation{Universit\'e de Montr\'eal, D\'epartement de Physique, IREX, Montr\'eal, QC H3C 3J7, Canada}

\author[0000-0001-5485-4675]{Ren\'e Doyon} % CONFIRMED AS CO-AUTHOR
\affiliation{Universit\'e de Montr\'eal, D\'epartement de Physique, IREX, Montr\'eal, QC H3C 3J7, Canada}
\affiliation{Observatoire du Mont-M\'egantic, Universit\'e de Montr\'eal, Montr\'eal H3C 3J7, Canada}

\author[0000-0002-9479-2744]{Mykhaylo Plotnykov} % CONFIRMED AS CO-AUTHOR
\affiliation{Department of Physics, University of Toronto, Toronto, ON M5S 3H4, Canada}

\author[0000-0001-5450-7067]{Guillaume H\'ebrard} % CONFIRMED AS CO-AUTHOR
\affiliation{Institut d'astrophysique de Paris, UMR7095 CNRS, Sorbonne Universit\'e, 98 bis bd Arago, 75014 Paris, France}

\author[0000-0003-0029-2835]{Farbod Jahandar} % CONFIRMED AS CO-AUTHOR
\affiliation{Universit\'e de Montr\'eal, D\'epartement de Physique, IREX, Montr\'eal, QC H3C 3J7, Canada}

\author[0000-0003-3506-5667]{\'Etienne Artigau} % CONFIRMED AS CO-AUTHOR
\affiliation{Universit\'e de Montr\'eal, D\'epartement de Physique, IREX, Montr\'eal, QC H3C 3J7, Canada}
\affiliation{Observatoire du Mont-M\'egantic, Universit\'e de Montr\'eal, Montr\'eal H3C 3J7, Canada}

\author[0000-0003-3993-4030]{Diana Valencia} % CONFIRMED AS CO-AUTHOR
\affiliation{Department of Physical \& Environmental Sciences, University of Toronto at Scarborough, Toronto, ON M1C 1A4, Canada}
\affiliation{David A. Dunlap Dept.\ of Astronomy \& Astrophysics, University of Toronto, 50 St. George Street, Toronto, Ontario, M5S 3H4, Canada}

\author[0000-0003-4166-4121]{Neil J. Cook} % CONFIRMED AS CO-AUTHOR
\affiliation{Universit\'e de Montr\'eal, D\'epartement de Physique, IREX, Montr\'eal, QC H3C 3J7, Canada}

\author[0000-0002-5084-168X]{Eder Martioli} % CONFIRMED AS CO-AUTHOR
\affiliation{Institut d'astrophysique de Paris, UMR7095 CNRS, Sorbonne Universit\'e, 98 bis bd Arago, 75014 Paris, France}
\affiliation{Laborat\'orio Nacional de Astrof\'isica, Rua Estados Unidos 154, Itajub\'a, MG 37504-364, Brazil}

\author[0000-0002-5922-8267]{Thomas Vandal} % CONFIRMED AS CO-AUTHOR
\affiliation{Universit\'e de Montr\'eal, D\'epartement de Physique, IREX, Montr\'eal, QC H3C 3J7, Canada}

\author[0000-0001-5541-2887]{Jean-Fran\c cois Donati} % CONFIRMED AS CO-AUTHOR
\affiliation{Universit\'e de Toulouse, CNRS, IRAP, 14 Avenue Belin, 31400 Toulouse, France}

\author[0000-0001-5383-9393]{Ryan Cloutier} % CONFIRMED AS CO-AUTHOR
\altaffiliation{Banting Fellow}
\affiliation{Center for Astrophysics $\vert{}$ Harvard \& Smithsonian, 60 Garden Street, Cambridge, MA, 02138, USA}

\author[0000-0001-8511-2981]{Norio Narita} % CONFIRMED AS CO-AUTHOR
\affiliation{Komaba Institute for Science, The University of Tokyo, 3-8-1 Komaba, Meguro, Tokyo 153-8902, Japan}
\affiliation{Astrobiology Center, 2-21-1 Osawa, Mitaka, Tokyo 181-8588, Japan}
\affiliation{Instituto de Astrof\'{i}sica de Canarias (IAC), 38205 La Laguna, Tenerife, Spain}

\author[0000-0002-4909-5763]{Akihiko Fukui} % CONFIRMED AS CO-AUTHOR
\affiliation{Komaba Institute for Science, The University of Tokyo, 3-8-1 Komaba, Meguro, Tokyo 153-8902, Japan}
\affiliation{Instituto de Astrof\'{i}sica de Canarias (IAC), 38205 La Laguna, Tenerife, Spain}

\author[0000-0003-3618-7535]{Teruyuki Hirano} % CONFIRMED AS CO-AUTHOR
\affiliation{Astrobiology Center, 2-21-1 Osawa, Mitaka, Tokyo 181-8588, Japan}
\affiliation{National Astronomical Observatory of Japan, 2-21-1 Osawa, Mitaka, Tokyo 181-8588, Japan}

\author[0000-0002-7613-393X]{Fran\c cois Bouchy} % CONFIRMED AS CO-AUTHOR
\affiliation{Departement d’astronomie, Universit\'e de Gen\`eve, Chemin Pegasi, 51, CH-1290 Versoix, Switzerland}

\author[0000-0001-6129-5699]{Nicolas B. Cowan} % CONFIRMED AS CO-AUTHOR
\affiliation{Department of Earth \& Planetary Sciences, McGill University, 3450 rue University, Montréal, QC H3A 0E8, Canada}
\affiliation{Department of Physics, McGill University, 3600 rue University, Montréal, QC H3A 2T8, Canada}

\author[0000-0002-9329-2190]{Erica J. Gonzales}  % CONFIRMED AS CO-AUTHOR
\affiliation{University of California Santa Cruz, Santa Cruz CA 95065, USA}

\author[0000-0002-5741-3047]{David R. Ciardi}  % CONFIRMED AS CO-AUTHOR
\affiliation{NASA Exoplanet Science Institute-Caltech/IPAC, Pasadena, CA 91125 USA}

\author[0000-0002-3481-9052]{Keivan G.\ Stassun} % CONFIRMED AS CO-AUTHOR
\affiliation{Department of Physics and Astronomy, Vanderbilt University, 6301 Stevenson Center Ln., Nashville, TN 37235, USA}
\affiliation{Department of Physics, Fisk University, 1000 17th Avenue North, Nashville, TN 37208, USA}

\author[0000-0002-0111-1234]{Luc Arnold} % CONFIRMED AS CO-AUTHOR
\affiliation{Canada-France-Hawaii Telescope, CNRS, Kamuela, HI 96743, USA}

\author[0000-0001-5578-1498]{Bj\"orn Benneke} % CONFIRMED AS CO-AUTHOR
\affiliation{Universit\'e de Montr\'eal, D\'epartement de Physique, IREX, Montr\'eal, QC H3C 3J7, Canada}

\author[0000-0002-1024-9841]{Isabelle Boisse} % CONFIRMED AS CO-AUTHOR
\affiliation{Aix Marseille Univ, CNRS, CNES, LAM, Marseille, France}

\author[0000-0001-9003-8894]{Xavier Bonfils} % CONFIRMED AS CO-AUTHOR
\affiliation{Univ.\ Grenoble Alpes, CNRS, IPAG, 38000 Grenoble, France}

\author[0000-0003-2471-1299]{Andr\'es Carmona} % CONFIRMED AS CO-AUTHOR
\affiliation{Univ.\ Grenoble Alpes, CNRS, IPAG, 38000 Grenoble, France}

\author[0000-0002-6174-4666]{P\'ia Cort\'es-Zuleta} % CONFIRMED AS CO-AUTHOR
\affiliation{Aix Marseille Univ, CNRS, CNES, LAM, Marseille, France}

\author[0000-0001-5099-7978]{Xavier Delfosse} % CONFIRMED AS CO-AUTHOR
\affiliation{Univ.\ Grenoble Alpes, CNRS, IPAG, 38000 Grenoble, France}

\author[0000-0003-0536-4607]{Thierry Forveille} % CONFIRMED AS CO-AUTHOR
\affiliation{Univ.\ Grenoble Alpes, CNRS, IPAG, 38000 Grenoble, France}

\author[0000-0002-1436-7351]{Pascal Fouqu\'e} % CONFIRMED AS CO-AUTHOR
\affiliation{Canada-France-Hawaii Telescope, CNRS, Kamuela, HI 96743, USA}
\affiliation{Universit\'e de Toulouse, CNRS, IRAP, 14 Avenue Belin, 31400 Toulouse, France}

\author[0000-0001-8056-9202]{Jo\~ao Gomes da Silva} % CONFIRMED AS CO-AUTHOR
\affiliation{Instituto de Astrofísica e Ciências do Espaço, Universidade do Porto, CAUP, Rua das Estrelas, 4150-762, Porto, Portugal}

\author[0000-0002-4715-9460]{Jon~M.~Jenkins} % CONFIRMED AS CO-AUTHOR
\affiliation{NASA Ames Research Center, Moffett Field, CA 94035, USA}

\author[0000-0001-9129-4929]{Flavien Kiefer} % CONFIRMED AS CO-AUTHOR
\affiliation{Institut d'astrophysique de Paris, UMR7095 CNRS, Sorbonne Universit\'e, 98 bis bd Arago, 75014 Paris, France}

\author[0000-0001-7157-6275]{\'Agnes K\'osp\'al} % CONFIRMED AS CO-AUTHOR
\affiliation{Konkoly Observatory, Research Centre for Astronomy and Earth Sciences, E\"otv\"os Lor\'and Research Network (ELKH), Konkoly-Thege Mikl\'os \'ut 15-17, 1121 Budapest, Hungary}
\affiliation{Max Planck Institute for Astronomy, K\"onigstuhl 17, 69117 Heidelberg, Germany}
\affiliation{ELTE E\"otv\"os Lor\'and University, Institute of Physics, P\'azm\'any P\'eter s\'et\'any 1/A, 1117 Budapest, Hungary}

\author[0000-0002-6780-4252]{David Lafreni\`ere} % CONFIRMED AS CO-AUTHOR
\affiliation{Universit\'e de Montr\'eal, D\'epartement de Physique, IREX, Montr\'eal, QC H3C 3J7, Canada}

\author[0000-0002-1532-9082]{Jorge H. C. Martins} % CONFIRMED AS CO-AUTHOR
\affiliation{Instituto de Astrofísica e Ciências do Espaço, Universidade do Porto, CAUP, Rua das Estrelas, 4150-762, Porto, Portugal}

\author[0000-0002-2842-3924]{Claire Moutou} % CONFIRMED AS CO-AUTHOR
\affiliation{Universit\'e de Toulouse, CNRS, IRAP, 14 Avenue Belin, 31400 Toulouse, France}

\author[0000-0001-7804-2145]{J.-D.~do~Nascimento,~Jr.} % CONFIRMED AS CO-AUTHOR
\affiliation{Universidade Federal do Rio Grande do Norte (UFRN), Departamento de F\'isica, 59078-970, Natal, RN, Brazil}
\affiliation{Center for Astrophysics $\vert{}$ Harvard \& Smithsonian, 60 Garden Street, Cambridge, MA, 02138, USA}

\author{Merwan Ould-Elhkim} % CONFIRMED AS CO-AUTHOR
\affiliation{Universit\'e de Toulouse, CNRS, IRAP, 14 Avenue Belin, 31400 Toulouse, France}

\author[0000-0002-8573-805X]{Stefan Pelletier} % CONFIRMED AS CO-AUTHOR
\affiliation{Universit\'e de Montr\'eal, D\'epartement de Physique, IREX, Montr\'eal, QC H3C 3J7, Canada}

\author[0000-0002-6778-7552]{Joseph D. Twicken} % CONFIRMED AS CO-AUTHOR
\affiliation{SETI Institute, Mountain View, CA  94043, USA}
\affiliation{NASA Ames Research Center, Moffett Field, CA  94035, USA}

\author[0000-0002-0514-5538]{Luke~G.~Bouma} % CONFIRMED AS CO-AUTHOR
\affiliation{Department of Astrophysical Sciences, Princeton University, 4 Ivy Lane, Princeton, NJ 08544, USA}

\author{Scott~Cartwright} % CONFIRMED AS CO-AUTHOR
\affiliation{Proto-Logic Consulting LLC, Washington DC, 20009 USA}

\author[0000-0002-7786-0661]{Antoine Darveau-Bernier} % CONFIRMED AS CO-AUTHOR
\affiliation{Universit\'e de Montr\'eal, D\'epartement de Physique, IREX, Montr\'eal, QC H3C 3J7, Canada}

\author[0000-0001-5707-8448]{Konstantin Grankin} % CONFIRMED AS CO-AUTHOR
\affiliation{Crimean Astrophysical Observatory, Department of Stellar Physics, Nauchny, 298409, Crimea}

\author[0000-0002-5658-5971]{Masahiro Ikoma} % CONFIRMED AS CO-AUTHOR
\affiliation{National Astronomical Observatory of Japan, 2-21-1 Osawa, Mitaka, Tokyo 181-8588, Japan}

\author[0000-0002-5331-6637]{Taiki Kagetani} % CONFIRMED AS CO-AUTHOR
\affiliation{Department of Multi-Disciplinary Sciences, Graduate School of Arts and Sciences, The University of Tokyo, 3-8-1 Komaba, Meguro, Tokyo 153-8902, Japan}

\author[0000-0003-1205-5108]{Kiyoe Kawauchi} % CONFIRMED AS CO-AUTHOR
\affiliation{Instituto de Astrof\'{i}sica de Canarias (IAC), 38205 La Laguna, Tenerife, Spain}
\affiliation{Departamento de Astrof\'{i}sica, Universidad de La Laguna (ULL), 38206 La Laguna, Tenerife, Spain}

\author[0000-0001-9032-5826]{Takanori Kodama} % CONFIRMED AS CO-AUTHOR
\affiliation{Komaba Institute for Science, The University of Tokyo, 3-8-1 Komaba, Meguro, Tokyo 153-8902, Japan}

\author[0000-0001-6181-3142]{Takayuki Kotani} % CONFIRMED AS CO-AUTHOR
\affiliation{Astrobiology Center, 2-21-1 Osawa, Mitaka, Tokyo 181-8588, Japan}
\affiliation{National Astronomical Observatory of Japan, 2-21-1 Osawa, Mitaka, Tokyo 181-8588, Japan}
\affiliation{Department of Astronomy, School of Science, The Graduate University for Advanced Studies (SOKENDAI), 2-21-1Osawa, Mitaka, Tokyo, Japan}

\author[0000-0001-9911-7388]{David~W.~Latham} % CONFIRMED AS CO-AUTHOR
\affiliation{Center for Astrophysics $\vert{}$ Harvard \& Smithsonian, 60 Garden Street, Cambridge, MA, 02138, USA}

\author{Kristen Menou} % CONFIRMED AS CO-AUTHOR
\affiliation{Department of Physical \& Environmental Sciences, University of Toronto at Scarborough, Toronto, ON M1C 1A4, Canada}
\affiliation{David A. Dunlap Dept.\ of Astronomy \& Astrophysics, University of Toronto, 50 St. George Street, Toronto, Ontario, M5S 3H4, Canada}
\affiliation{Department of Physics, University of Toronto, Toronto, ON M5S 3H4, Canada}

\author[0000-0003-2058-6662]{George~Ricker} % CONFIRMED AS CO-AUTHOR
\affiliation{MIT Kavli Institute for Astrophysics and Space Research, Massachusetts Institute of Technology, Cambridge, MA 02139, USA}
\affiliation{MIT Department of Physics, Massachusetts Institute of Technology, Cambridge, MA 02139, USA}

\author[0000-0002-6892-6948]{Sara~Seager} % CONFIRMED AS CO-AUTHOR
\affiliation{MIT Kavli Institute for Astrophysics and Space Research, Massachusetts Institute of Technology, Cambridge, MA 02139, USA}
\affiliation{Earth and Planetary Sciences, Massachusetts Institute of Technology, 77 Massachusetts Avenue, Cambridge, MA 02139, USA}
\affiliation{Department of Aeronautics and Astronautics, MIT, 77 Massachusetts Avenue, Cambridge, MA 02139, USA}

\author[0000-0002-6510-0681]{Motohide Tamura} % CONFIRMED AS CO-AUTHOR
\affiliation{Department of Astronomy, Graduate School of Science, The University of Tokyo, 7-3-1 Hongo, Bunkyo, Tokyo 113-0033, Japan}
\affiliation{Astrobiology Center, 2-21-1 Osawa, Mitaka, Tokyo 181-8588, Japan}
\affiliation{National Astronomical Observatory of Japan, 2-21-1 Osawa, Mitaka, Tokyo 181-8588, Japan}

\author[0000-0001-6763-6562]{Roland~Vanderspek}
\affiliation{MIT Kavli Institute for Astrophysics and Space Research, Massachusetts Institute of Technology, Cambridge, MA 02139, USA}

\author[0000-0002-7522-8195]{Noriharu Watanabe}
% CONFIRMED AS CO-AUTHOR
\affiliation{Department of Multi-Disciplinary Sciences, Graduate School of Arts and Sciences, The University of Tokyo, 3-8-1 Komaba, Meguro, Tokyo 153-8902, Japan}

\begin{abstract}

Exploring the properties of exoplanets near or inside the radius valley provides insights on the transition from the rocky super-Earths to the larger, hydrogen-rich atmosphere mini-Neptunes. Here, we report the discovery of TOI-1452\,b, a transiting super-Earth ($R_{\rm p} = 1.67 \pm 0.07$\,R$_{\oplus}$) in an 11.1--day temperate orbit ($T_{\rm eq} = 326 \pm 7$\,K) around the primary member ($H = 10.0$, $T_{\rm eff} = 3185 \pm 50$\,K) of a nearby visual binary M dwarf. The transits were first detected by TESS, then successfully isolated between the two $3\farcs2$ companions with ground-based photometry from OMM and MuSCAT3. The planetary nature of TOI-1452\,b was established through high-precision velocimetry with the near-infrared SPIRou spectropolarimeter as part of the ongoing SPIRou Legacy Survey. The measured planetary mass ($4.8 \pm 1.3$\,M$_{\oplus}$) and inferred bulk density ($5.6^{+1.8}_{-1.6}$\,g/cm$^3$) is suggestive of a rocky core surrounded by a volatile-rich envelope. More quantitatively, the mass and radius of \hbox{TOI-1452\,b}, combined with the stellar abundance of refractory elements (Fe, Mg and Si) measured by SPIRou, is consistent with a core mass fraction of $18\pm6$\% and a water mass fraction of $22^{+21}_{-13}$\%. The water world candidate \hbox{TOI-1452\,b} is a prime target for future atmospheric characterization with JWST, featuring a Transmission Spectroscopy Metric similar to other well-known temperate small planets such as LHS 1140\,b and K2-18\,b. The system is located near Webb's northern Continuous Viewing Zone, implying that is can be followed at almost any moment of the year.
\end{abstract}

\section{Introduction} \label{sec:intro}

Over the past decade, it has become increasingly clear that the typical extrasolar planetary system is quite different from our Solar System. Exoplanets are usually found in a much more compact orbital configuration \citep{Howard_2010} and the majority of systems have at least one planet with a size intermediate between the Earth and Neptune (\citealt{Howard_2012}; \citealt{Fressin_2013}). Population studies based on the \textit{Kepler} sample have shown that the occurrence rate distribution of close-in ($P < 100$\,days) exoplanets displays a valley/gap near 1.5--2.0\,R$_{\oplus}$ (\citealt{Fulton_2017}; \citealt{Fulton-Petigura_2018}; \citealt{Mayo_2018}; \citealt{Hardegree-Ullman_2020}). This radius valley most likely separates scaled-up, rocky versions of the Earth (super-Earths) and hydrogen-rich planets reminiscent of Neptune, but smaller (mini-Neptunes). This transition is known to be period-dependent (\citealt{VanEylen_2018}; \citealt{Martinez_2019}) and to vary with the host star properties such as metallicity (\citealt{Petigura_2018}; \citealt{Owen_2018}), mass (\citealt{McDonald_2019}; \citealt{Cloutier-Menou_2020}), and age (\citealt{Berger_2020}; \citealt{David_2021}). The existence of a radius valley was rapidly attributed to total or partial photoevaporation of the atmosphere by highly energetic photons during the first 100 Myr, when the host star is more active (\citealt{Owen_2013}, \citeyear{Owen_2017}; \citealt{Lopez_2014}; \citealt{Lopez_2018}; \citealt{Wu_2019}). However, another atmospheric erosion mechanism is plausible, involving mass loss caused by the release of energy from the planet core, accumulated during formation and slowly cooling down over Gyr timescales (\citealt{Ginzburg_2018}; \citealt{Gupta_2019}, \citeyear{Gupta_2020}). More recently, \cite{Lee_2021} have shown that the radius valley can be sculpted as a feature of formation, involving gas-poor accretion and supporting the hypothesis of a primordial bimodal distribution, rather than the result of subsequent atmospheric erosion. In order to identify which mechanism dominates, \cite{Rogers_2021} predict that the number of well-characterized small exoplanets must reach $\gtrsim 5000$. Such characterization requires the precise knowledge of planetary radii ($\lesssim5\%$ uncertainty) and, if possible, the planet mass. The combination of the two measurements leads to the mean density of the objects, a way to determine whether their internal structure is compatible with a rocky, gaseous, or intermediate bulk composition.

Identifying new small planets transiting nearby bright stars is the primary objective of the ongoing NASA Transiting Exoplanet Survey Satellite (TESS) mission \citep{Ricker_2015}. In operation since 2018, TESS has observed 85\% of the celestial sphere, staring for at least $\sim$27\,days at over 50 sectors covered so far ($24\arcdeg\times 96\arcdeg$ per sector). The TESS survey has already unveiled more than 5000 candidate exoplanets of which more than two hundred have been confirmed as new transiting planetary systems, including small planets around M-dwarf hosts (e.g., TOI-270, \citealt{Gunther_2019}; LP 791-18, \citealt{Crossfield_2019}; L~98-59, \citealt{Cloutier_2019};
LTT~1445\,A, \citealt{Winters_2019};
LTT~3780, \citealt{Cloutier_2020a}; TOI-1235, \citealt{Cloutier_2020b}; TOI-700, \citealt{Gilbert_2020}; TOI-1266, \citealt{Demory_2020}; LP 714-47, \citealt{Dreizler_2020}; TOI-776, \citealt{Luque_2021}).

M dwarfs represent prime targets not only for TESS, but in exoplanetary science in general. They are the most abundant stars in the solar neighborhood \citep{Reyle_2021} and host on average 2.5$\pm$0.2 planets per M dwarf with radii 1--4 R$_\oplus$ \citep{Dressing_2015}. Their smaller size compared to Sun-like stars facilitate the detection and characterization of new exoplanets by producing deeper transits for planets of a given size. The larger planet-to-star mass ratio amplifies the planetary radial velocity (RV) signal, allowing easier mass determination. Lastly, their lower luminosity results in a closer-in Habitable Zone (HZ), with orbital periods typically of one or two weeks adequately sampled by TESS.

The James Webb Space Telescope (JWST) is poised to revolutionize the field of exoplanet atmospheres \citep{Bean_2018} by offering a collecting area more than 6 times larger than the Hubble Space Telescope (HST) and spectral coverage from the visible to the mid-infrared (0.6--28\,$\mu$m). JWST will allow simultaneous identification of many chemical species with large absorption cross section in the infrared (e.g., H$_2$O, CH$_4$, CO, CO$_2$, NH$_3$), as well as probe the atmosphere of terrestrial planets with unprecedented sensitivity. One key objective of TESS is to discover the best transiting exoplanets amenable for atmospheric characterization with JWST \citep{Kempton_2018}.

Here, we report the discovery of a new small exoplanet around the nearby M dwarf TOI-1452. The planet was first detected by TESS, then characterized via follow-up efforts including RV monitoring with the SPIRou spectropolarimeter. The complete set of observations is described in Section~\ref{sec:obs}. The host star properties and physical parameters are derived in Section~\ref{sec:stellar_char}. Our data analysis and results are presented in Section~\ref{sec:analysis}. The implications of this discovery and prospects for follow-up characterization are discussed in Section~\ref{sec:discussion}, followed by concluding remarks in Section~\ref{sec:conclusion}.

\section{Observations} \label{sec:obs}

\subsection{TESS photometry} \label{sec:phot_tess}

\begin{figure}[t]
 \begin{center}
    \includegraphics[width=0.75\linewidth]{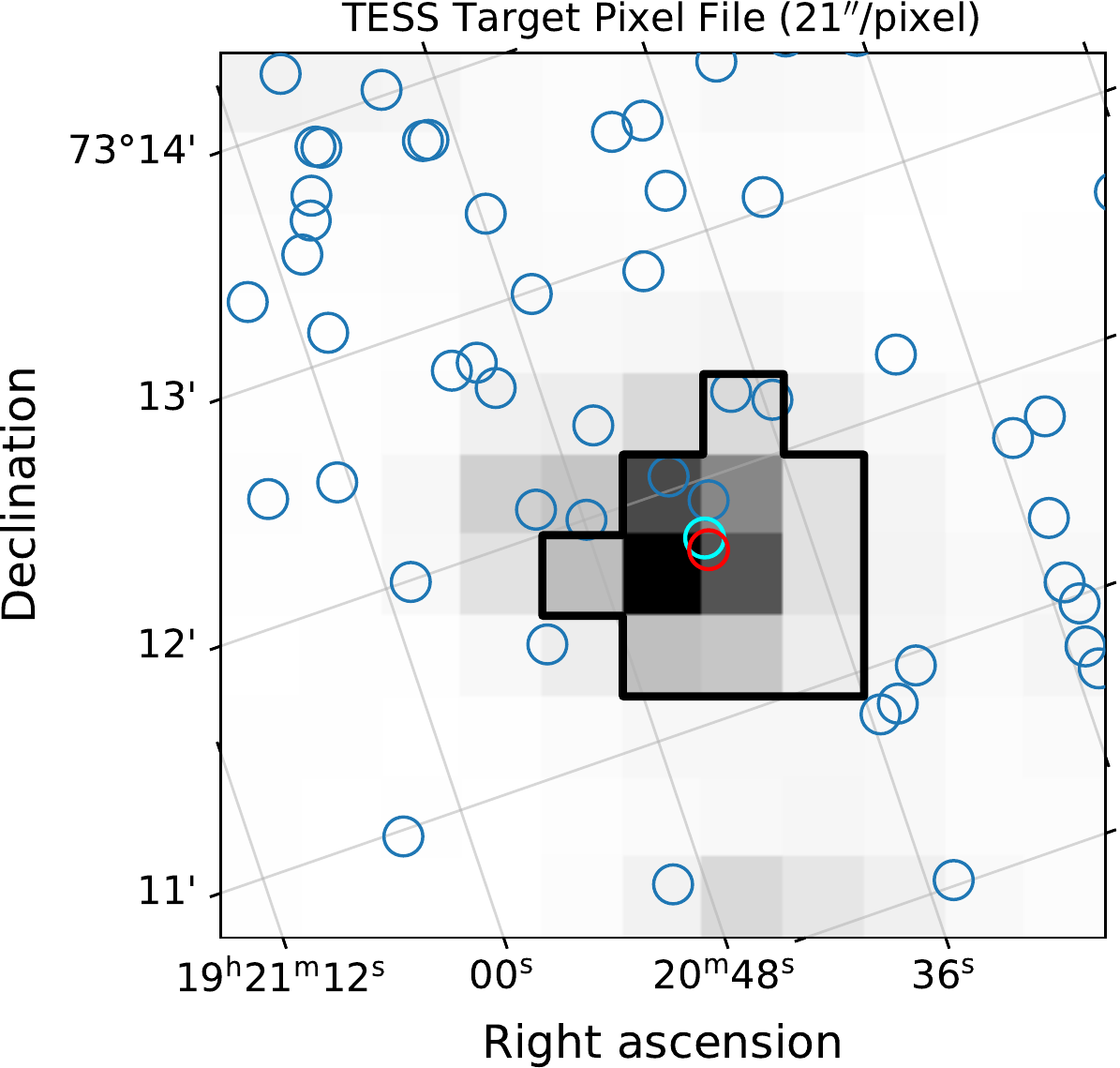}
    
    \vspace{0.25cm}
    
    \includegraphics[width=0.75\linewidth]{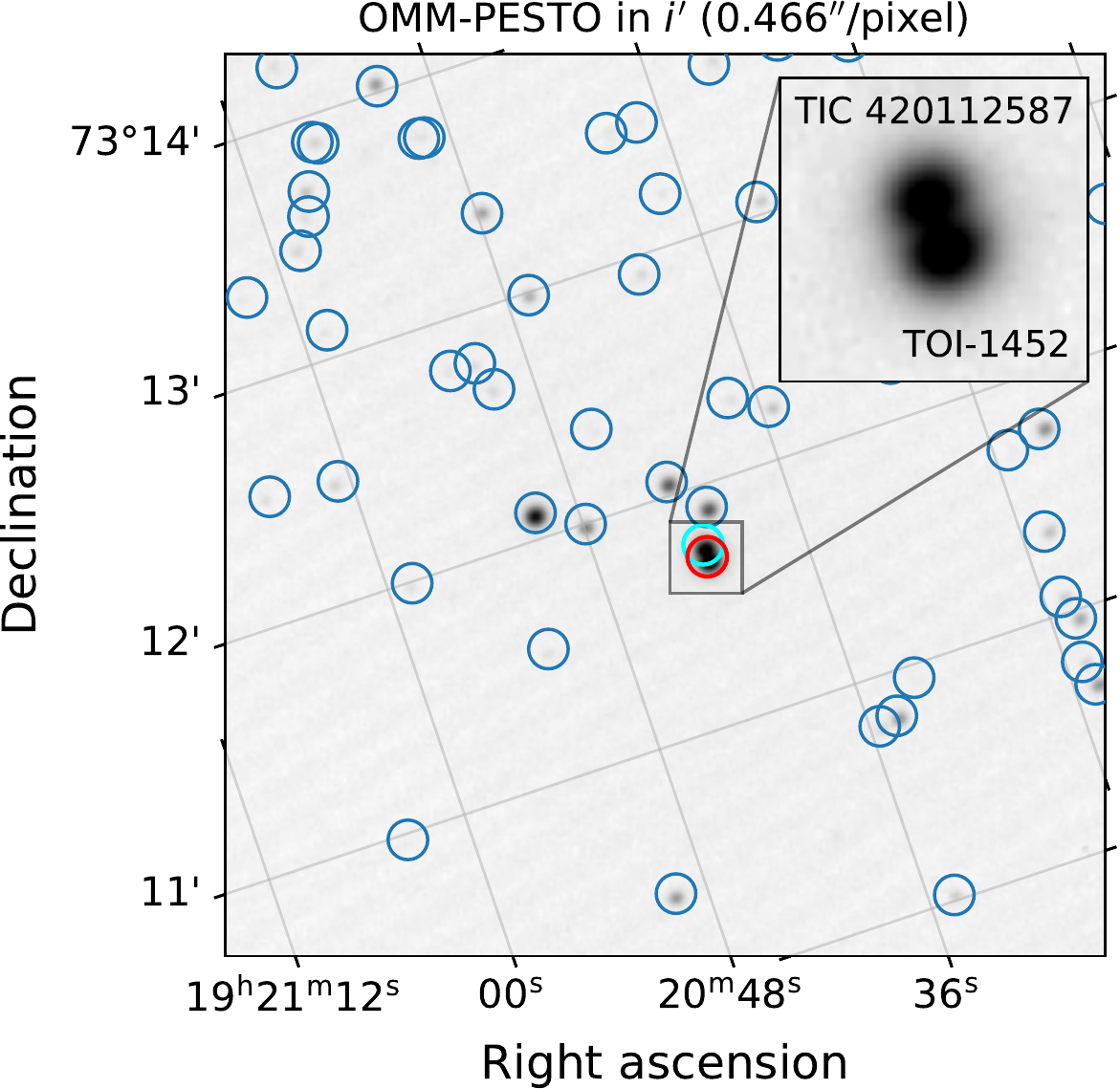}
  \caption{\textit{Upper panel}: TESS Target Pixel File of TOI-1452 from sector 14 ($11\times11$ pixels sub-region). The optimal aperture to extract the \texttt{PDCSAP} photometry is delimited by a black line. Nearby \textit{Gaia} EDR3 sources are represented with blue circles. TOI-1452 and its $3\farcs2$ companion (TIC 420112587) are shown with a red and a cyan circle respectively. \textit{Lower panel}: Same region of the sky observed with OMM-PESTO 1.6\,m on 2020-02-22 (see Sect.\ \ref{sec:phot_ground}). The visual binary was partially resolved.}
  \label{fig:TESS_fov}
 \end{center}
\end{figure}

\begin{figure*}[ht!]
\minipage{0.5\textwidth}
  \centering
  \includegraphics[width=1\linewidth]{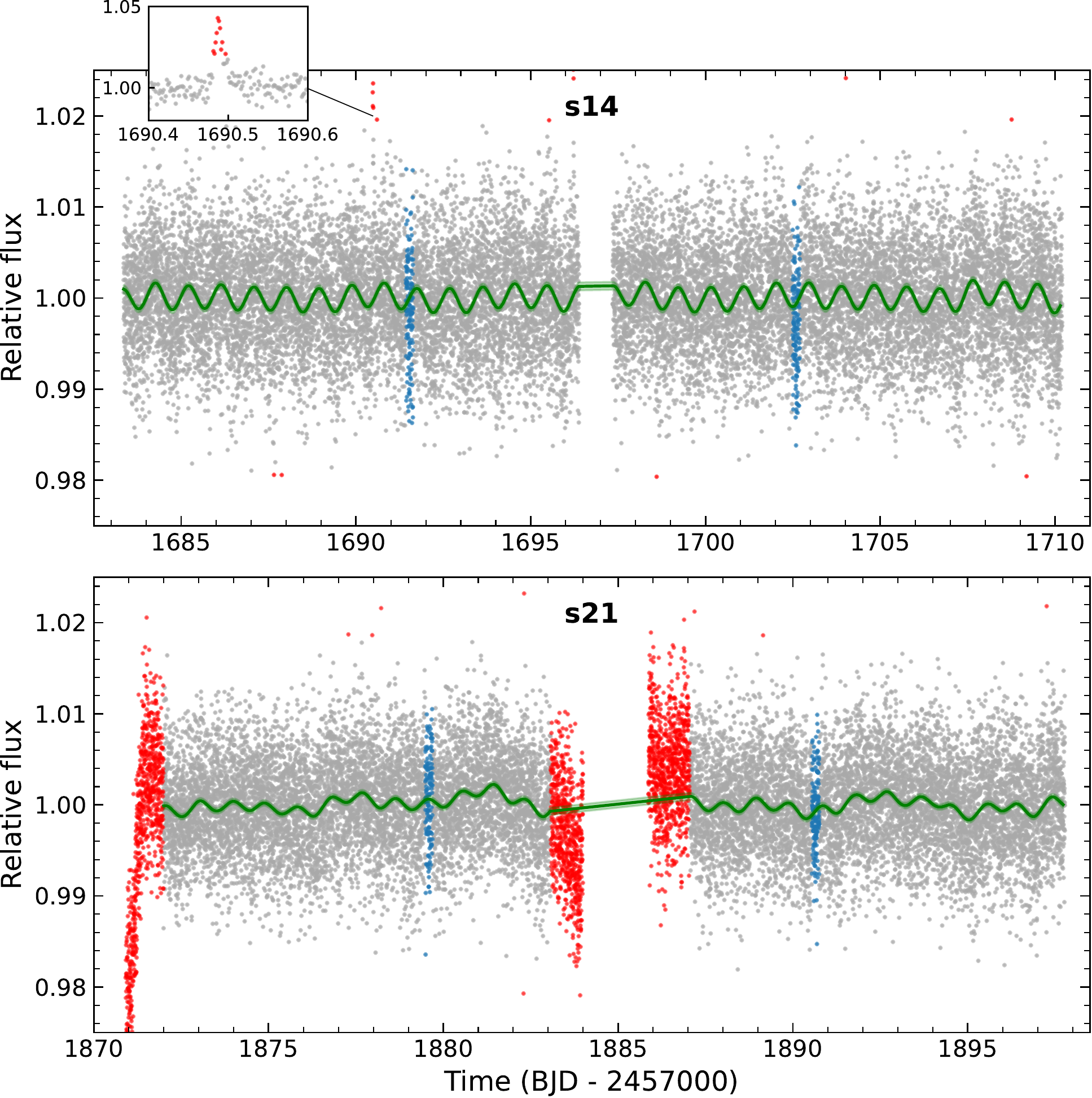}
  \endminipage\hfill
  \minipage{0.48\textwidth}
  \centering
  \includegraphics[width= 1\linewidth]{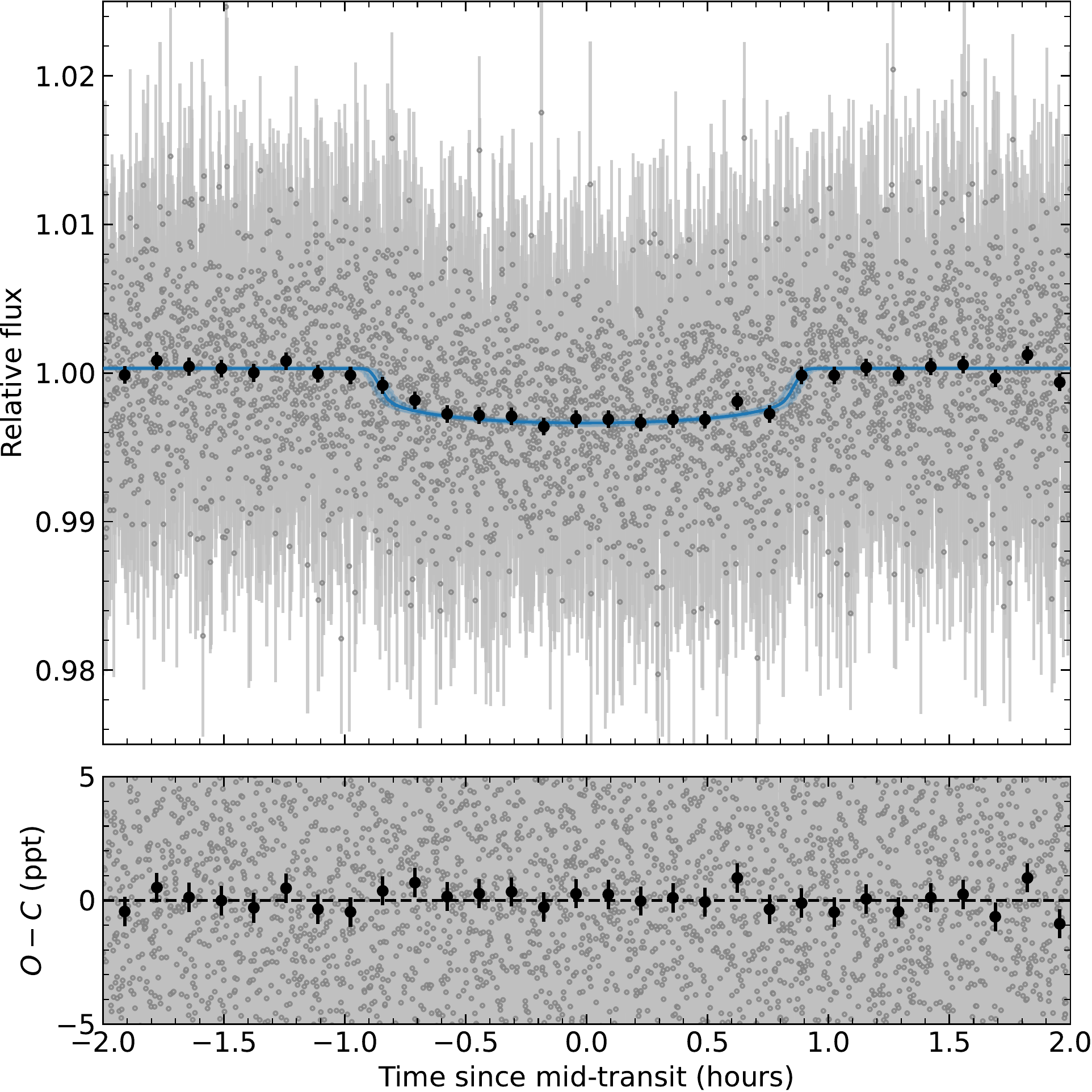}
  \endminipage\hfill
  \caption{\textit{Left panels}: Normalized \texttt{PDCSAP} light curve of TOI-1452 from sectors 14 and 21, featuring transits (blue data points), a $\sim$5\% stellar flare event (zoomed in sub-panel), and outliers (red data points) either rejected by sigma clipping (3.5\,$\sigma$ clip) or manually (sector 21). A quasi-periodic Gaussian Process model is depicted with the green curve (details in Sect.\ \ref{sec:tess_analysis}). The remaining sectors are presented in Fig.~\ref{fig:TESS_lc_complete}. \textit{Right panel}: TESS phase-folded corrected transits (32) from sectors 14--26, 40--41, and 47. Binned photometry (8 min phase bin) is represented with black points. The blue curve shows the best-fit transit model (described in Sect.\ \ref{sec:jointfit}), with the 68\% confidence interval envelope in light blue. The residuals of this fit are shown below.}
\label{fig:TESS_lc_phase}
\end{figure*}

TOI-1452 (TIC 420112589) was observed by TESS in sectors 14 through 26 (except 18), thus almost continuously from July 18, 2019 to July 4, 2020, in sectors 40--41 from June 25 to August 20, 2021, and finally in sector 47 from December 31, 2021 to January 27, 2022 (details in Table~\ref{table:tessobs}). TOI-1452 was sampled at the TESS 2-minute ``short'' cadence, as the star is part of the Cool Dwarf List \citep{Muirhead_2018}, a specially curated list of high-priority late-K and M dwarfs added to the TESS Input Catalog (TIC, \citealt{Stassun_2018b}, \citeyear{Stassun_2019}). We used the publicly available\footnote{Mikulski Archive for Space Telescopes (MAST): \href{https://archive.stsci.edu/tess/}{\texttt{archive.stsci.edu/tess/}}} per-sector light curves produced by the TESS Science Processing Operations Center (SPOC, \citealt{Jenkins_2016}) at NASA Ames, more specifically their Presearch Data Conditioning Simple Aperture Photometry (\texttt{PDCSAP}, \citealt{Smith_2012}; \citealt{Stumpe_2012}, \citeyear{Stumpe_2014}). The \texttt{PDCSAP} light curves are corrected for both instrumental systematic trends seen across stars in the same sector/camera/CCD and for flux contamination from nearby stars located within a few TESS pixels (21$\arcsec$). Flux dilution reduces the observable transit depth, resulting in an underestimation of the planetary radius if not accounted for. This was particularly important for TOI-1452 because a companion star (TIC 420112587, see Sect.\ \ref{sec:bound}) is separated by only $3\farcs2$ and has a similar magnitude in the TESS band ($\Delta T = 0.204$). A new background correction was implemented for the TESS extended mission (starting with sector 27). We followed the procedure outlined in the TESS Data Release 38 notes\footnote{\href{https://archive.stsci.edu/tess/tess_drn.html}{\texttt{archive.stsci.edu/tess/tess\_drn.html}}} to correct our \texttt{PDCSAP} fluxes from the primary mission (sectors 14--26), adjusting the baseline level and reducing the inferred transit depth by $\sim$1.7\%. This ensures that the primary and extended mission data produce the same estimate of the planetary radius. Figure~\ref{fig:TESS_fov} shows a 11$\times$11 pixels sub-region around TOI-1452 from TESS sector 14 and the same region of the sky observed from the ground. This illustrates how TESS alone cannot resolve the source of a transit between TOI-1452 and the nearby companion TIC 420112587. The normalized \texttt{PDCSAP} light curve of TOI-1452 from sectors 14 and 21 is presented in Figure~\ref{fig:TESS_lc_phase}, while the remaining sectors are shown in Figure~\ref{fig:TESS_lc_complete}.

\begin{deluxetable}{ccccc}
\tablecaption{TESS observations of TOI-1452}
\tablehead{
\colhead{Sector} & \colhead{Camera} & \colhead{CCD} & \colhead{UT Start Date} & \colhead{UT End Date}
}
\startdata
14 & 3 & 2 & 2019-07-18 & 2019-08-14\\
15 & 3 & 2 & 2019-08-15 & 2019-09-10\\
16 & 2 & 1 & 2019-09-12 & 2019-10-06\\
17 & 4 & 2 & 2019-10-08 & 2019-11-02\\
19 & 4 & 1 & 2019-11-28 & 2019-12-23\\
20 & 4 & 1 & 2019-12-24 & 2019-01-20\\
21 & 4 & 1 & 2020-01-21 & 2020-02-18\\
22 & 4 & 4 & 2020-02-19 & 2020-03-17\\
23 & 4 & 4 & 2020-03-19 & 2020-04-15\\
24 & 3 & 4 & 2020-04-16 & 2020-05-12\\
25 & 3 & 3 & 2020-05-14 & 2020-06-08\\
26 & 3 & 3 & 2020-06-09 & 2020-07-04\\
40 & 3 & 2 & 2021-06-25 & 2021-07-23\\
41 & 3 & 2 & 2021-07-24 & 2021-08-20\\
47 & 4 & 1 & 2021-12-31 & 2022-01-27
\enddata
\tablenocomments{}
\label{table:tessobs}
\end{deluxetable}

A search of the sectors 14--16 with an adaptive, wavelet-based matched filter (\citealt{Jenkins_2002}; \citealt{Jenkins_2010}, \citealt{Jenkins_2020}) first identified transit signatures for TOI-1452. The Data Validation Reports (DVR; \citealt{Twicken_2018}; \citealt{Li_2019}) fitted a limb-darkened transit model with a signal-to-noise ratio (SNR) of 8.0, a period of 11.06409 days, and an average uncontaminated transit depth of 3.778 parts per thousand (ppt), corresponding to a preliminary planetary radius of $1.83 \pm 0.30$\,R$_{\oplus}$. This led to the announcement of the planet candidate TOI-1452.01 \citep{Guerrero_2021} by the TESS Science Office on October 26, 2019. Simultaneously, the TESS mission announced the candidate TOI-1760.01 around the companion star TIC 420112587 sharing the same ephemeris as TOI-1452.01. Ground-based photometry was able to isolate the transit signal, originating from TOI-1452 (see Sect.\ \ref{sec:psf}). 
The latest available DVR from sectors 14--41 includes 30 transits and reports a period of 11.06196 days along with a radius $R_{\rm p} = 1.60 \pm 0.42$\,R$_{\oplus}$. Our complete reanalysis of the TESS light curve presented in Section~\ref{sec:analysis} has resulted in a more precise planetary radius, in agreement with previous estimates. Figure~\ref{fig:TESS_lc_phase} shows the phase-folded 32 transits from currently available sectors. We note that TESS is expected to continue monitoring TOI-1452 during 2022.

\subsection{OMM-PESTO transit monitoring} \label{sec:phot_ground}

Due to the coarse image sampling of TESS (21\arcsec \ per pixel), the origin of a transit signal may be ambiguous when several stars are located inside the aperture (e.g., Fig.\ \ref{fig:TESS_fov}). For this reason, TESS planet candidates are prone to false positives, occasionally attributed to a nearby eclipsing binary (NEB) contaminating the light curve \citep{Sullivan_2015}. Ground-based follow-up with arc-second angular resolution is therefore necessary to validate on-target transit and reject the NEB scenario. For TOI-1452, a particular challenge was to determine the signal's provenance between the target and its $3\farcs2$ companion. 

Two transit events of TOI-1452.01 were observed using the PESTO camera installed on the 1.6\,m telescope of the Observatoire du Mont-Mégantic (OMM), Québec, Canada. PESTO features a $1024\times 1024$ pixel \hbox{EMCCD} detector with a pixel scale of $0\farcs466$, providing a field of view (FOV) of $7\farcm95\times7\farcm95$. We scheduled the two observing sequences with the \texttt{TESS Transit Finder} (\texttt{TTF}), a customized version of the \texttt{Tapir} software package \citep{Jensen_2013}, and have used \texttt{AstroImageJ} (\texttt{AIJ}; \citealt{Collins_2017}) to perform image calibrations, including bias subtractions and flat field corrections, and differential aperture photometry. 

\begin{figure*}[t!]
\minipage{0.5\textwidth}
  \centering
  \includegraphics[width=1\linewidth]{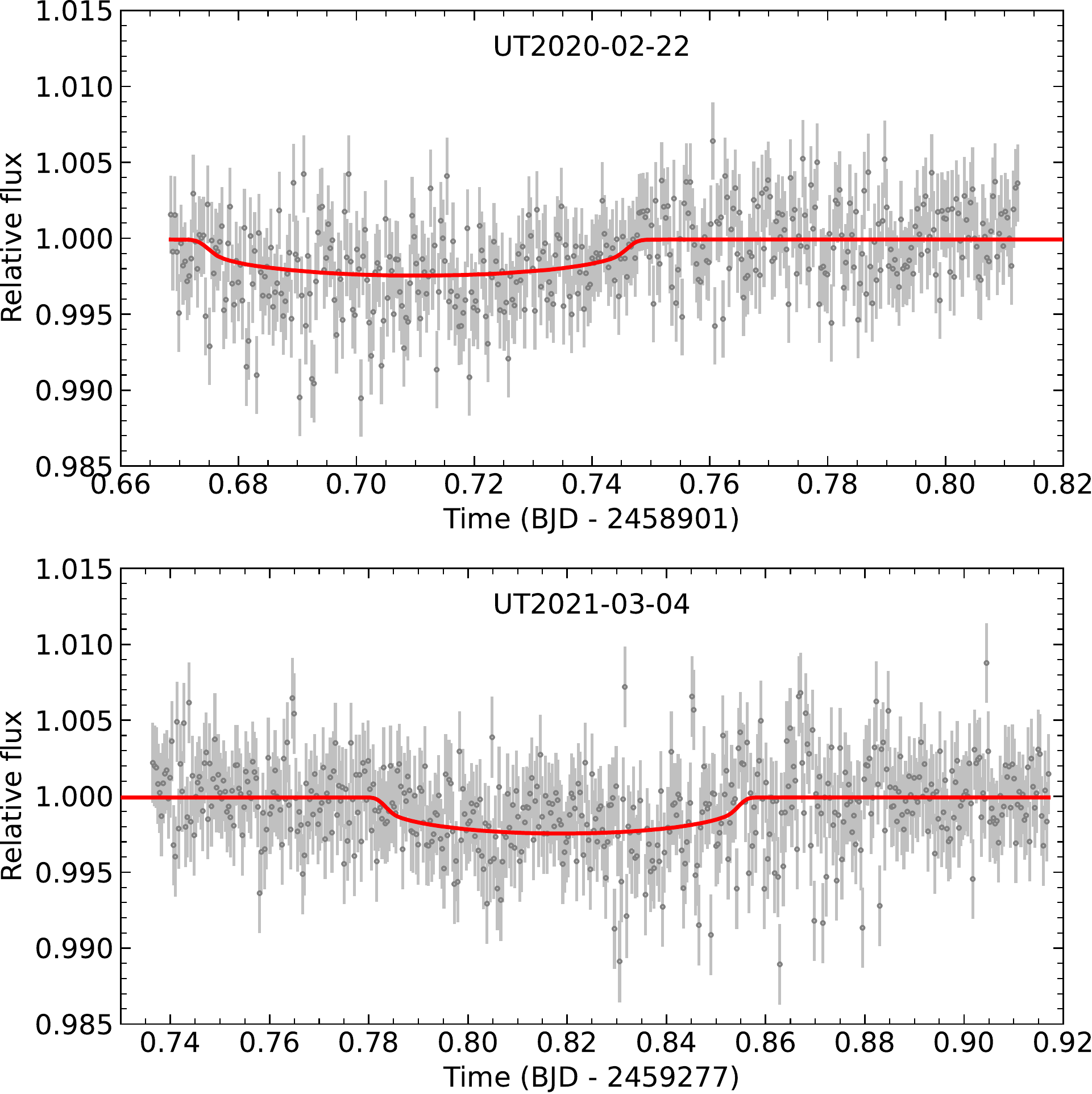}
  \endminipage\hfill
  \minipage{0.5\textwidth}
  \centering
  \includegraphics[width=1\linewidth]{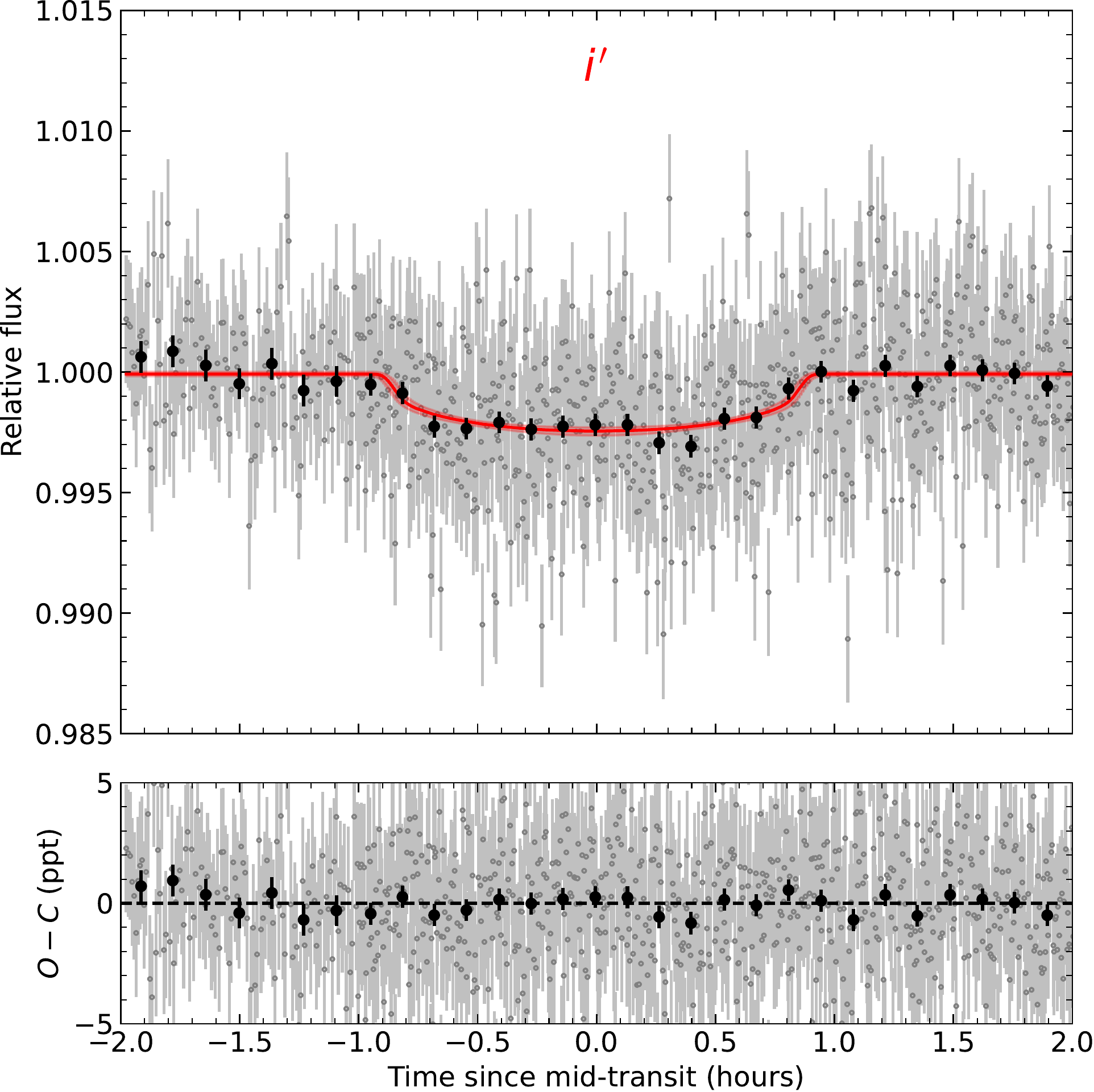}
  \endminipage\hfill
  
  \caption{Ground-based transit monitoring of TOI-1452.01 with the PESTO camera installed at Observatoire du Mont-Mégantic. \textit{Left panels:} Individual transit observations in the $i^{\prime}$ band on 2020 February 22 UT and 2021 March 4 UT. \textit{Right panel:} Transits from the left panels folded in phase, with black points representing the binned photometry (8 min phase bin). The red curve shows the best-fit transit model (described in Sect.\ \ref{sec:jointfit}), with the 68\% confidence interval envelope in transparent red. The residuals are shown below the phase-folded transit.}
\label{fig:OMM-PESTO_lc}
\end{figure*}

A first full transit was observed on February 22, 2020 in the $i^{\prime}$ filter with a sequence of 30\,s exposure time. As seen in the lower panel of Figure~\ref{fig:TESS_fov}, \hbox{TOI-1452} and TIC\,420112587 were partially resolved. Using a circular aperture of 7$\farcs$9 containing both stars, the transit was detected 53\,min earlier than predicted by the \texttt{TTF} (2.2$\sigma$ early), causing us to miss observing a proper pre-ingress baseline (see Fig.\ \ref{fig:OMM-PESTO_lc}, upper left). Additional TESS data later confirmed that the period was slightly overestimated by SPOC (sectors 14--16 only) at the time of observations, explaining why the transit arrived ``early''. The transit timing was also later confirmed by TESS sector 22 data, which was contemporaneous to this dataset. Even without a proper baseline, this transit was particularly valuable because it allowed us to reject the NEB false positive scenario and to determine, using point spread function (PSF) fitting (see Sect.\ \ref{sec:psf}), that the signal originated from TOI-1452.

A second full transit of TOI-1452.01 was observed on March 4, 2021 in $i^{\prime}$, using a 10\,s exposure time sequence. With a combined aperture of 8$\farcs$4, the transit was detected on time according to the \texttt{TTF}. In addition to a standard airmass linear detrending (also performed for the first transit), we used the Width (mean of the x- and y-direction FWHM) detrending option in \texttt{AIJ}. This was necessary to account for flux loss when the seeing was worse for certain exposures in the sequence, without increasing the aperture radius and dealing with flux contamination from a third star. This made sure that the transit depth was consistent with the one derived from the first PESTO observation, when the overall seeing was better (see Table~\ref{table:gbobs}). 

The OMM-PESTO observations are summarized in Table~\ref{table:gbobs}. The resulting aperture photometry transits are shown in Figure~\ref{fig:OMM-PESTO_lc}, and were included with the 32 TESS transits in our joint analysis (transit and RV datasets) presented in Section~\ref{sec:jointfit}.

\subsection{MuSCAT3 transit monitoring} \label{sec:phot_ground_muscat}

A full transit of TOI-1452.01 was observed on September 8, 2021 with the multiband imager MuSCAT3 \citep{2020SPIE11447E..5KN} on the 2\,m Faulkes Telescope North (FTN) of Las Cumbres Observatory (LCO) at Haleakala observatory, Hawaii. MuSCAT3 has four optical channels, each of which is equipped with a 2k $\times$ 2k CCD camera with a pixel scale of 0\farcs266 pixel$^{-1}$, enabling $g^{\prime}$-, $r^{\prime}$-, $i^{\prime}$-, and $z_{\rm s}$-band simultaneous imaging. For transit monitoring, significant chromaticity in the transit depths could indicate a diluted eclipsing binary. The exposure times were set at 35, 12, 6, and 3\,s for the $g^{\prime}$, $r^{\prime}$, $i^{\prime}$, and $z_{\rm s}$ bands, respectively. The observations were performed in-focus to spatially resolve the host star from the nearby companion star at $3\farcs2$, resulting in the FWHM of stellar point spread function of 3--8 pixels ($0\farcs8$--$2\farcs0$) depending on the airmass and band (see Table~\ref{table:gbobs}).

\begin{figure*}[t]
\includegraphics[width=1\linewidth]{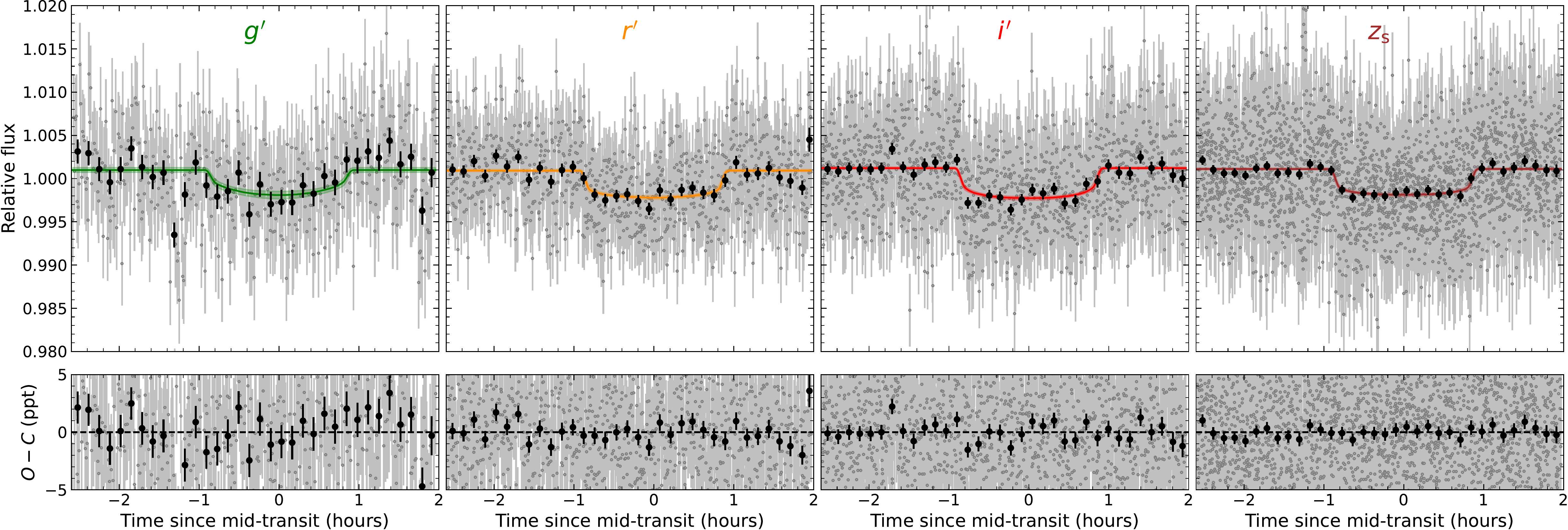}
  \caption{Ground-based transit follow-up of TOI-1452.01 on 2021 September 8 UT with the multi-filter ($g^{\prime}$, $r^{\prime}$, $i^{\prime}$, $z_{\rm s}$) MuSCAT3 instrument installed on LCO-FTN at Haleakala Observatory. For each corresponding filter, the black points depict the binned photometry (8 min temporal bin). The color-coded curves correspond to each filter's best-fit transit model (described in Sect.\ \ref{sec:jointfit}), with their respective 68\% confidence interval envelope in lighter shade. The residuals are shown below each phase-folded transit.}
  \label{fig:MuSCAT3_lc}
\end{figure*}

The obtained images were calibrated by the {\tt BANZAI} pipeline \citep{curtis_mccully_2018_1257560}. We performed aperture photometry on the calibrated images using a custom pipeline \citep{2011PASJ...63..287F} with aperture radii of 5 pixels, or $1\farcs3$, for all bands, which is almost free from flux contamination from the nearby companion star. For each band, we extracted the light curve using different sets of comparison stars, but have found that using only the companion (TIC 420112587) as a reference produced the minimum point-to-point dispersion. Since both the companion and the target stars have a similar color and are close to each other, the attenuation by the atmosphere is almost identical, so that we can safely assume that any second-order extinction effect (airmass dependent) is almost negligible. We clearly detected the transit on the target star in all bands, as shown in Figure~\ref{fig:MuSCAT3_lc}, providing further unambiguous evidence that TOI-1452 hosts a transiting object. A summary of this dataset is provided in Table~\ref{table:gbobs}.

\begin{deluxetable}{ccccc}
\tablecaption{Summary of the ground-based transit monitoring of TOI-1452}
\tablehead{
\colhead{UT Date} & \colhead{Camera} & \colhead{Filter} & \colhead{PSF} & \colhead{Aperture}\\[-0.1cm]
 & & & FWHM ($^{\prime \prime}$) & Size ($^{\prime \prime}$)
}
\startdata
\multicolumn{5}{c}{\textit{OMM 1.6\,m}}\\
2020-02-22 & PESTO & $i^{\prime}$ & 2.9 & 7.9$^\dag$\\
2021-03-04 & PESTO & $i^{\prime}$ & 3.9 & 8.4$^\dag$\\
\hline
\multicolumn{5}{c}{\textit{LCO-FTN 2\,m}}\\
2021-09-08 & MuSCAT3 & $g^{\prime}$ & 1.8 & 1.3\\
2021-09-08 & MuSCAT3 & $r^{\prime}$ & 1.3 & 1.3\\
2021-09-08 & MuSCAT3 & $i^{\prime}$ & 1.3 & 1.3\\
2021-09-08 & MuSCAT3 & $z_{\rm s}$ & 1.0 & 1.3\\
\enddata
\tablecomments{$^\dag$Using an aperture containing TOI-1452 and TIC 420112587}
\label{table:gbobs}
\end{deluxetable}

\subsection{Keck\,II/NIRC2 high-resolution imaging} \label{sec:imaging}

\begin{figure}[h]
\centering
    \includegraphics[width=1\linewidth]{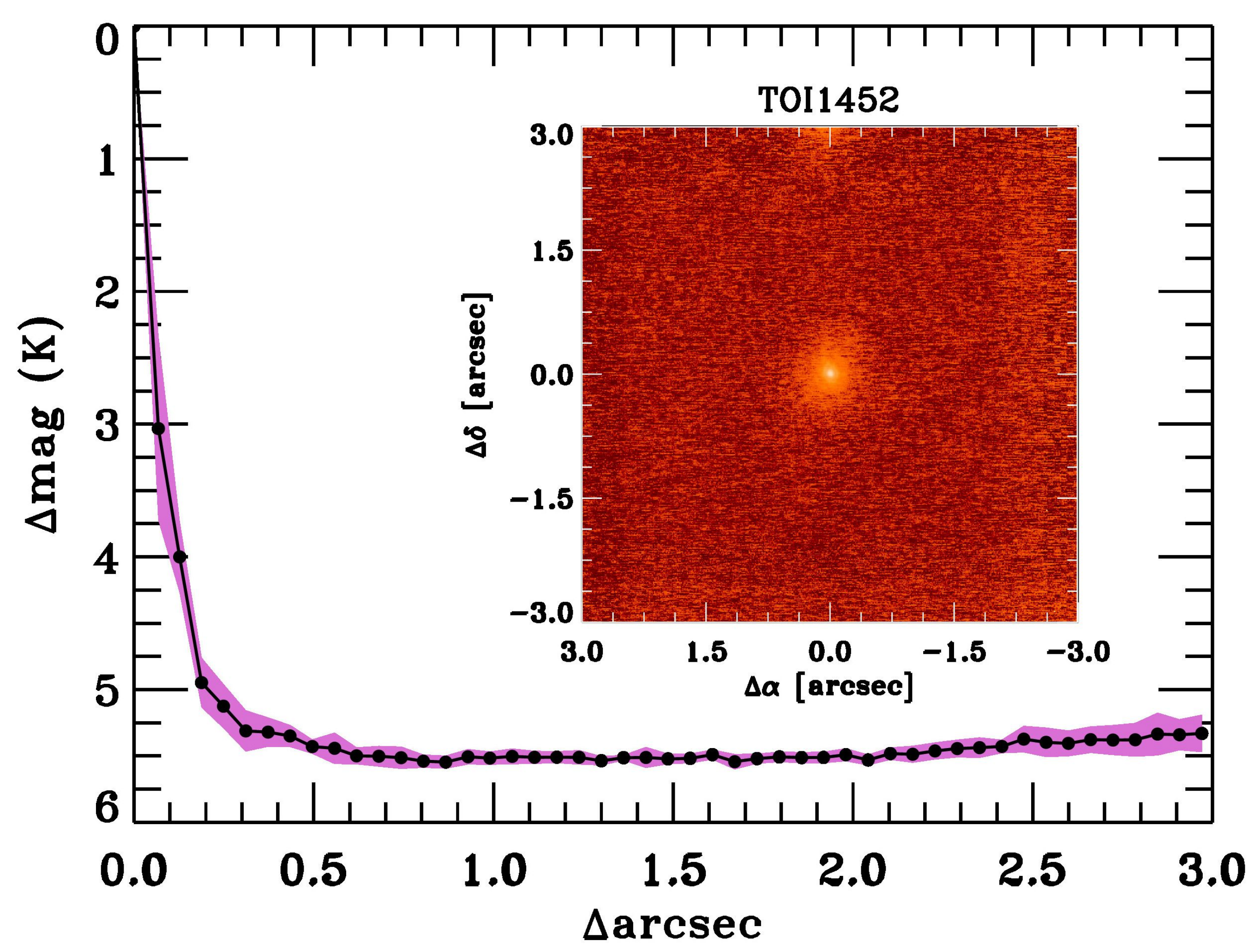}
    \linespread{1}
  \caption{$K$-band 5$\sigma$ contrast curve of TOI-1452 from KeckII/NIRC2 adaptive optics imaging. No close companion is detected.}
  \label{fig:Keck_imaging}
\end{figure}

One or more unresolved sources not in \textit{Gaia} EDR3 could still be located close to TOI-1452, whether gravitationally bound or not. A blended eclipsing binary (BEB) could indicate a false positive detection, and any other flux source would lead to underestimate the size of the transiting object in the TESS, PESTO and MuSCAT3 light curves. For these reasons, we searched for sub-arcsecond sources around TOI-1452 with the NIRC2 adaptive optics imaging camera installed on the 10\,m Keck\,II telescope. The images were acquired on May 28, 2020 in the $K$ band with a spatial resolution of 0$\farcs$01 per pixel, integration time per coadd of 1.6\,s, mean PSF FWHM of 0$\farcs$061, and airmass of 1.69. Figure~\ref{fig:Keck_imaging} shows the 5$\sigma$ contrast curve of TOI-1452, revealing that no additional companion is detected with a contrast ratio $\Delta K \leq 5.429$ for separation greater than $0\farcs5$. Although in the $K$ band, this contrast limit is similar to the difference in magnitude required ($\Delta T = 5.55$) for a 50\% depth BEB to mimic a 3\,ppt transit in the TESS light curve. Following the procedure of \cite{Lillo-Box_2014}, we calculated the probability of contamination from a blended source due to a random alignment inside $0\farcs5$. For this, we simulated the galactic stellar population in a region near the target with \texttt{TRILEGAL} \citep{Girardi_2012}, using their default bulge, halo, disk (thin and thick) parameters and the log-normal initial mass function of \cite{Chabrier_2001}. The probability of an undetected source with $\Delta K \leq 5.55$ inside $0\farcs5$ is less than 0.04\%, so we can safely assume that the transit signal is not produced by a BEB or significantly diluted by a background star.

\subsection{SPIRou velocimetry} \label{sec:nIR_rv}

TOI-1452 was observed at 53 epochs from June 4, 2020 to October 8, 2020 with the near-infrared (0.98--2.5\,$\mu$m) SPIRou spectropolarimeter (\citealt{Donati_2018}; \citeyear{Donati_2020}) mounted on the 3.6\,m Canada-France-Hawaii telescope (CFHT). The observations were conducted as part of the ongoing SPIRou Legacy Survey (SLS; \citealt{Donati_2020}), more precisely its Transit Follow-up program (SLS-WP2), which aims to characterize exoplanets orbiting low mass stars revealed by photometric surveys such as TESS. SLS-WP2 has thus far allowed the characterization of the brown dwarf TOI-1278\,B \citep{Artigau_2021}, the sub-Neptune TOI-1759\,b \citep{Martioli_2022} transiting M dwarfs, and the studies of the transiting planets HD 189733\,b (\citealt{Moutou_2020}; \citealt{Boucher_2021}) and AU Mic\,b \citep{Martioli_2020}. 

SPIRou offers simultaneous high resolution spectroscopy and polarimetry, with a spectral resolving power $R \sim 70\,000$. Each epoch measurement consisted of four consecutive 15-min exposures, i.e., a polarimetric sequence, with two rotating Fresnel rhombs varying positions between the exposures. During such a sequence, the two science fibers, A and B, each receive orthogonal polarization states, giving access to the circular polarization and total intensity of the light beam (Stokes V and I). A total of 212 spectra were collected, with SNR per spectral element ($\sim$2.2\,km/s/pixel for SPIRou) between 20 and 65 (median of 55) near 1.6\,$\mu$m. Four individual spectra were rejected; one due to loss of guiding, three others because of high extinction (clouds). A single polarimetric sequence of four 15-min exposures was also acquired on the $3\farcs2$ companion (TIC 420112587) on April 22, 2021, principally to check its rotation profile and magnetic activity level (see Section~\ref{sec:polarimetry}). The diameter of the SPIRou fiber is $1\farcs33$ and the typical seeing during the observations of TOI-1452 was $0\farcs8$. We measure no correlation between the radial velocity residuals (Keplerian and activity models described in Sect.\ \ref{sec:jointfit}) and the seeing, suggesting that any effect of contamination from the companion was negligible. Per-epoch RV measurements consisted of taking the error weighted mean of the individual observations within a polarimetric sequence. The data analysis presented in Section~\ref{sec:jointfit} was performed on the unbinned RVs, but we show the per-epoch average to facilitate visualization.

The SPIRou data were reduced with \texttt{APERO} v0.7.194 (Cook et al., in prep.). In brief, \texttt{APERO} starts by correcting known H4RG infrared detector defects \citep{Artigau_2018}, then proceeds to identify bad pixels, locate each spectral order on the image, calculate the shape of the instrument pupil slicer \citep{Micheau_2018}, and finally determine from nightly calibration sequences the flat and blaze corrections to apply. Once this preprocessing and calibration step is completed, \texttt{APERO} performs an optimal flux extraction \citep{Horne_1986} in both science channels, separately (fibers A and B) and together (AB), as well as in the simultaneous calibration channel (fiber C). The extracted 2D images (49 orders by 4088 pixels) are then spectral flat fielded, as well as corrected for thermal background and for any leakage from the calibration channel to the science ones. A nightly pixel-to-wavelength solution is applied using a combination of a UNe hollow-cathode lamp and a Fabry-Perot (FP), as described in \cite{Hobson_2021}. \texttt{APERO} uses the simultaneous FP measurements from fiber C to calculate drifts between individual science frames relative to the nightly wavelength solution (typically below 2\,m/s). Finally, a telluric absorption and night-sky emission correction is applied in a two-step process. The science frames are first pre-cleaned with a TAPAS \citep{Bertaux_2014} absorption model that leaves percent-level residuals for deep ($>$50\%) H$_2$O and dry absorption features (e.g., CH$_4$, O$_2$, CO$_2$, N$_2$O and O$_3$). Then, a telluric residuals model with 3 degrees of freedom per pixel (optical depths for the H$_2$O and dry components and a constant) is fitted to the pre-cleaned spectra. The grid of telluric models was generated from a set of rapidly rotating hot stars observed with SPIRou at various airmass, water columns, and dry absorptions, producing telluric corrected spectra with final residuals at the level of the PCA-based method of \cite{Artigau_2014}.

Radial velocity measurements were obtained from the telluric-corrected spectra using the novel line-by-line (LBL) method \citep{Artigau_2022}. The LBL formalism is based on the \cite{Bouchy_2001} framework, in which
Doppler shifts are inferred for individual spectral lines ($\sim$16\,000 for an M dwarf observed with SPIRou) as opposed to a given spectral range.  As in \cite{Bouchy_2001}, such calculations require a noiseless template since velocities are derived from the comparison between the residuals (observed spectrum minus template) and the derivative of the template. For a given observed star, one uses in practice a high SNR combined spectrum as a template, so that any remaining noise is small compared to that of an individual spectrum. For TOI-1452, the combined spectrum produced by \texttt{APERO} did not reach a SNR as high as other bright standard stars observed in the SLS. Moreover, TOI-1452 is located near the North ecliptic pole, meaning its yearly Barycentric Earth Radial Velocity (BERV) variation is small. Our observations with SPIRou covered BERV excursions between 1.7 and 4.8\,km/s, which is not ideal to filter out tellurics lines (i.e., stellar lines do not move a lot with respect to the telluric lines), producing a template that still contains some telluric artefacts. For these reasons, we used the template of Gl 699, a standard star monitored with SPIRou for 2.5\,years with a spectral type (M4V) similar to that of TOI-1452 (M4$\pm$0.5, see Sect.\ \ref{sec:toi_1452}) and a good BERV coverage ($\pm 26$\,km/s). 

For each spectrum, the LBL algorithm combines thousands of per-line velocities into a single RV measurement, with per-line uncertainties varying from 50 m/s for the strongest features to tens of km/s for the shallow ones. This is achieved using a simple mixture model: per-line velocities either originate from a Gaussian distribution centered on the mean velocity, with a standard deviation derived from \cite{Bouchy_2001}, or they arise from another distribution, namely that of high-sigma outliers, whose plausible causes are diverse (persisting bad pixels, cosmic rays, telluric residuals, etc.). Lastly, the LBL RVs are corrected for the instrumental day-to-day drift measured by the FP and for a long-term zero point obtained with a Gaussian process regression using the most observed stars in the SLS. This zero point calibration is similar to \cite{Courcol_2015} for the SOPHIE spectrograph, but will be described in more details in a forthcoming publication (Vandal et al., in prep.). The comparison between the LBL and other methods such as the cross-correlation function and template matching is discussed in \cite{Martioli_2022} and in \cite{Artigau_2022}. The final SPIRou radial velocities of TOI-1452 are listed in Table~\ref{table:spirou_rv}, with typical precision of 8.0\,m/s per exposure, or 4.0\,m/s per epoch.

\subsection{IRD velocimetry} \label{sec:IRD_rv}

Seven high-resolution spectra of TOI-1452 were obtained with the InfraRed Doppler (IRD) spectrograph on the Subaru 8.2\,m telescope \citep{2012SPIE.8446E..1TT, 2018SPIE10702E..11K} between September 26, 2020 and June 25, 2021. IRD covers the near-infrared wavelengths between 970\,nm and 1730\,nm, with a spectral resolution $R \sim 70\,000$. For accurate RV measurements, stellar spectra were obtained simultaneously with the reference spectra of the laser-frequency comb (LFC). The integration times were set to 600--1500 sec, depending on the available observing time slots and sky conditions. The IRD fiber has a $0\farcs48$ diameter, so that flux contamination from the companion star is not an issue.

The raw IRD data were reduced following the standard procedure of \cite{2020PASJ...72...93H}. We extracted wavelength-calibrated one-dimensional spectra for TOI-1452, as well as for the simultaneously injected LFC. The typical SNR of the TOI-1452 extracted spectra was 60--70 per pixel around 1000\,nm. To measure precise RVs for TOI-1452, the reduced spectra were put into the RV analysis pipeline for IRD \citep{2020PASJ...72...93H}. This pipeline fits each small spectral segment of the observed spectra by the forward-modeling technique, taking into account the instantaneous variations of Earth's atmospheric features as well as the instrumental profile of the spectrograph (which is estimated based on each laser-comb spectrum). The seven IRD RV measurements have an overall precision of 4.0\,m/s and are given in Table~\ref{table:spirou_rv}.

\section{Stellar Characterization} \label{sec:stellar_char}

\subsection{TOI-1452 (TIC 420112589)} \label{sec:toi_1452}

The star TOI-1452 (TIC 420112589) is a nearby M dwarf at a distance of 30.504 $\pm$ 0.013 pc \citep{Gaia_Collaboration_2021}. This star does not belong to any known young stellar moving groups, with a very high probability ($>$99.9\%) of being a field star \citep{Gagne_2018}. The presence of flares and short-period sinusoidal signal in the TESS \texttt{PDCSAP} data (see Fig.\ \ref{fig:TESS_lc_phase} and \ref{fig:TESS_lc_complete}) cannot be attributed with certainty to TOI-1452, due to flux contamination from multiple nearby objects. An analysis of the photometric variations is presented in Section~\ref{sec:tess_analysis}, but we note that the polarimetric data from SPIRou reveal no important surface magnetic field variations (see Sect.\ \ref{sec:polarimetry}), suggesting that TOI-1452 is relatively quiet, with a rotation period probably much longer than the modulation seen in the TESS light curve.

As discussed in Section~\ref{sec:stellar_char_spirou}, we measure an effective temperature of $3185 \pm 50$\,K for TOI-1452 using the SPIRou combined spectrum, from which a spectral type (SpT) between M4 and M4.5 is inferred based on Table 5 of \cite{Pecaut_Mamajek_2013}. We also considered the \textit{Gaia} DR2 color to SpT relation of \cite{Kiman_2019}, more specifically the $G$\,--\,$G_{\rm RP}$ relationship, for an independent SpT determination. From this relationship, the \textit{Gaia} magnitudes and their respective uncertainties, we obtain a SpT of M$3.7 \pm 0.6$. The same M4 spectral type was derived through a visual comparison of the SPIRou combined spectrum, degraded to a lower resolution ($R\sim5\,000$), with spectral type standards of the IRTF spectra library (\citealt{Cushing_2005}; \citealt{Rayner_2009}). Considering all these estimates, we adopt a SpT of M4 $\pm$ 0.5.

The mass of TOI-1452 was inferred from the \cite{Mann_2019} absolute $K_{\rm s}$ magnitude ($M_{K_{\rm s}}$) to $M_{\star}$ relation for M dwarfs. Taking into consideration the dispersion of this relation, the $K_{\rm s}$ magnitude, the distance, and their corresponding uncertainties, a mass of $0.249 \pm 0.008$\,M$_{\odot}$ is obtained. A similar approach was used to measure the stellar radius, this time using the $M_{K_{\rm s}}$--$R_{\star}$ relationship of \cite{Mann_2015}, from which we derive $R_{\star} = 0.275 \pm 0.009$\,R$_{\odot}$. Other physical parameters such as the surface gravity ($\log g$), the mean density ($\rho_{\star}$), and the luminosity ($L_{\star}$) were determined from the $M_{\star}$, $R_{\star}$, and $T_{\rm eff}$ estimates. The stellar parameters of TOI-1452 are summarized in Table~\ref{table:stellarparams}.

\begin{deluxetable}{ccc}
\tablecaption{TOI-1452 stellar properties}
\tablehead{
\colhead{Parameter} & \colhead{Value} & \colhead{Ref.}
}
\startdata
\multicolumn{3}{c}{\textit{Designations}}\\
TIC & 420112589 & 1\\
TOI & 1452 & 1\\
2MASS & J19204172+7311434 & 2\\
UCAC4 & 816-023943 & 3\\
\textit{Gaia} EDR3 & 2264839957167921024 & 4\\
\hline
\multicolumn{3}{c}{\textit{Astrometry}}\\
RA (J2016.0) & 19:20:41.75 & 4\\
DEC (J2016.0) & +73:11:42.35 & 4\\
$\mu_{\alpha} \cos \delta$ (mas/yr) & 7.800 $\pm$ 0.017 & 4\\
$\mu_{\delta}$ (mas/yr) & -74.076 $\pm$ 0.017 & 4\\
$\pi$ (mas) & 32.7823 $\pm$ 0.0140 & 4\\
Distance (pc) & 30.5043 $\pm$ 0.0130 & 4\\
\hline
\multicolumn{3}{c}{\textit{Stellar parameters}}\\
$T_{\rm eff}$ (K) & 3185 $\pm$ 50 & 5\\
SpT & M4 $\pm$ 0.5 & 5\\
$\left[ {\rm M/H} \right]$ & $-0.07$ $\pm$ 0.02 & 5\\
$M_{\star}$ (M$_{\odot}$) & 0.249 $\pm$ 0.008 & 5\\
$R_{\star}$ (R$_{\odot}$) & 0.275 $\pm$ 0.009 & 5\\
log $g$ (dex) & 4.95 $\pm$ 0.03 & 5\\
$\rho_{\star}$ (g/cm$^3$) & 16.8 $\pm$ 1.9 & 5\\
$L_{\star}$ (L$_{\odot}$) & 0.0070 $\pm$ 0.0006 & 5\\
\hline
\multicolumn{3}{c}{\textit{Photometry}}\\
$B$ & 15.94 $\pm$ 0.03 & 1\\
$V$ & 14.35 $\pm$ 0.12 & 1\\
$G_{\rm BP}$ & 15.222 $\pm$ 0.004 & 4\\
$G$ & 13.598 $\pm$ 0.003 & 4\\
$G_{\rm RP}$ & 12.362 $\pm$ 0.004 & 4\\
$T$ & 12.295 $\pm$ 0.007 & 1\\
$g$ & 15.580 $\pm$ 0.002 & 6\\
$r$ & 14.383 $\pm$ 0.007& 6\\
$i$ & 12.873$^*$ & 6\\
$z$ & 12.272$^*$ & 6\\
$y$ & 11.875 $\pm$ 0.020& 6\\
$J$ & 10.604 $\pm$ 0.058 & 2\\
$H$ & 10.026 $\pm$ 0.058 & 2\\
$K_{\rm s}$ & 9.740 $\pm$ 0.046 & 2\\
$W1$ & 8.938 $\pm$ 0.023$^\dag$ & 7\\
$W2$ & 8.760 $\pm$ 0.019$^\dag$ & 7\\
$W3$ & 8.686 $\pm$ 0.023$^\dag$ & 7\\
$W4$ & 8.46 $\pm$ 0.29$^\dag$ & 7\\
\enddata
\tablecomments{$^*$The uncertainty was not indicated.\\ 
$^\dag$WISE magnitudes include the flux from TOI-1452 and TIC 420112587.}
\tablerefs{(1) TIC \citep{Stassun_2019}. (2) 2MASS \citep{Skrutskie_2006}. (3) UCAC4 \citep{Zacharias_2013}. (4) \textit{Gaia} EDR3 \citep{Gaia_Collaboration_2021}. (5) This work. (6) Pan-STARRS1 DR2 \citep{Chambers_2016}. (7)\ AllWISE \citep{Wright_2010}.}
\label{table:stellarparams}
\end{deluxetable}

\subsection{Bound companion (TIC 420112587)} \label{sec:bound}

TOI-1452 has a resolved companion (TIC 420112587) with several comparable photometric and astrometric measurements. The two objects have similar \textit{Gaia} EDR3 magnitudes of \hbox{$G = 13.598 \pm 0.003$} and \hbox{$G = 13.830 \pm 0.003$} for \hbox{TOI-1452} and TIC 420112587 respectively. Their \textit{Gaia} EDR3 parallaxes are identical (within the errors), $32.782 \pm 0.014$\,mas for TOI-1452 and $32.791 \pm 0.014$\,mas for TIC 420112587, indicating a very similar distance to these stars. Their projected angular separation is $3\farcs182$, which corresponds to a projected physical separation of $\sim$97\,au, using a common approximate distance of 30.5\,pc. The proper motion of TIC 420112587 is similar to that of TOI-1452, with $\mu_{\alpha} \cos \delta$ and $\mu_{\delta}$ within 15\% for the two stars (see Tables~\ref{table:stellarparams} and \ref{table:stellarparams2}). 
TOI-1452 and TIC 420112587 most likely form a visual binary, i.e., a resolved gravitationally bound system, which was previously reported in the TOI visual-binary catalog of \cite{Mugrauer_2020}, as well as in the binary catalog based on \textit{Gaia} EDR3 of \cite{El-Badry_2021}. From our single visit on the companion star with SPIRou, we measure an RV offset between TOI-1452 and TIC 420112587 of $-6.9$\,km/s. The SPIRou template spectrum of TIC 420112587 combining only four individual spectra at the same epoch and BERV does not allow for a similar spectral analysis as the one presented for TOI-1452 in Section~\ref{sec:stellar_char_spirou}.

Using the empirical relationships of \cite{Mann_2015, Mann_2019}, we obtain a mass of 0.226 $\pm$ 0.006\,M$_{\odot}$ and a radius of 0.254 $\pm$ 0.008\,R$_{\odot}$ for TIC 420112587. The mass ratio of the binary system is close to unity ($q = 0.91 \pm 0.04$), with TOI-1452 as the primary member. 
The projected physical separation and masses of the two stars imply an orbital period of about 1400 years. The radial velocity variation expected from such an orbital motion and for a circular orbit is under $\sim$1.5\,m/s over the span of our SPIRou RV observations. According to Table 5 of \cite{Pecaut_Mamajek_2013} and a $T_{\rm eff} = 3060 \pm 50$\,K derived from the spectral energy distribution (see analysis below), TIC 420112587 has an M5 spectral type. The \cite{Kiman_2019} $G$\,--\,$G_{\rm RP}$ relationship yields a SpT of M$4.0\pm0.6$, so we adopt an intermediate spectral type of M4.5 $\pm$ 0.5. A summary of the stellar properties of TIC 420112587 is presented in Table~\ref{table:stellarparams2}.

\begin{deluxetable}{ccc}
\tablecaption{TIC 420112587 stellar properties}
\tablehead{
\colhead{Parameter} & \colhead{Value} & \colhead{Ref.}
}
\startdata
\multicolumn{3}{c}{\textit{Designations}}\\
TIC & 420112587 & 1\\
TOI & 1760 & 1\\
2MASS & J19204172+7311467 & 2\\
\textit{Gaia} EDR3 & 2264839952875245696 & 3\\
\hline
\multicolumn{3}{c}{\textit{Astrometry}}\\
RA (J2016.0) & 19:20:41.76 & 3\\
DEC (J2016.0) & +73:11:45.53 & 3\\
$\mu_{\alpha} \cos \delta$ (mas/yr) & 6.845 $\pm$ 0.017 & 3\\
$\mu_{\delta}$ (mas/yr) & -82.216 $\pm$ 0.017 & 3\\
$\pi$ (mas) & 32.7913 $\pm$ 0.0141 & 3\\
Distance (pc) & 30.4959 $\pm$ 0.0131 & 3\\
\hline
\multicolumn{3}{c}{\textit{Stellar parameters}}\\
$T_{\rm eff}$ (K) & 3060 $\pm$ 50 & 4\\
SpT & M4.5 $\pm$ 0.5 & 4\\
$M_{\star}$ (M$_{\odot}$) & 0.226 $\pm$ 0.006 & 4\\
$R_{\star}$ (R$_{\odot}$) & 0.254 $\pm$ 0.008 & 4\\
log $g$ (dex) & 4.98 $\pm$ 0.03 & 4\\
$\rho_{\star}$ (g/cm$^3$) & 19.5 $\pm$ 1.8 & 4\\
$L_{\star}$ (L$_{\odot}$) & 0.0051 $\pm$ 0.0005 & 4\\
\hline
\multicolumn{3}{c}{\textit{Photometry}}\\
$B$ & 15.76 $\pm$ 0.17 & 1\\
$V$ & 13.99 $\pm$ 0.2 & 1\\
$G_{\rm BP}$ & 15.512 $\pm$ 0.005 & 3\\
$G$ & 13.830 $\pm$ 0.003 & 3\\
$G_{\rm RP}$ & 12.576 $\pm$ 0.004 & 3\\
$T$ & 12.499 $\pm$ 0.008 & 1\\
$g$ & 15.890 $\pm$ 0.002 & 5\\
$r$ & 14.659 $\pm$ 0.003 & 5\\
$i$ & 13.153 $\pm$ 0.002 & 5\\
$z$ & 12.456 $\pm$ 0.021 & 5\\
$y$ & 12.111 $\pm$ 0.007 & 5\\
$J$ & 10.795 $\pm$ 0.027 & 2\\
$H$ & 10.257 $\pm$ 0.031 & 2\\
$K_{\rm s}$ & 9.944 $\pm$ 0.023 & 2\\
\enddata
\tablerefs{(1) TIC \citep{Stassun_2019}. (2) 2MASS \citep{Skrutskie_2006}. (3) \textit{Gaia} EDR3 \citep{Gaia_Collaboration_2021}. (4) This work. (5) Pan-STARRS1 DR2 \citep{Chambers_2016}.}
\label{table:stellarparams2}
\end{deluxetable}

\subsection{Spectral energy distribution fit} \label{sec:sed_fit}

\begin{figure}[ht!]
\centering
    \includegraphics[width=1\linewidth]{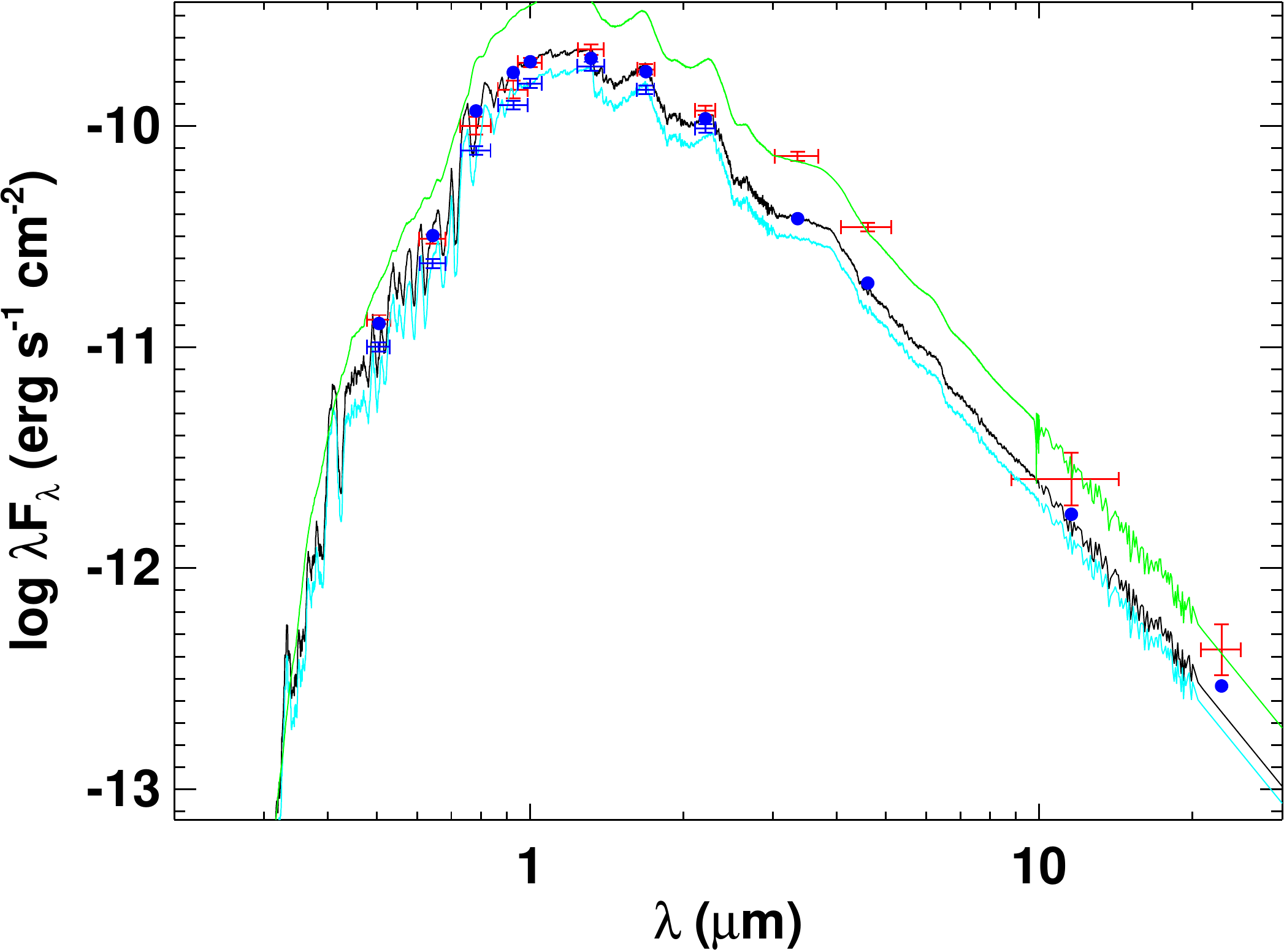}
    \linespread{1}
  \caption{Spectral Energy Distributions (SED) of TOI-1452 and TIC 420112587 fitted with a NextGen atmospheric model, respectively depicted with a black and a cyan line. Photometric measurements are represented with red (TOI-1452) and blue (TIC 420112587) error bars, where the horizontal bars represent the effective widths of the bandpasses and the small blue dots represent the model fluxes of TOI-1452 for comparison to the red symbols. The WISE measurements (3.4, 4.6, 12, and 22\,$\mu$m), excluded from this analysis for including both star fluxes, agree with the sum of the two SED (green). %\todo{Blue points in red to match the color of TOI-1452 photometric bands?}
  }
  \label{fig:SED}
\end{figure}

As an independent determination of the basic stellar parameters, as well as to estimate the contaminating flux from the nearby companion star, we performed an analysis of the broadband spectral energy distribution (SED) of the stars together with the \textit{Gaia}  EDR3 parallax \citep[with no systematic offset applied;  e.g.,][]{Stassun_2021}, following the procedures described in \cite{Stassun_2016} and \cite{Stassun_2017,Stassun_2018a}. For both stars, we pulled the $JHK_S$ magnitudes from 2MASS, the W1--W4 magnitudes from WISE, and the $grizy$ magnitudes from Pan-STARRS. Together, the available photometry spans the full stellar SED over the wavelength range 0.4--10~$\mu$m (see Figure~\ref{fig:SED}). We excluded the WISE photometry from the initial fitting because the two stars are blended in WISE, such that the catalog photometry in fact represents the sum of the fluxes of both stars. 

For each star, we performed a fit using NextGen stellar atmosphere models \citep{Hauschildt_1999}, with the free parameters being the effective temperature ($T_{\rm eff}$) and metallicity ([Fe/H]). The remaining free parameter is the extinction $A_V$, which we fixed at zero due to the stars' proximity. The resulting fit for TOI-1452 (Figure~\ref{fig:SED}) has a reduced $\chi^2$ of 1.8 with $T_{\rm eff} = 3100 \pm 50$~K and [Fe/H] = $0.0 \pm 0.5$. Integrating the model SED gives the bolometric flux at Earth, $F_{\rm bol} = 2.34 \pm 0.11 \times 10^{-11}$ erg~s$^{-1}$~cm$^{-2}$. Taking the $F_{\rm bol}$ and $T_{\rm eff}$ together with the {\it Gaia\/} parallax, gives the stellar radius, $R_\star = 0.286 \pm 0.011$~R$_\odot$. This independent radius measurement is consistent, although slightly less precise, with the one derived using \citealt{Mann_2015} ($R_\star = 0.275 \pm 0.009$~R$_\odot$).

Similarly, the resulting parameters for the companion star from the SED fit are $T_{\rm eff} = 3060 \pm 50$\,K, [Fe/H]~$ = 0.0 \pm 0.5$, and $R_\star = 0.263 \pm 0.010$\,R$_\odot$. This radius estimate is again fully consistent with the value derived from empirical relation ($R_\star = 0.254 \pm 0.008$\,R$_\odot$). The sum of the two stellar models is compared to the combined WISE fluxes in Figure~\ref{fig:SED}, showing good agreement. Integrating the companion SED within the TESS bandpass yields a flux ratio (companion relative to TOI-1452) of $0.77 \pm 0.03$. Note that the flux ratio derived strictly from the $T$ magnitudes from the TIC is $0.829 \pm 0.002$. In the event that the \texttt{PDCSAP} overestimated the dilution correction for TIC 420112587, this difference in flux ratio would imply a $\sim$1.7\% overestimation of the planetary radius.

\subsection{Stellar parameters from SPIRou spectra} \label{sec:stellar_char_spirou}

\begin{figure}
    \centering
    \includegraphics[width=1\linewidth]{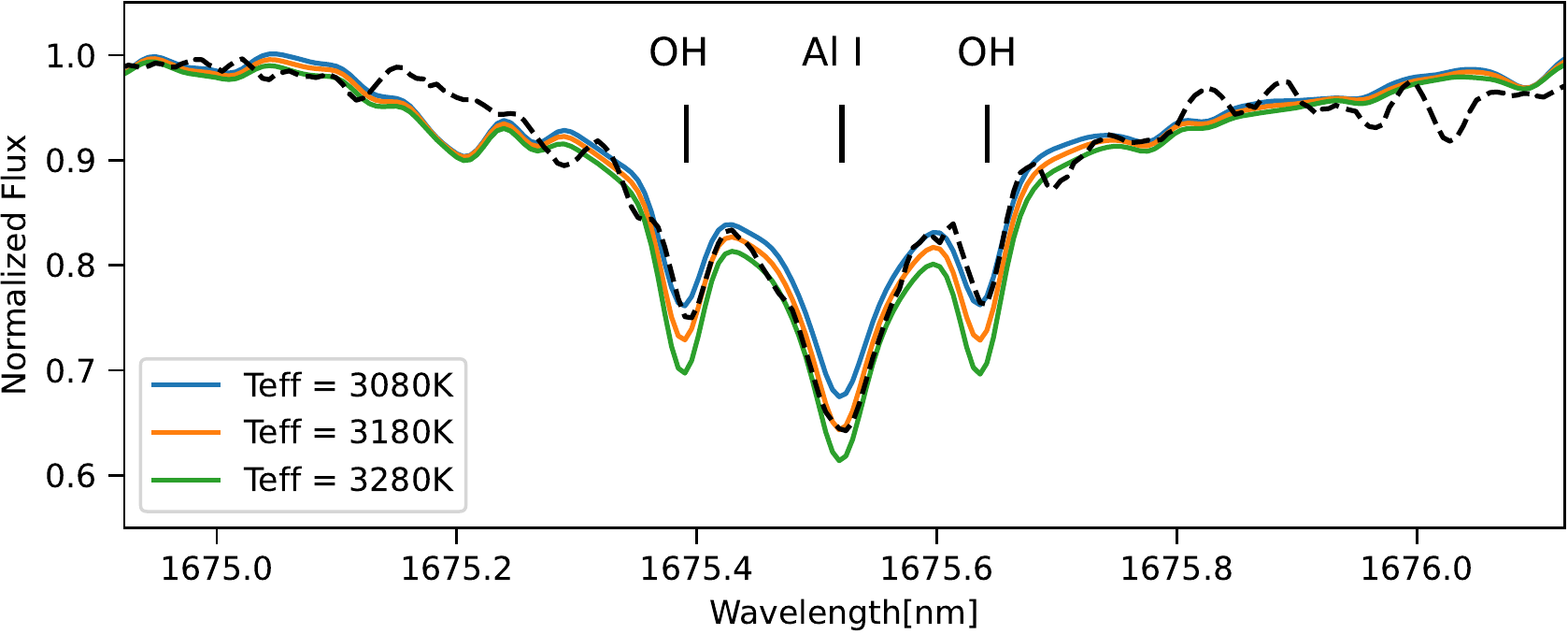}
    \includegraphics[width=1\linewidth]{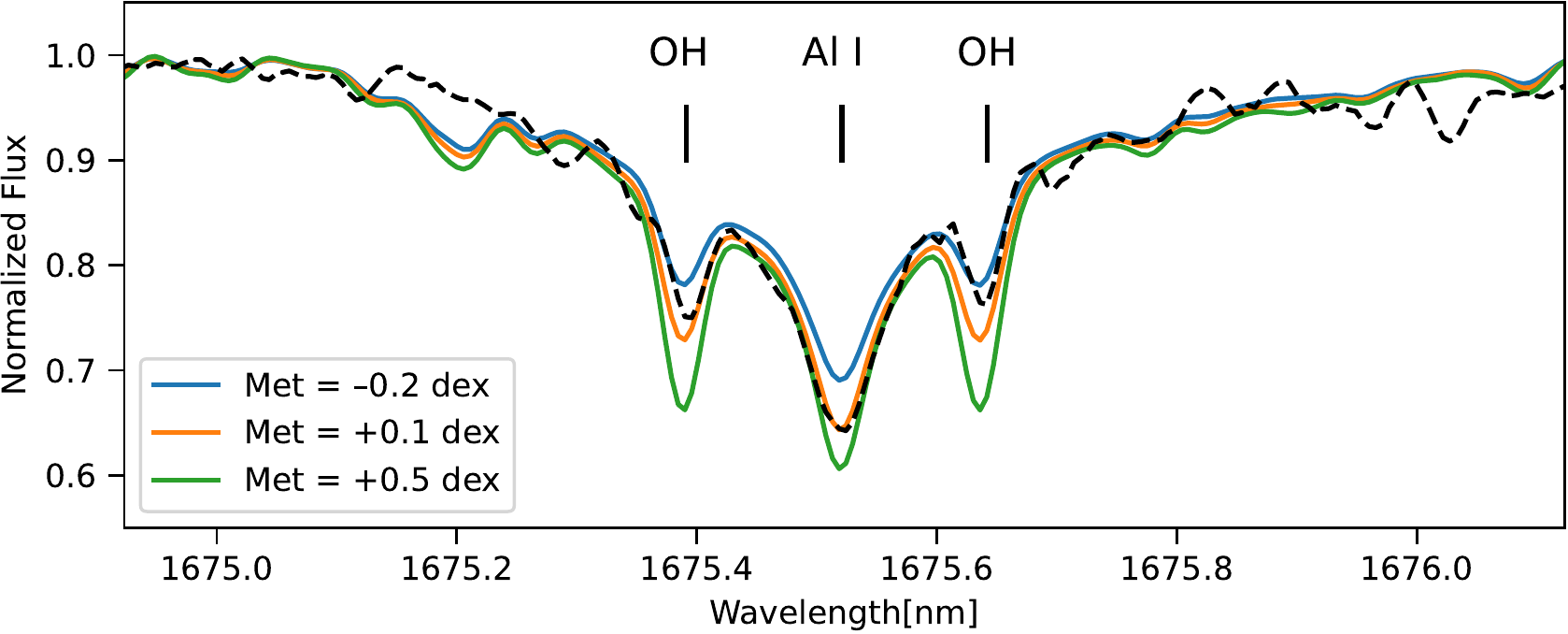}
    \caption{SPIRou observation of the Al I line (1675.514\,nm) of TOI-1452 (black dashed line). \textit{Top panel}: The solid lines represent the ACES models for a fixed metallicity of 0.1\,dex and $T_{\rm eff}$ values of 3060\,K, 3160\,K, and 3260\,K.  \textit{Bottom panel}: Same as top panel, but the ACES models have a fixed $T_{\rm eff} = 3160$\,K and metallicity values of $-0.2$\,dex, 0.1\,dex and 0.5\,dex. These plots illustrate  the good sensitivity of near-IR high-resolution spectroscopy for constraining both the metallicity and effective temperature of M dwarfs.}
    \label{fig:teff_met}
\end{figure}

The high-resolution combined spectrum of TOI-1452 from SPIRou lets us determine $T_{\rm eff}$ and the abundance of several elements with relatively good accuracy. This work follows the methodology of Jahandar et al.\ (in prep.), which we briefly summarize here. Because models and observations can often show significant discrepancies (e.g., continuum mismatch in the $Y$ and $J$ bands), we only select for the fitting analysis a subset of relatively strong lines that are matching the models. The selected lines are then divided into several groups of 15 lines, each analyzed independently through a chi-squared fitting routine to infer both $T_{\rm eff}$ and [M/H] for all groups. The spectrum is compared with a grid of ACES models (\citealt{allard2012models}; \citealt{husser2013new}). The advantage of this method is that it yields several (typically 15) independent measurements that can be used to characterize the inherent uncertainties associated with the fitting procedure. This analysis applied to the TOI-1452 spectrum yields $T_{\rm eff} = 3185\pm50$\,K and [M/H]=$-0.07\pm0.02$, in good agreement with the parameters derived from the SED fitting analysis. The quoted uncertainty for $T_{\rm eff}$ is internal to our fitting methodology and ignore potential systematic differences with bolometric $T_{\rm eff}$ estimates based on interferometric measurements. While our $T_{\rm eff}$ estimates have yet to be calibrated with bolometric $T_{\rm eff}$, it is empirically demonstrated that 50--60\,K is a typical uncertainty derived from atmosphere models (e.g., \citealt{Mann2013}; \citeyear{Mann_2015}). We thus adopt 50\,K for our $T_{\rm eff}$ uncertainty, a conservative value given that the temperature derived from the SPIRou spectrum is inferred from several independent measurements. An illustration of the temperature and abundance sensitivity for an Al I line (at 1675.514\,nm) is shown in Figure~\ref{fig:teff_met}. In practice, several tens of lines are used to derive $T_{\rm eff}$. 

\begin{figure}
    \centering
    \includegraphics[width=1\linewidth]{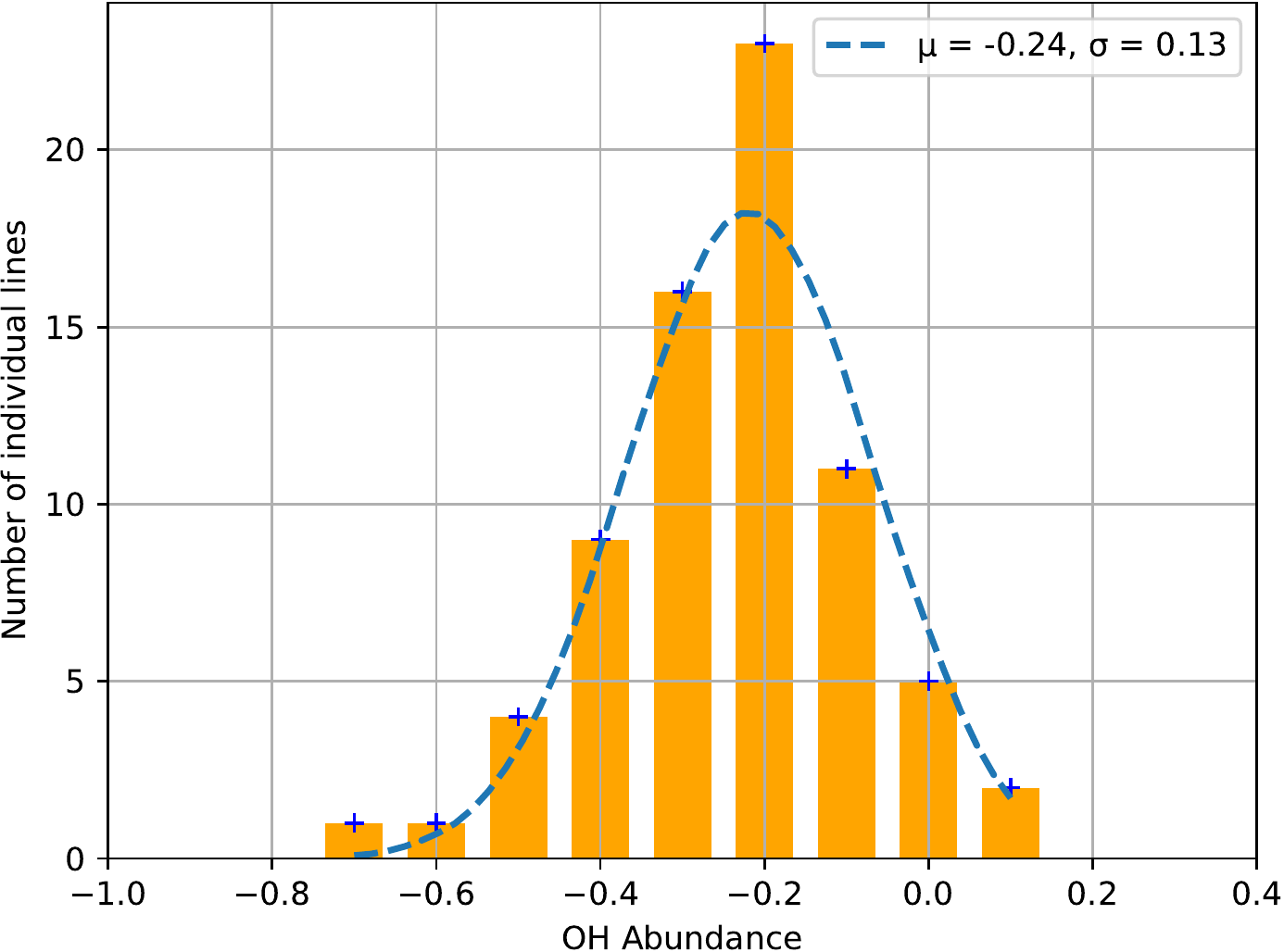}
    \caption{Distribution of OH abundance measurements from 72 isolated OH lines in the SPIRou combined spectrum of TOI-1452. The inherent, per-line uncertainty (0.13 dex) is inferred from a Gaussian fit of this distribution. This uncertainty does not take into account possible systematic errors of the stellar atmosphere models.}
    \label{fig:OHdist}
\end{figure}

\begin{figure}
    \centering
    \includegraphics[width=1\linewidth]{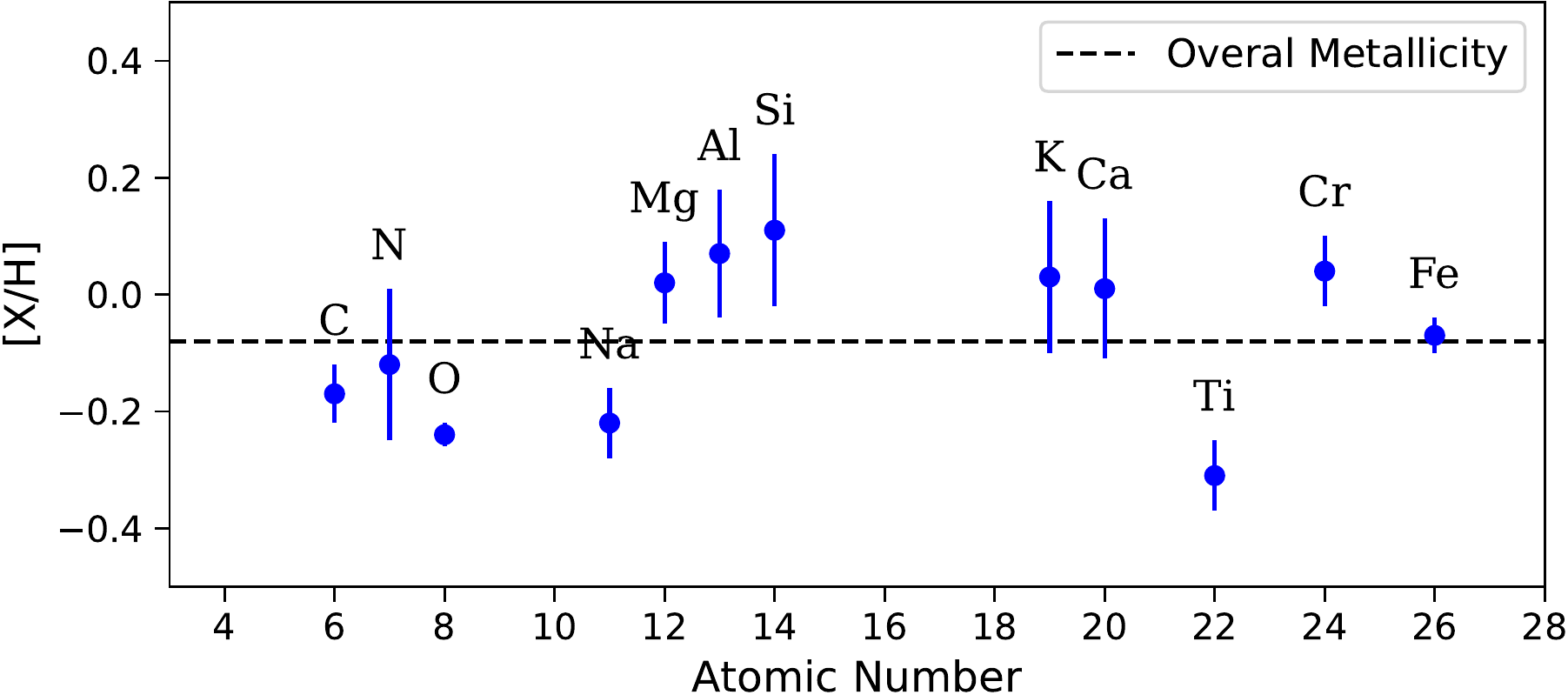}
    \caption{Chemical abundances of 12 different elements via line-by-line spectroscopy of 151 spectral features. The dashed line represents the average metallicity of the star corresponding to the average of all elements.}
    \label{fig:abundances}
\end{figure}

Once $T_{\rm eff}$ is determined, one can then proceed, through a similar procedure, to determine the abundance of all individual lines of a given element.
The high-resolution near-infrared spectrum of an M dwarf is characterized by several hundreds of relatively strong OH lines. By selecting only those that are well isolated, i.e., with no known spectral features within a few pixels using the PHOENIX/BT-Settl (\citealt{Allard_2012}; \citeyear{Allard_2013}) and NIST \citep{Ralchenko_2010} line lists,
we find 72 OH lines, whose individual abundance can be used to quantify the inherent, per-line uncertainty of this method. This uncertainty obviously does not consider any possible systematic errors associated with the ACES atmosphere models. 
The 72 independent OH abundance measurements are presented in Figure~\ref{fig:OHdist}, showing a good match with a Gaussian distribution with standard deviation 0.13\,dex. 
For all elements and molecules detected in TOI-1452, we list the average abundance of all lines in Table~\ref{table:abundances} (see also Figure~\ref{fig:abundances}). For chemical species with only one line, we adopt an uncertainty of 0.13\,dex from the OH distribution. We report abundances for Fe, Mg and Si that constitute the bulk material of an exoplanet core and mantle. The overall metallicity ([M/H]) and its corresponding error are determined by averaging the final abundance of each element in Table~\ref{table:abundances}, assuming a common uncertainty for all elements taken as the median of all individual uncertainties. This approach is chosen to avoid putting too much weight on the oxygen abundance characterized by a small uncertainty.

\begin{deluxetable}{cccc}
\tablecaption{Stellar abundance of TOI-1452 for various chemical species measured by SPIRou}
\tablehead{
\colhead{Element} & \colhead{[X/H]} & \colhead{$\sigma$} & \colhead{\# of lines}}
\startdata
Fe I      & $-0.07$               & 0.03  & 38           \\
Al I      & 0.07                  & 0.11  & 4           \\
Mg I      & $0.02$                & 0.07  & 5        \\
Si I      & $0.11$                & 0.13  & 1         \\
Ti I      & $-0.31$               & 0.06  & 10           \\
Ca I      & $0.01$                & 0.12  & 2        \\
Cr I      & $0.04$                & 0.06  & 4           \\
K I       &  0.03                 & 0.13  & 1           \\
O I$^*$   & $-0.24$               & 0.02  & 72         \\
C I       & $-0.17$               & 0.05  & 11           \\
N I       & $-0.12$               & 0.13  & 1           \\
Na I      & $-0.22$               & 0.06  & 2           \\
$<>^{\dag}$         & $-0.07$               & 0.02  & -- 
\enddata
\tablecomments{$^*$The oxygen abundance is inferred from OH lines.\\ $^{\dag}$Average abundance of all elements.}
\label{table:abundances}
\end{deluxetable}

\subsection{Spectropolarimetry with SPIRou} \label{sec:polarimetry}

The combination of the four exposures within a polarimetric sequence obtained with SPIRou yields the circular polarization profile at the surface of the star \citep{Donati_2020}. The intensity (Stokes I), circular (Stokes V), and null polarization spectra were generated in \texttt{APERO} following the \texttt{spirou-polarimetry} code\footnote{\href{https://github.com/edermartioli/spirou-polarimetry}{\texttt{github.com/edermartioli/spirou-polarimetry}}}. We applied the Least-Square Deconvolution (LSD) method of \cite{Donati_1997}, also outlined in \cite{Martioli_2020}, to compute the average I and V profiles. We used the VALD database \citep{Piskunov_1995} and a MARCS atmosphere model \citep{Gustafsson_2008} with $T_{\rm eff} = 3000$\,K and log $g$ = 5.0\,dex to search for valid atomic features. Lines deeper than 3\,\% and with a known Landé factor were selected to produce a line mask of 955 atomic lines, used in this LSD analysis of TOI-1452. An estimate of the longitudinal magnetic field ($B_{\ell}$) at the stellar surface can then be obtained using Equation\,5 of \cite{Donati_1997}, combining the Stokes I and V LSD profiles, the mean Landé factor of 1.24, and the mean wavelength of 1604.59\,nm. By doing this over multiple epochs, one can monitor the large-scale surface magnetic field, expected to vary with the rotation of the star. The polarimetric capabilities of SPIRou can thus serve as a useful activity tracer simultaneous to the RV measurements, as demonstrated in \cite{Martioli_2022}, where the rotation period of the moderately active M0 star TOI-1759 ($P_{\rm rot} = 35.65^{+0.17}_{-0.15}$\,days) was determined from the $B_{\ell}$ time series. We obtained independent and consistent values for the $B_{\ell}$ of TOI-1452 using the Libre-Esprit pipeline (\citealt{Donati_1997}, \citealt{Donati_2020}), but present below the values from the \texttt{APERO} pipeline.
 
\begin{figure}[ht!]
\centering
    \includegraphics[width=0.9\linewidth]{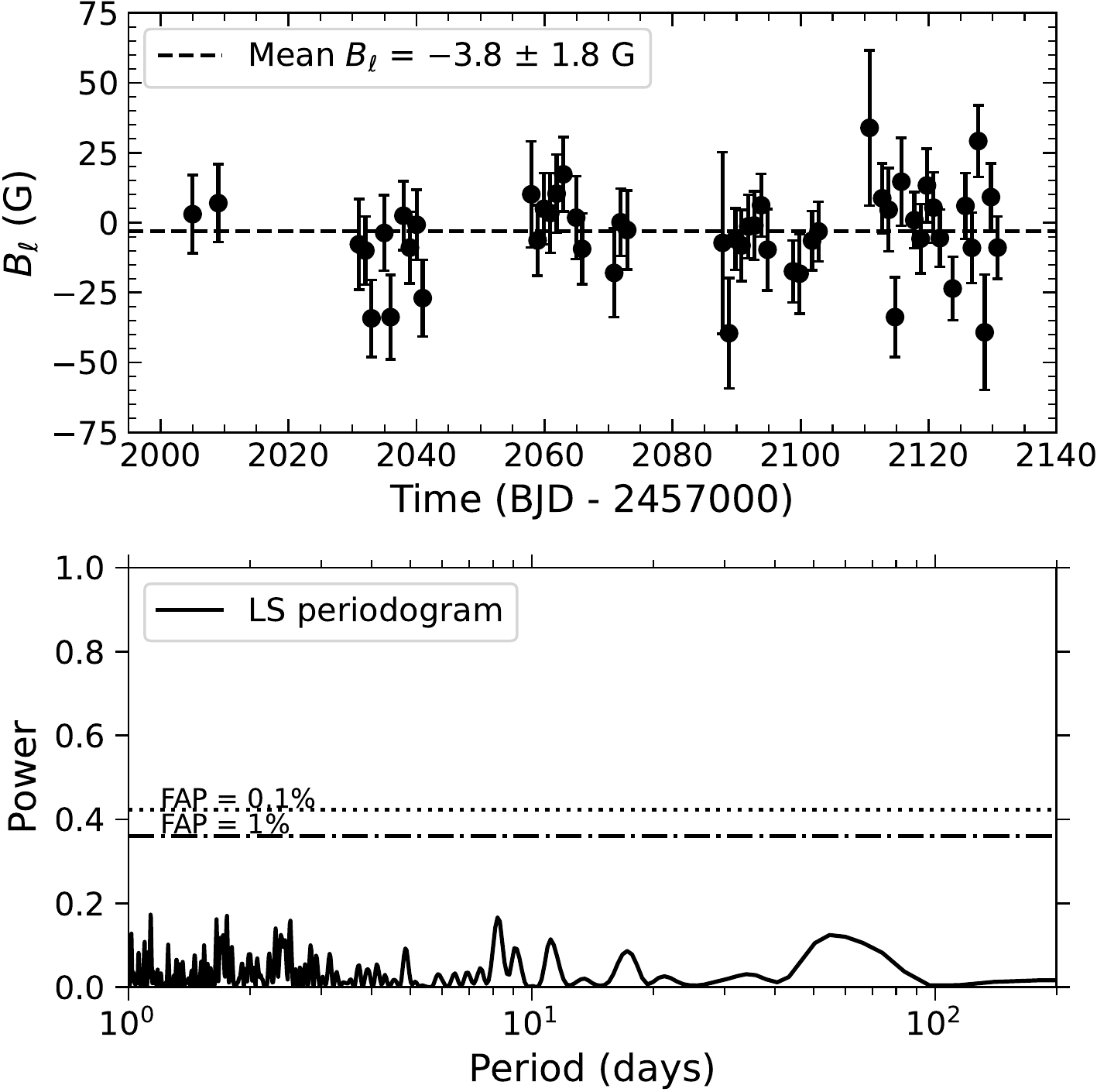}
    \linespread{1}
  \caption{\textit{Top panel}: Monitoring of the longitudinal magnetic field ($B_{\ell}$) of TOI-1452 with SPIRou. \textit{Bottom panel}: Lomb-Scargle periodogram of the $B_{\ell}$ time series. No clear periodic signal is detected, suggesting a relatively quiet star.}
  \label{fig:Blong}
\end{figure}

The $B_{\ell}$ time series of TOI-1452 is presented in Figure~\ref{fig:Blong}. A simple Lomb-Scargle periodogram analysis shows no obvious periodicity. The $B_{\ell}$ data do not favor a sinusoidal model, which could be associated with stellar rotation, over a constant magnetic field (mean $B_{\ell} = -3.8 \pm 1.8$\,G). The small variation of $B_{\ell}$ suggests that the field is intrinsically weak (quiet star), or that it is strongly axisymmetric with respect to the rotation axis. Alternatively, the rotation period of TOI-1452 could be longer than the 4-month span of our observations, but this is close to the largest known period for M dwarfs \citep{Newton_2018}.

This LSD analysis was also applied on the companion star TIC 420112587 using the single polarimetric sequence acquired with SPIRou. We report a polarimetric signal and a $B_{\ell}$ consistent with a null value, indicating that the companion is also probably inactive.

\section{Data Analysis \& Results} \label{sec:analysis}

\subsection{Determining the transit origin with PSF photometry} \label{sec:psf}

The objective of the first OMM-PESTO transit follow-up was to establish the origin of the TESS signal, particularly between the target (TOI-1452) and its companion (TIC 420112587). Standard aperture photometry ruled out any NEB in the FOV, but was unable to isolate the transit between the two stars, as they were only partially resolved. We therefore had to rely on a different method using point spread function (PSF) fitting
to extract the relative flux of both stars. We used the  \texttt{photutils} \citep{Larry_Bradley_2020} package to perform the DAOPHOT \citep{Stetson_1987} PSF photometry algorithm. This was achieved by fitting the PSFs with an effective PSF model (ePSF) generated in \texttt{photutils} using the 6 stars with the highest SNR in the FOV (excluding our targets), then integrating the best-fit models over pixels containing the stars' signal.

\begin{figure}[h]
 \begin{center}
    \includegraphics[width=1\linewidth]{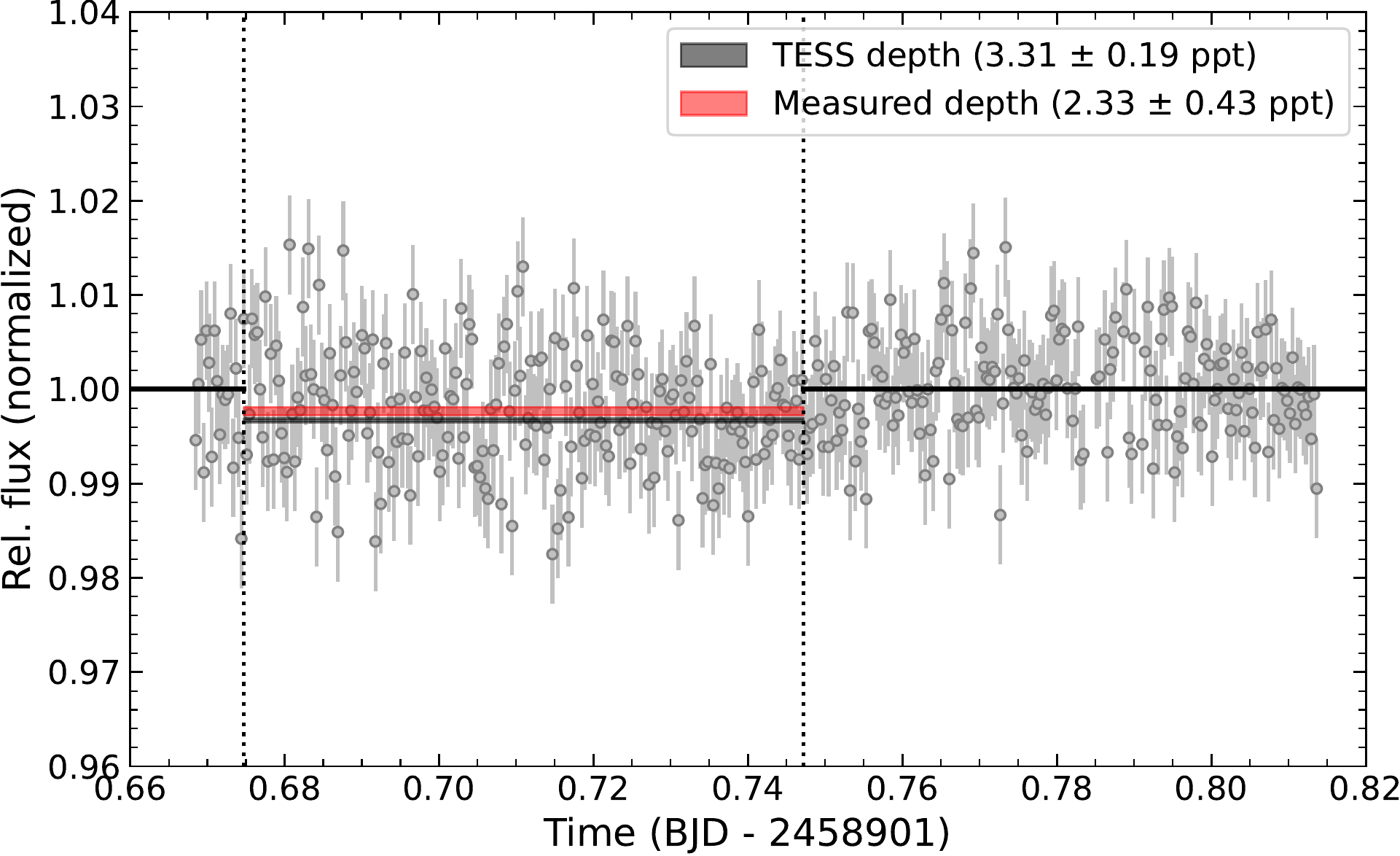}
  \caption{PSF photometry relative light curve (TOI-1452 to TIC 420112587 flux ratio) from OMM-PESTO on 2020-02-22. The dotted black lines represent the ingress and egress of the transit estimated from contemporaneous TESS sector 22 data. The light curve is normalized with the out-of-transit median. A flux deficit during transit with a depth comparable to TESS is detected on TOI-1452.}
  \label{fig:PSFphot}
 \end{center}
\end{figure}

The intent here was not to produce a precise uncontaminated light curve, but rather to detect any flux deficit (or excess) that would indicate from which star the transit originates. We thus inspected the TOI-1452 to TIC 420112587 flux ratio as a function of time, normalized to unity outside of transit. The resulting relative light curve is presented in Figure~\ref{fig:PSFphot} and shows a flux deficit on TOI-1452 during transit. We did not fit a transit model on this light curve, as it is less precise than the one obtained using a combined circular aperture (Fig.\ \ref{fig:OMM-PESTO_lc}). We nonetheless measure a mean relative flux deficit of $2.33 \pm 0.43$\,ppt, which is an approximation of the uncontaminated transit depth. This flux deficit is comparable in amplitude to the diluted corrected TESS depth ($3.31 \pm 0.19$\,ppt) and was detected with a confidence level sufficiently high ($>$5$\sigma$) to conclude that TOI-1452 was the source of the transit and justify an RV monitoring campaign on this star, starting with SPIRou in June 2020. Later, the MuSCAT3 photometry was able to resolve TOI-1452 and TIC 420112587 and unambiguously identify that the former star hosts a transiting object.

\subsection{TESS light curve analysis} \label{sec:tess_analysis}

The TOI-1452 \texttt{PDCSAP} light curve (Fig.\ \ref{fig:TESS_lc_phase} and \ref{fig:TESS_lc_complete}) features stellar flares with amplitude of a few percents and ppt-level sinusoidal variations. A strong peak at 0.93\,days appears in the Lomb-Scargle periodogram of the multi-year light curve, as well as in all individual sectors. However, computing the autocorrelation function, which is more reliable for accurate photometric rotation period determination \citep{McQuillan_2013}, would often find a period of 1.9\,days (2$\times$0.93\,days) depending on the sector. Since the \texttt{PDCSAP} data are corrected for systematic trends, it is unlikely that such corrections significantly perturb those short-term flares and sinusoids. Regardless of the origin of these signals (TOI-1452, TIC 420112587, or any contaminating star), it is crucial to remove the periodic variations to accurately measure the transit parameters. To accomplish this, we adopted a sequential approach where we first correct the \texttt{PDCSAP} data using a Gaussian Process (GP), then fit the 32 corrected transits with a model. The details of the GP regression are presented below, while the transit modeling is described in Section~\ref{sec:jointfit}. 

We started by masking the epochs of transit and removing outliers from the \texttt{PDCSAP} light curve with sigma clipping. It was determined that a 3.5\,$\sigma$ clipping was robust enough to remove both obvious outliers and stellar flares. This sigma clipping removed less than 0.2\% of the out-of-transit data. Parts of sectors 21 and 47 coinciding with TESS momentum dump events show large amplitude variations; those were considered to be non-astrophysical and were manually rejected. We also rejected data points in sectors 40 and 41, as they are isolated and have a median considerably different than unity. The data not considered in this analysis are either displayed in blue (transits) or in red (rejected) in Figures~\ref{fig:TESS_lc_phase} and \ref{fig:TESS_lc_complete}. The \textit{cleaned} out-of-transit \texttt{PDCSAP} dataset was too large ($N = 237\,634$) to be efficiently modeled with a GP. We therefore binned the data and instead used the corresponding 1-hour effective cadence light curve ($N = 7924$).

The GP regression was done with \texttt{celerite2} (\citealt{celerite1_2017}; \citeyear{celerite2_2018}). We selected its \texttt{RotationTerm} kernel because it was specifically designed to model a range of quasi-periodic variability, from stellar rotation to pulsations. This kernel is the sum of two stochastically-driven, damped harmonic oscillator (SHO) terms (\texttt{SHOTerm}) capturing both primary ($P_{\rm GP}$) and secondary ($P_{\rm GP}/2$) modes in Fourier space. The Fourier transform of the covariance function, known as the power spectral density (PSD), takes the following form:
\begin{equation}
\begin{split}
S(\omega) = \sqrt{\frac{2}{\pi}} & \frac{S_1\,\omega_1^4}        {(\omega^2-\omega_1^2)^2 + \omega_1^2\,\omega^2/Q_1^2}\\
& + \sqrt{\frac{2}{\pi}} \frac{S_2\,\omega_2^4} {(\omega^2-\omega_2^2)^2 + \omega_2^2\,\omega^2/Q_2^2}
\end{split}
\end{equation}
where each \texttt{SHOTerm} PSD is described by their respective power $S_{1}$, $S_{2}$ at $\omega = 0$, their undamped angular frequency $\omega_{1}$, $\omega_{2}$, and their own quality factor $Q_{1}$, $Q_{2}$. 

Since the periods of the two oscillators are separated by a factor of 2 ($\omega_{2} = 2 \omega_{1}$), the parametrization below reduces by one the number of free parameters:
\begin{gather}
    \sigma_{1} = \sqrt{S_1 \omega_1 Q_1}\\
    \sigma_{2} = \sqrt{S_2 \omega_2 Q_2}\\
    \tau_1 = \frac{2 Q_1}{\omega_1}\\
    \tau_2 = \frac{2 Q_2}{\omega_2}\\
    P_{\rm GP} = \frac{2 \pi}{\omega_1} = \frac{4 \pi}{\omega_2}
\end{gather}
where $\sigma_1$, $\sigma_2$ are the standard deviations (amplitudes) of the primary and secondary modes, $\tau_1$, $\tau_2$ are the damping timescales of the primary and secondary oscillations, and $P_{\rm GP}$ is the undamped period of the primary mode. Note that these parameters differ slightly from the default \texttt{RotationTerm} kernel parametrization by making no assumptions on the relative amplitudes and quality factors between the two modes.

Our \texttt{PDCSAP} GP model consisted of the five hyperparameters above, plus an excess white noise term $s$. We sampled the posterior distributions of the parameters in their logarithmic form $\{\ln \sigma_1$, $\ln \sigma_2$, $\ln \tau_1$, $\ln \tau_2$, $\ln P_{\rm GP}$, $\ln s \}$ using the Markov chain Monte Carlo (MCMC) package \texttt{emcee} \citep{Foreman-Mackey_2013} and a Bayesian formalism. We employed 100 walkers and performed 100\,000 steps with a burn-in of 10\,000. The number of steps was greater than 50 times the autocorrelation timescale for each parameter, which usually indicates a sufficient level of convergence (\citealt{Sokal_1997}; \citealt{Foreman-Mackey_2019}). The adopted prior distributions and the posteriors median, 16$^{\rm th}$ and 84$^{\rm th}$ percentiles are reported in Table~\ref{table:lightcurveparams}. The resulting mean GP prediction is shown in Figures~\ref{fig:TESS_lc_phase} and \ref{fig:TESS_lc_complete} superimposed on the original \texttt{PDCSAP} cadence. 
Even though the sinusoidal variations visually appear to repeat every $\sim$0.93 day, our model converged to a very well constrained primary oscillation of $1.8680 \pm 0.0004$\,days, thus indicating significant power at the second harmonic.

\begin{deluxetable}{ccc}
\tablecaption{Prior and posterior distributions of the quasi-periodic GP model of the TOI-1452 TESS light curve (details in Sect.\ \ref{sec:tess_analysis})}
\tablehead{
\colhead{Parameter} & \colhead{Prior} & \colhead{Posterior}}
\startdata
$\ln \sigma_1$ & $\mathcal{U}\left(-10, 0\right)$ & $-7.5^{+0.5}_{-0.5}$\\
$\ln \sigma_2$ & $\mathcal{U}\left(-10, 0\right)$ & $-7.1^{+0.5}_{-0.4}$\\
$\ln \left[\tau_1 / \textrm{days} \right]$ & $\mathcal{U}\left(-10, 10\right)$ & $-1.42^{+0.13}_{-0.14}$\\
$\ln \left[\tau_2 / \textrm{days} \right]$ & $\mathcal{U}\left(-10, 10\right)$ & $7.1^{+1.1}_{-0.8}$\\
$\ln \left[P_{\rm GP} / \textrm{days} \right]$ & $\mathcal{U}\left(-2, 5\right)$ & 0.6248$^{+0.0002}_{-0.0002}$\\
$\ln s$ & $\mathcal{U}\left(-15, 0\right)$ & $-12.7^{+1.6}_{-1.5}$\\
\enddata
\tablecomments{$\mathcal{U}\left(a,b \right)$ is the uniform distribution between value $a$ and $b$.}
\label{table:lightcurveparams}
\end{deluxetable}

It is beyond the scope of this study to assess the exact cause of this strong and persistent signal, but we showed earlier that the SPIRou magnetic field constraints of TOI-1452 are inconsistent with a fast rotator and active object. Moreover, a 1.9-day rotation period for TOI-1452 would correspond to a $v \sin i$ of $\sim$7\,km/s, readily detectable in the SPIRou combined spectrum. Instead, the mean line profile FWHM measured from the cross-correlation function (CCF) calculated in \texttt{APERO} suggests a slow rotator (i.e., $v \sin i < 2$\,km/s). We repeated this step for the companion star from the single visit with SPIRou, and also measured a FWHM consistent with $v \sin i < 2$\,km/s. Thus, the rotation of the companion star is also most probably not causing this photometric signal.

\subsection{Joint transit-RV fit} \label{sec:jointfit}

In order to constrain the physical and orbital parameters of TOI-1452\,b, we conducted a joint analysis of the transits (TESS, OMM-PESTO, and MuSCAT3) and the RV data (SPIRou and IRD). The joint fit was performed using the \texttt{juliet} \citep{Espinoza_2019} package, which utilizes \texttt{batman} \citep{Kreidberg_2015} to generate transit models and \texttt{radvel} \citep{Fulton_2018} to compute Keplerian RV models. The \texttt{juliet} framework implements nested sampling algorithms to sample posterior distributions, while also enabling model comparison via evaluations of the Bayesian log-evidence ($\ln Z$). We chose the \texttt{dynesty} \citep{Speagle_2020} dynamic nested sampling option in \texttt{juliet}. Standard nested sampling \citep{Skilling_2006} was designed to estimate evidences, not posteriors, and thus struggles with parameter estimation for complex distributions. Dynamic nested sampling \citep{Higson_2019}, on the other hand, adapts the number of live points based on the structure of the posteriors, providing parameter estimation comparable to MCMC algorithms.

The transit and RV components of the joint fit have four parameters in common: the orbital period $P$, the time of inferior conjunction $t_0$, the eccentricity $e$, and the argument of periastron $\omega$.
For the transit modeling, we followed the parametrization from \cite{Espinoza_2018} of the impact parameter $b$ and the planet-to-star radius ratio $p = R_{\rm p} / R_{\star}$ to efficiently sample physically plausible values in ($b$,~$p$) space. Instead of fitting the scaled semi-major axis $a/R_{\star}$, we used the stellar density $\rho_{\star}$ parameterization available in \texttt{juliet}. Fitting $\rho_{\star}$ takes into account any prior information on the stellar mass and radius. We adopted a Gaussian prior on $\rho_{\star}$ using the value and uncertainty in Table~\ref{table:stellarparams}. Stellar limb-darkening effects in TESS, OMM-PESTO, and MuSCAT3 transits were modeled using per-instrument and per-filter quadratic $q_1$ and $q_2$ parameters defined in \cite{Kipping_2013}. For each instrument, we included in \texttt{juliet} a flux dilution factor $D$, a baseline flux $M$, and an extra jitter term $\sigma$. We set $D_{\rm TESS}$ to 1 (no dilution), as the \texttt{PDCSAP} data are already corrected for crowding effects. The OMM-PESTO light curve combines the flux of TOI-1452 and TIC 420112587, which requires an adequate $D_{\rm PESTO}$ factor to compensate for contamination. We thus constructed a Gaussian prior on $D_{\rm PESTO}$ with a mean value calculated with Equation 6 of \cite{Espinoza_2019} and flux ratio derived from TOI-1452 and TIC 420112587 magnitudes in the $i$ band (see Tables~\ref{table:stellarparams} and \ref{table:stellarparams2}). The adopted prior on $D_{\rm PESTO}$ was $\mathcal{N}\left(0.564, 0.0564^2\right)$, that is with a 10\% standard deviation to account for errors on the magnitudes and deviations between $i$ and $i^{\prime}$. We also explored fixing $D_{\rm PESTO}$ to 0.564, while letting $D_{\rm TESS}$ vary freely between 0 and 2. Both approaches yielded a consistent measurement of the planetary radius (within 1-$\sigma$), indicating that the \texttt{PDCSAP} fluxes were in all likelihood properly corrected for contamination. The dilution in the MuSCAT3 light curves was a priori unknown. However, it is expected that the $g^{\prime}$ transit was more affected by dilution, as the seeing was worse for this filter (see Table \ref{table:gbobs}). We adopted a conservative approach where a different $D_{\rm MuSCAT3}$ is applied for each filter, with uniform priors between 0.5 (twice the flux) and 1. 

The parameters specific to the RV Keplerian component were the semi-amplitude $K$, per-instrument offsets $\gamma$ and extra white noise terms $\sigma$. We explored adding a global GP to model common stellar activity signal in the SPIRou and IRD data. For this, we used the GP implementation in \texttt{juliet} that runs \texttt{celerite} \citep{celerite1_2017}. We chose the Matérn-3/2 approximation kernel, which takes the following form:
\begin{equation}
k_{i,j}\left( \tau \right) = A_{\rm GP}^2 \left[ \left( 1 + 1/\epsilon \right) e^{-\left(1 - \epsilon \right) w} \left(1 - 1/\epsilon \right) e^{-\left(1 + \epsilon \right) w}\right]
\end{equation}
where $\tau = |t_i - t_j|$ is the time interval between data points $i$ and $j$, $A_{\rm GP}$ is the amplitude of the GP, $w = \sqrt{3} \tau/\ell_{\rm GP}$, with $\ell_{\rm GP}$ the timescale of the GP, and $\epsilon$ is set to 0.01 (when $\epsilon \rightarrow 0$, $k_{i,j}$ converges to a Matérn-3/2 kernel). We did not fit a per-instrument $A_{\rm GP}$ and $\ell_{\rm GP}$ due to the limited number of RV measurements from IRD. We also considered choosing a quasi-periodic kernel in \texttt{celerite} instead (Equation 56 of \citealt{celerite1_2017}). Since no clear periodicity was detected in the $B_{\ell}$ time series, or other activity indicators from the LBL such as the dLW metric \citep{Zechmeister_2018} or chromatic velocity slope changes, we applied a uniform prior on the stellar rotation period, namely $\mathcal{U}(0.1, 120)$\,days. We found that the Matérn-3/2 kernel gave equivalent results with fewer hyperparameters needed (2 instead of 4) and that the quasi-periodic GP did not converge to a specific rotation period, showing no preference for a period of 0.93\,days (or 2$\times$0.93\,days) as seen in TESS photometry. This is another indication that the sinusoidal signal in the out-of-transit \texttt{PDCSAP} data is probably not associated with TOI-1452 stellar activity.

We examined the change in Bayesian log-evidence for a suite of joint models ($\mathcal{M}$), all having an identical transit component. The ``zero'' planet model ($\mathcal{M}_{\rm 0p}$) has a $K$ fixed to 0\,m/s, with only the RV offsets and extra white noise terms allowed to vary. This model tests whether the RV dispersion can be fully explained by white noise only, without questioning the transit detection. Single planet models can either be with circular ($\mathcal{M}_{\rm 1cp}$; $e = 0$, $\omega = 90^{\circ}$) or eccentric orbits ($\mathcal{M}_{\rm 1ep}$; free $e, \omega$). Two additional models include a global RV activity GP ($\mathcal{M}_{\rm 1cp+GP}$ and $\mathcal{M}_{\rm 1ep+GP}$). To objectively assess the contribution from the IRD observations, we decided to apply this framework first on the SPIRou data individually, then using the full RV dataset (SPIRou + IRD).

For two competing models, the difference in log-evidence ($\Delta \ln Z$) informs on the probability that one model matches the data better than the other. To interpret the significance of the $\Delta \ln Z$ and select the ``best'' model, we followed the empirical scale introduced in Table 1 of \cite{Trotta_2008}. A $\Delta \ln Z > 5$ translates into ``strong" evidence towards the model with the highest $\ln Z$. A $2.5 < \Delta \ln Z < 5$ corresponds to ``moderate'' evidence, while $\Delta \ln Z \leq 2.5$ shows ``weak'' evidence at best, i.e., neither model should be favoured in that case.

\begin{figure}[ht!]
 \centering
  \includegraphics[width= 1\linewidth]{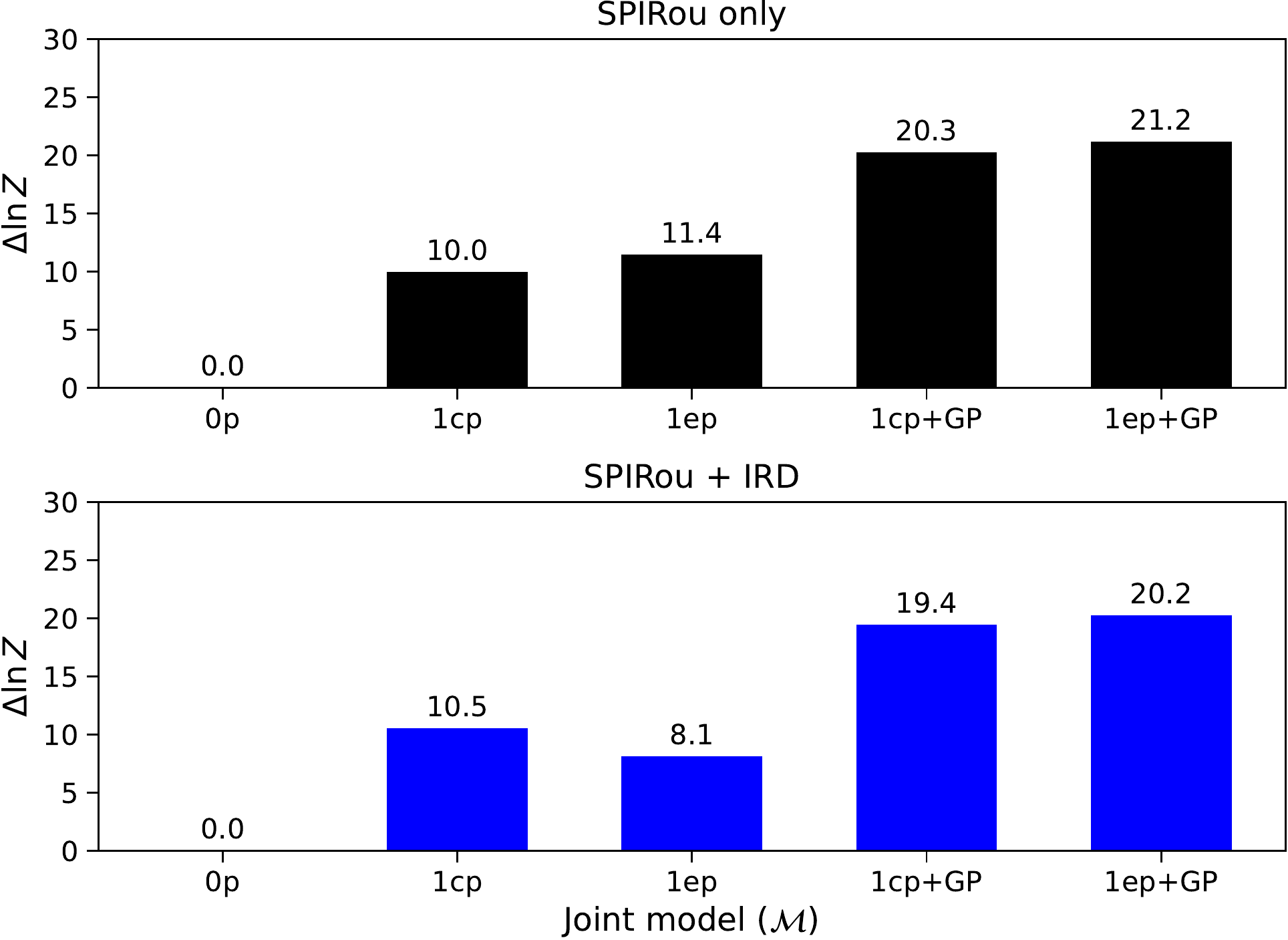}
  \caption{Bayesian log-evidence ($\ln Z$) for different joint transit-RV models ($\mathcal{M}$) and RV datasets. The typical uncertainty on the $\Delta \ln Z$ is $0.7$. The ``zero" planet model $\mathcal{M}_{\rm 0p}$ has a fixed $K = 0$\,m/s. Single planet models $\mathcal{M}_{\rm 1cp}$ and $\mathcal{M}_{\rm 1ep}$ correspond to an RV component with a circular ($e = 0$, $\omega = 90^{\circ}$) or eccentric (free $e, \omega$) orbits respectively. Models $\mathcal{M}_{\rm 1cp+GP}$ and $\mathcal{M}_{\rm 1ep+GP}$ add a Gaussian process to remove correlated noise in the RV data.}
\label{fig:modelselect}
\end{figure}

\begin{figure*}[b!]
\centering
    \includegraphics[width=1\linewidth]{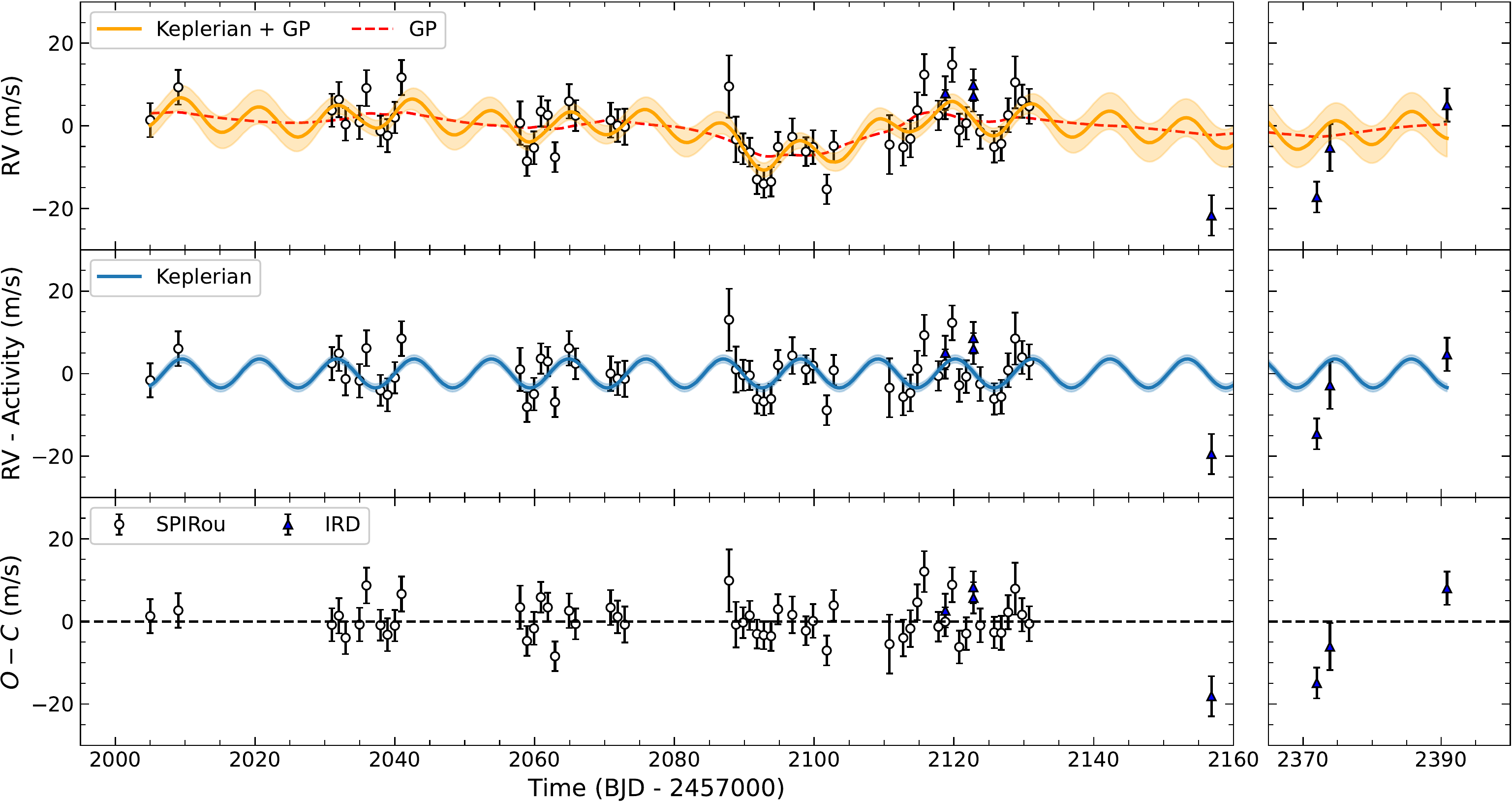}
    \linespread{1}
  \caption{RV time series from SPIRou and IRD with the best-fit Keplerian + activity GP (orange curve), activity GP only (red dashed curve), and Keplerian only (blue curve) models overplotted. The residuals below show an overall agreement between the SPIRou errors and the RV dispersion, which is difficult to assess for IRD given the limited number of measurements.}
  \label{fig:RV_analysis}
\end{figure*}

Figure~\ref{fig:modelselect} shows the Bayesian log-evidence for different joint models and datasets. Note that the typical errors on the $\ln Z$ computed by \texttt{dynesty} were 0.5, so that the $\Delta \ln Z$ presented in Figure~\ref{fig:modelselect} have associated uncertainties of 0.7. 
We first observe that all planetary models are strongly favoured ($\Delta \ln Z > 5$) compared to the ``zero'' planet solution ($\mathcal{M}_{\rm 0p}$), providing quantitative evidence that the TOI-1452\,b Keplerian signal is detected in velocimetry, in phase with transit. There is also compelling evidence for models with an RV activity GP ($\mathcal{M}_{\rm 1cp+GP}$ and $\mathcal{M}_{\rm 1ep+GP}$), increasing $\ln Z$ by approximately 10 relative to $\mathcal{M}_{\rm 1cp}$ and $\mathcal{M}_{\rm 1ep}$. However, considering only the SPIRou data yields similar or slightly larger $\Delta \ln Z$ for all $\mathcal{M}$ compared to joint fits that include the seven IRD RV measurements.
This suggests that the IRD observations do not significantly contribute to improve the Keplerian solution for TOI-1452\,b. The median RV uncertainty from IRD (4.03\,m/s) is nearly identical to SPIRou (4.00\,m/s), but the point-to-point scatter (RMS) is much larger: respectively 12.71\,m/s for IRD and 5.76\,m/s for SPIRou. The planetary models fail to capture the extra RMS in the IRD data, and are instead converging to solutions with white noise term comparable to the overall scatter (see Table~\ref{table:rvcomponent}). This is apparent in Figure~\ref{fig:RV_analysis} showing the RV component of the joint fit (model $\mathcal{M}_{\rm 1cp+GP}$) using the full dataset, with each instrument having their original error bar plotted. The IRD radial velocities were produced using the template spectrum of TOI-1452. As previously mentioned, this star has a small BERV excursion, which is not ideal for filtering out telluric lines. This may explain the increased dispersion in the resulting RVs, in this case, at a level much larger than the Keplerian signal. For this reason, we opted to present below the results using only the SPIRou RVs. We nonetheless provide all the relevant parameters of the RV modeling for the SPIRou only and SPIRou+IRD datasets in Table~\ref{table:rvcomponent}.

\begin{figure}[h]
\centering
    \includegraphics[width=1\linewidth]{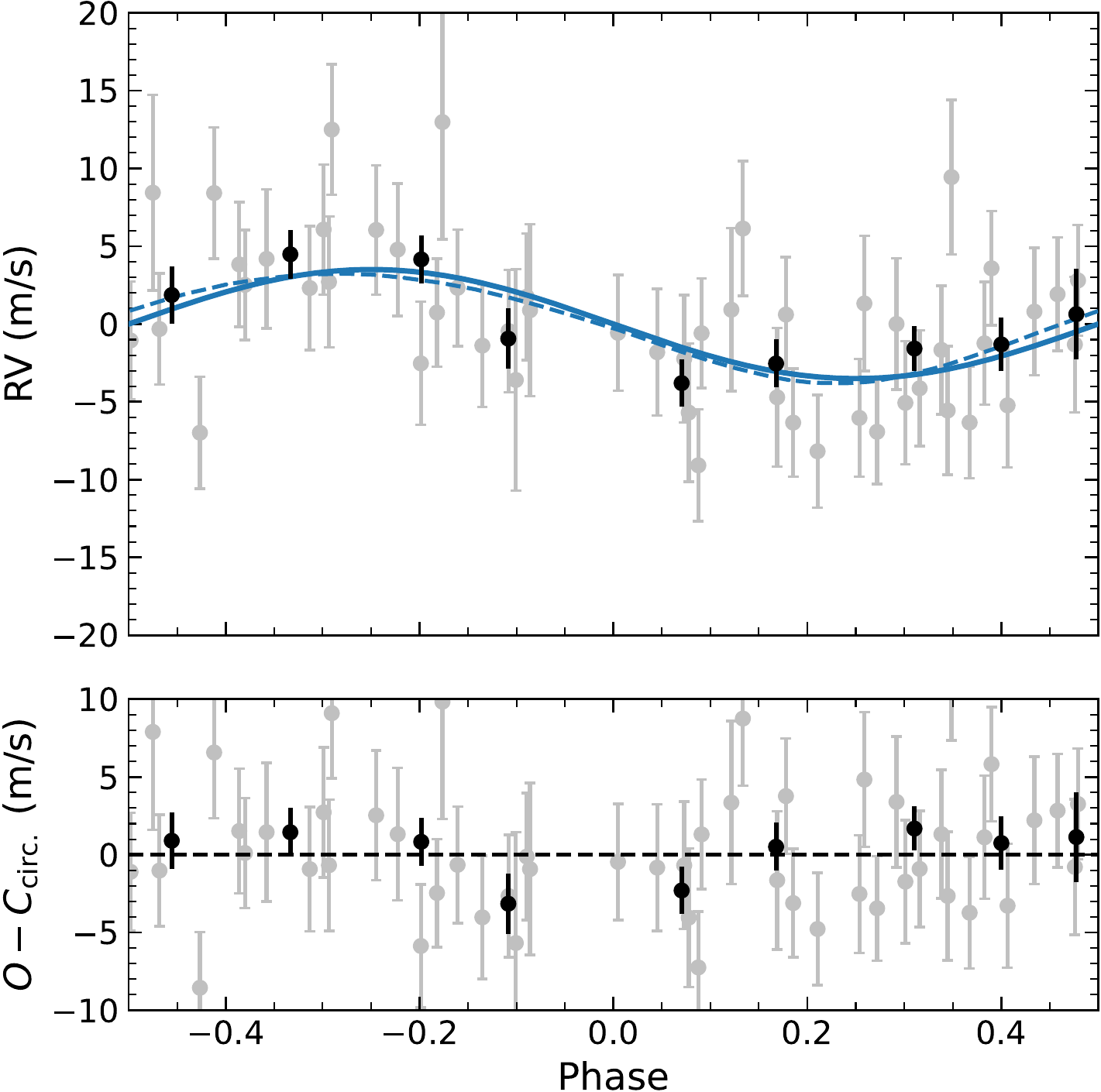}
    \linespread{1}
  \caption{Phase-folded SPIRou RV curve, with systemic velocity and activity GP removed. Binned RV measurements (0.1 phase bin) are marked with black circles. The best-fit Keplerian models are depicted with a solid (circular orbit) and a dashed (eccentric orbit) blue curves. The residuals of the circular fit are presented below. The circular solution yields smaller residuals RMS (3.58\,m/s) compared to the eccentric model (3.66\,m/s), yet both Bayesian evidences ($\ln Z$) are indistinguishable (see Fig.\ \ref{fig:modelselect}).}
  \label{fig:RV_phase}
\end{figure}

The eccentric model $\mathcal{M}_{\rm 1ep+GP}$ produced the highest $\ln Z$ (Fig.\ \ref{fig:modelselect}), but with a Bayesian evidence indistinguishable from the circular model $\mathcal{M}_{\rm 1cp+GP}$ ($\Delta \ln Z = 0.9$). We report an eccentricity of $0.12^{+0.12}_{-0.08}$, with $e < 0.32$ at a 95\% confidence, but argue that the simpler, circular model should be preferred at this point. The adopted priors and resulting posteriors of the $\mathcal{M}_{\rm 1cp+GP}$ fit are summarized in Table~\ref{table:modelparams}. The MuSCAT3 photometric parameters are given in another table (Table~\ref{table:muscat3params}) to facilitate comparison between filters. We measure a dilution factor $D$ consistent with no dilution for the $i^{\prime}$ transit, with moderate level of contamination ($\sim$30\%) in the $g^{\prime}$ band. Even if we assume instead that the flux dilution was exactly zero for all MuSCAT3 filters, the uncorrected transit depths ($\delta_{\rm uncorr.}$) presented in Table~\ref{table:muscat3params} show no sign of strong chromaticity.
The best-fit transit models of the TESS, OMM-PESTO, and MuSCAT3 photometry are shown in Figures~\ref{fig:TESS_lc_phase}, \ref{fig:OMM-PESTO_lc}, and \ref{fig:MuSCAT3_lc} respectively. The best circular ($\mathcal{M}_{\rm 1cp+GP}$) and eccentric ($\mathcal{M}_{\rm 1ep+GP}$) RV orbital fits of TOI-1452\,b are depicted in Figure~\ref{fig:RV_phase} in a phase-folded format.

\begin{deluxetable*}{lccr}
\tablecaption{Prior and posterior distributions of the joint transit-RV fit for model $\mathcal{M}_{\rm 1cp+GP}$ (details in Sect.\ \ref{sec:jointfit}) using only the SPIRou radial velocities}
\tablehead{
\colhead{Parameter} & \colhead{Prior$^{\rm a}$} & \colhead{Posterior} & \colhead{Description}
}
\startdata
\multicolumn{4}{c}{\textit{Fitted parameters}}\\
$\rho_{\star}$ (g/cm$^3$) & $\mathcal{N}\left(16.8, 1.9^2\right)$ & 16.8$^{+1.0}_{-1.4}$ & Stellar density\\
$P$ (days) & $\mathcal{U}\left(11.0, 11.1\right)$ & 11.06201 $\pm$ 0.00002 & Orbital period\\
$t_0$ (BJD - 2\,457\,000) & $\mathcal{U}\left(1691.4, 1691.6\right)$ & 1691.5321 $\pm$ 0.0015 & Time of inferior conjunction\\
$r_1$ & $\mathcal{U}\left(0, 1\right)$ & 0.46$\pm$0.08 & Parametrization$^{\rm b}$\,for $R_{\rm p}/R_{\star}$ and $b$\\
$r_2$ & $\mathcal{U}\left(0, 1\right)$ & 0.0555 $\pm$ 0.0014 & Parametrization$^{\rm b}$\,for $R_{\rm p}/R_{\star}$ and $b$\\
$K$ (m/s) & $\mathcal{U}\left(0, 10\right)$ & 3.50 $\pm$ 0.94 & RV semi-amplitude\\
$e$ & 0 (fixed) & 0 & Orbital eccentricity\\
$\omega$ ($^{\circ}$) & 90 (fixed) & 90 & Argument of periastron\\
$A_{\rm GP}$ (m/s) & $\mathcal{LU}\left(10^{-2}, 100\right)$& 4.5$^{+2.0}_{-1.2}$ & Amplitude of the GP\\
$\ell_{\rm GP}$ (days) & $\mathcal{LU}\left(10^{-2}, 100\right)$ & 11.3$^{+12.0}_{-6.4}$ & Timescale of the GP\\
$\gamma_{\rm SPIRou}$ (m/s) & $\mathcal{U}\left(-33995,-33975\right)$ & $-33985$ $\pm$ 2 & SPIRou RV systemic component\\
$\sigma_{\rm SPIRou}$ (m/s) & $\mathcal{U}\left(0, 10\right)$ & 2.3 $\pm$ 1.3 & SPIRou RV extra white noise$^{\rm c}$ \\
$q_{\rm 1,TESS}$ & $\mathcal{U}\left(0, 1\right)$ & 0.35$^{+0.27}_{-0.19}$ & TESS limb-darkening parameter$^{\rm d}$\\
$q_{\rm 2,TESS}$ & $\mathcal{U}\left(0, 1\right)$ & 0.37$^{+0.33}_{-0.24}$ & TESS limb-darkening parameter$^{\rm d}$\\
$D_{\rm TESS}$ & 1.0 (fixed) & 1.0 & TESS dilution factor\\
$M_{\rm TESS}$ & $\mathcal{N}\left(0, 0.1^2\right)$ & -0.00032 $\pm$ 0.00010 & TESS baseline flux\\
$\sigma_{\rm TESS}$ (ppm)& $\mathcal{LU}\left(1, 10\,000\right)$ & 15$^{+70}_{-12}$ & TESS extra white noise\\
$q_{\rm 1,PESTO}$ & $\mathcal{U}\left(0, 1\right)$ & 0.67$^{+0.21}_{-0.26}$ & PESTO limb-darkening parameter$^{\rm d}$\\
$q_{\rm 2,PESTO}$ & $\mathcal{U}\left(0, 1\right)$ & 0.46$^{+0.28}_{-0.25}$ & PESTO limb-darkening parameter$^{\rm d}$\\
$D_{\rm PESTO}$ & $\mathcal{N}\left(0.564, 0.0564^2\right)$ & 0.586 $\pm$ 0.040 & PESTO dilution factor\\
$M_{\rm PESTO}$ & $\mathcal{N}\left(0, 0.1^2\right)$ & 0.00013 $\pm$ 0.00018 & PESTO baseline flux\\
$\sigma_{\rm PESTO}$ (ppm)& $\mathcal{LU}\left(1, 10\,000\right)$ & 2287 $\pm$ 67 & PESTO extra white noise
\\
\multicolumn{4}{c}{$\mathbf{\vdots}$}\\
\multicolumn{4}{c}{MuSCAT3 photometric parameters in Table~\ref{table:muscat3params}}\\
\hline
\multicolumn{4}{c}{\textit{Derived parameters}}\\
$R_{\rm p}$ (R$_{\oplus}$) & --- & 1.672 $\pm$ 0.071 & Planetary radius\\
$M_{\rm p}$ (M$_{\oplus}$) & --- & 4.82 $\pm$ 1.30 & Planetary mass\\
$\rho$ (g/cm$^3$) & --- & 5.6$^{+1.8}_{-1.6}$ & Planetary bulk density\\
$a$ (au) & --- & 0.061 $\pm$ 0.003 & Orbital semi-major axis\\
$\delta \equiv \left(R_{\rm p}/R_{\star} \right)^2$ & --- & 3.09 $\pm$ 0.16 & Transit depth\\
$b$ & --- & 0.19 $\pm$ 0.13 & Transit impact parameter\\
$i$ ($^{\circ}$) & --- & 89.77 $\pm$ 0.16 & Orbital inclination\\
$T_{\rm eq}$ (K) & & & Equilibrium temperature\\
\hspace{0.25cm} $\left[A_{\rm B} = 0\right]$ & --- & 326 $\pm$ 7 & \\
\hspace{0.25cm} $\left[A_{\rm B} = 0.3\right]$ & --- & 298 $\pm$ 6 & \\
\hspace{0.25cm} $\left[A_{\rm B} = 0.77\right]$ & --- & 226 $\pm$ 5 & \\
$S$ (S$_{\oplus}$) & --- & 1.8 $\pm$ 0.2 & Insolation\\
\enddata
\tablecomments{$^{\rm a}\mathcal{U}\left(a,b \right)$ is the uniform distribution between value $a$ and $b$. $\mathcal{LU}\left(a,b \right)$ is the log-uniform (Jeffreys) distribution between value $a$ and $b$. $\mathcal{N}\left(\mu, \sigma^2 \right)$ is the normal distribution with mean $\mu$ and variance $\sigma^2$.\\
$^{\rm b}$Parametrization from \cite{Espinoza_2018}.\\
$^{\rm c}$White noise term for single exposures within polarimetric sequences. \\
$^{\rm d}\{q_1, q_2\}$ are linked to the quadratic limb-darkening coefficients $\{u_1, u_2\}$ through the transformations outlined in \cite{Kipping_2013}.}
\label{table:modelparams}
\end{deluxetable*}

\begin{deluxetable*}{cccccc}
\tablecaption{Prior and posterior distributions of the MuSCAT3 photometric parameters for model $\mathcal{M}_{\rm 1cp+GP}$ (details in Sect.\ \ref{sec:jointfit}) using only the SPIRou radial velocities}
\tablehead{
\colhead{Parameter} & \colhead{Prior} & \colhead{$g^{\prime}$} & \colhead{$r^{\prime}$} & \colhead{$i^{\prime}$} & \colhead{$z_{\rm s}$}}
\startdata
\multicolumn{6}{c}{\textit{Fitted parameters}}\\
$q_1$ & $\mathcal{U}\left(0, 1\right)$ & 0.60$^{+0.26}_{-0.31}$ & 0.22$^{+0.27}_{-0.15}$ & 0.46$^{+0.32}_{-0.27}$ & 0.24$^{+0.29}_{-0.16}$\\
$q_2$ & $\mathcal{U}\left(0, 1\right)$ & 0.59$^{+0.27}_{-0.33}$ & 0.37$^{+0.35}_{-0.24}$ & 0.24$^{+0.28}_{-0.16}$ & 0.27$^{+0.33}_{-0.19}$ \\
$D$ & $\mathcal{U}\left(0.5, 1\right)$ & 0.71 $\pm$ 0.12 & 0.88$^{+0.07}_{-0.09}$ & 0.96$^{+0.03}_{-0.05}$ & 0.85 $\pm$ 0.07\\
$M$ & $\mathcal{N}\left(0, 0.1^2\right)$ & -0.00141 $\pm$ 0.00036 & -0.00106 $\pm$ 0.00016 & -0.00130 $\pm$ 0.00014 & -0.00131 $\pm$ 0.00013\\
$\sigma$ (ppm)& $\mathcal{LU}\left(1, 10\,000\right)$ & 2748 $\pm$ 319 & 46$^{+427}_{-43}$ & 1837$^{+191}_{-206}$ & 29$^{+227}_{-26}$ \\
\hline
\multicolumn{6}{c}{\textit{Derived parameters}}\\
$\delta_{\rm uncorr.}$ (ppt) & --- & 2.18$^{+0.40}_{-0.36}$ & 2.72$^{+0.21}_{-0.24}$ & 2.94 $\pm$ 0.17 & 2.64 $\pm$ 0.20
\enddata
\tablenocomments{}
\label{table:muscat3params}
\end{deluxetable*}

\section{Discussion} \label{sec:discussion}

\subsection{Planet composition} \label{sec:composition}

\begin{figure}[ht!]
\centering
    \includegraphics[width=1\linewidth]{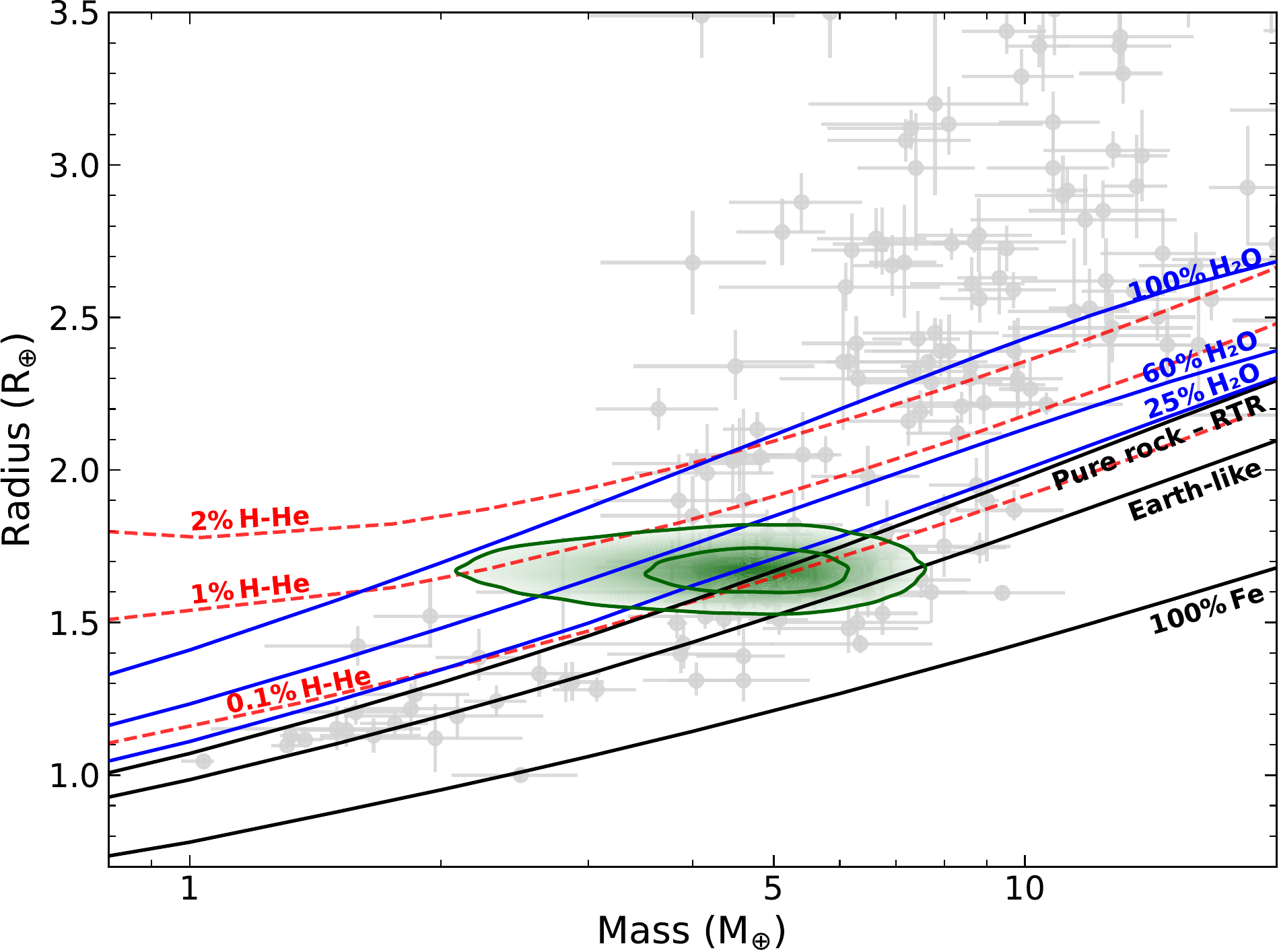}
  \caption{Mass--radius diagram of exoplanets (gray points). Only planets with mass and radius known with precision better than 30\% are shown. Various theoretical composition curves are plotted, using interior structure models from \citealt{Valencia2007,Valencia2013,Guillot1995}. All models assume no atmosphere, except the H-He (red dashed lines) models that correspond to Earth-like interior surrounded by a solar composition envelope at $T = 300$\,K. The mass and radius posteriors (green 1 and 2-$\sigma$ contours) of TOI-1452\,b are consistent with either a water-rich interior with $\sim$25\% H$_2$O by mass, a pure rock (rocky threshold radius curve including relevant phase transitions), or a terrestrial planet with a thin hydrogen envelope of $\sim$0.1\% H-He by mass.}
  \label{fig:MR}
\end{figure}

The analysis of the transit and RV data yields a mass of $4.82 \pm 1.30$\,M$_{\oplus}$ and a radius of $1.672 \pm 0.071$\,R$_{\oplus}$, which together convert into a planetary bulk density of 5.6$^{+1.8}_{-1.6}$\,g/cm$^3$. A density similar to that of the Earth (5.5\,g/cm$^3$) for a planet that has more mass is indicative of an object composed of lighter material. We placed these measurements in a mass-radius diagram (Fig.\ \ref{fig:MR}), along with various theoretical composition curves obtained by the interior structure model of \citealt{Valencia2007, Valencia2013,Plotnykov2020}. We populated this diagram, and the figures in the discussion below, with data from the NASA Exoplanet Archive \citep{Akeson_2013} using the \texttt{exofile} tool\footnote{\href{https://github.com/AntoineDarveau/exofile}{\texttt{github.com/AntoineDarveau/exofile}}}.

By comparing the mass and size of TOI-1452\,b to theoretical M-R curves in Figure~\ref{fig:MR}, we see three possibilities for the nature of this planet: (1) an ocean planet, (2) a bare rock with an iron content less than that of Earth, or (3) a terrestrial planet with a thin, low molecular weight atmosphere (e.g., H-He). The water world hypothesis is supported by a temperate equilibrium temperature for TOI-1452\,b of $298 \pm 6$\,K assuming an Earth-like Bond albedo ($A_{\rm B} = 0.3$), and between 226 and 326\,K for extreme $A_{\rm B}$ of 0.77 (Venus-like) and 0 (pure absorber). The insolation level of TOI-1452\,b is about 80\% higher than the Earth ($S = 1.8 \pm 0.2$\,S$_{\oplus}$), similar to Venus ($S = 1.91$\,S$_{\oplus}$).

Focusing on the first two possibilities, we used an MCMC approach (\texttt{emcee}, \citealt{Foreman-Mackey_2013}) coupled to an interior structure model \citep{Valencia2007} to obtain the distribution of mass fractions of iron and water that are consistent with the data. The details of this modeling can be found in \citet{Plotnykov2020}. The H$_2$O layer is described by the equation of state from \citealt{Hemley1987, Stewart_2005}. We report below the 16$^{\rm th}$, 50$^{\rm th}$, and 84$^{\rm th}$ percentiles of the posterior distributions (available in Appendix~\ref{mcmc:summary}). 

\begin{figure}[ht!]
\centering
    \includegraphics[width=1\linewidth]{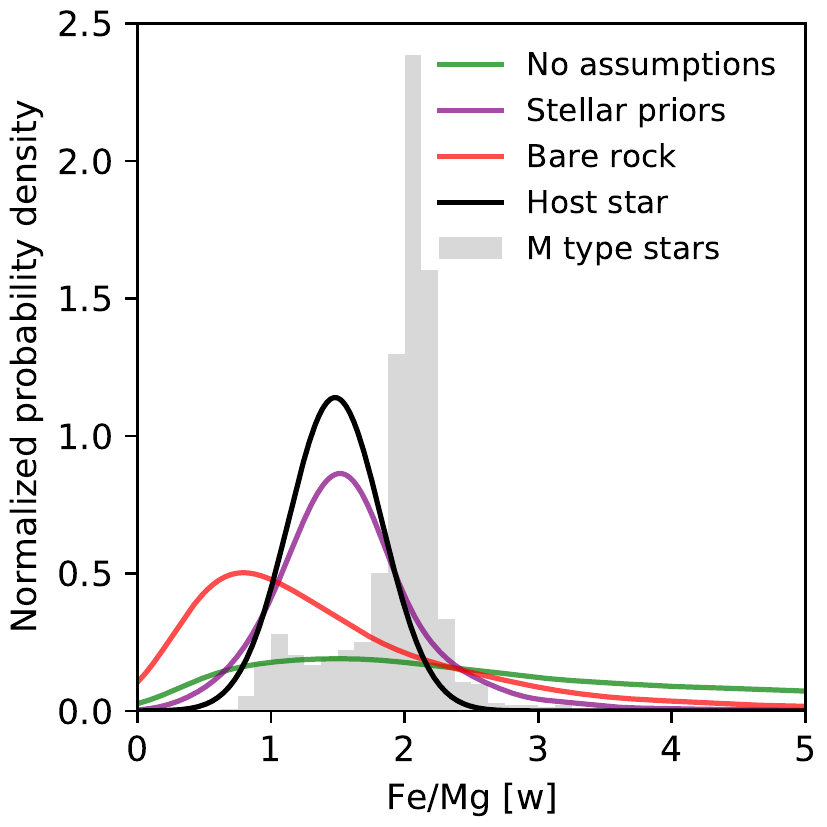}
  \caption{Fe/Mg distribution for TOI-1452\,b according to different model assumptions (Table~\ref{table:ratios}) compared to that of its host star (black). We used kernel density estimate to draw the probabilities from the posteriors. We also include the M dwarfs distribution of Fe/Mg from the APOGEE database (gray histogram, sample size of $\sim$1000, \citealt{Majewski_2016}; \citealt{Ahumada_2020}) for comparison. 
  }
  \label{fig:Fe2Mg}
\end{figure}

From the chemical analysis of the star (Sect.\ \ref{sec:stellar_char_spirou}), we obtain chemical ratios relevant to the planetary interior. Notably, TOI-1452 might have a slightly lower Fe/Mg weight ratio compared to the Sun (see Table~\ref{table:ratios}), but
in fair agreement with that of a sample of $\sim$1000 M dwarfs from the APOGEE (DR16) catalog (\citealt{Majewski_2016}; \citealt{Ahumada_2020}) with known chemical ratios (see Fig.\ \ref{fig:Fe2Mg}). The APOGEE abundances are derived from high-resolution near-infrared spectroscopy ($R \sim 22\,000$), with typical uncertainty on [Fe/H] and [Mg/H] of 0.02\,dex. Obtaining the Si ratios for TOI-1452 was difficult given the scarcity of spectral lines. In addition, we derive a C/O weight ratio consistent with the solar value. Our chemical abundance ratios for TOI-1452 are summarized in Table~\ref{table:ratios}.

To infer the planet's composition for scenarios 1 and 2, we can either make no assumptions on the refractory ratios, and thus, obtain all possible compositions that fit the mass-radius data; or, assume that the refractory ratios of planet and star are related, and use the star's ratios as priors in the Bayesian analysis. Given that the refractory ratios of super-Earths seem to span a larger range than that of stars \citep{Plotnykov2020}, we applied both methods here. In both cases, we kept the Mg/Si ratio in the planetary mantle the same as the star. 
This assumption should not affect the results considerably, given that the mantle minerals formed by different Mg/Si ratios have similar equations of state and thus, the relative content of Mg to Si is not constrained by planetary mass and radius data \citep{Plotnykov2020}. 

In the case where we make no assumptions, we obtain a water-mass fraction (WMF) of $0.27^{+0.20}_{-0.15}$, and core-mass fraction (CMF) of $0.30^{+0.20}_{-0.17}$. These values translate to $ \mathrm{Fe/Si} = 2.9^{+5.4}_{-1.9}$ and $\mathrm{Fe/Mg} = 3.4 ^{+6.3}_{-2.2}$ by weight. The planetary refractory ratios are particularly large primarily due to large uncertainty in planetary mass but also the degeneracy that ensues when considering water.  The tail of the Fe/Mg distribution is long because this case allows for little to no mantle in the planet (see Fig.\ \ref{fig:corner}).
If instead, we assume the refractory ratios of the star as priors, we obtain a lower water and core mass fractions: WMF = $0.22^{+0.21}_{-0.13}$, CMF = $0.18 \pm 0.06$, resulting in Fe/Si = $1.3 \pm 0.4$ and Fe/Mg = $1.5 \pm 0.4$. Thus, both abundance scenarios yield a non-zero, yet poorly-constrained, WMF. The large uncertainty on both the CMF and WMF is rooted to the modest mass constraint (3.7$\sigma$). Better mass measurements are needed to confirm that TOI-1452\,b has a significant WMF.

A different possibility (scenario 2 discussed above) is that this planet is a bare rock with no significant atmosphere, perhaps because it lost any acquired water through atmospheric evaporation during the high insolation phase of the M dwarf host star (\citealt{Bolmont2017,Barnes2013}). In this case, we constrain the physical model solutions but make no assumptions on the refractory ratios. The results show a $\mathrm{CMF} = 0.19^{+0.18}_{-0.12}$, with Fe/Si = $1.0^{+1.0}_{-0.5}$ and Fe/Mg = $1.2^{+1.2}_{-0.6}$ by weight. This CMF indicates a planet not as dense as the Earth, with refractory ratios still consistent at the 1-$\sigma$ level with those of its host star (see Table \ref{table:ratios} and Fig.\ \ref{fig:Fe2Mg}).
However, the maximum a posteriori estimate of Fe/Mg ($\sim$0.8) for this bare rock scenario is lower than that of its host star. Forming planets that are iron poor with respect to their host star is difficult \citep{Scora2020}. Therefore, based on our current knowledge of planet formation and the chemical characteristics of the star, this scenario is less likely.

A summary of this interior modeling is presented in Table~\ref{table:ratios} and Figure~\ref{fig:Fe2Mg}. The posterior distributions for the three models (no assumptions, stellar priors and bare rock) are shown in Appendix~\ref{mcmc:summary}.

Outside of these scenarios, our observations do not rule out other structures with a low molecular weight atmosphere (scenario 3) such as an Earth-like interior surrounded by an H-He envelope at $T = 300\,K$ containing $\sim$0.1\% of the total mass (up to 0.5\% H-He at 1-$\sigma$). 
One way to firmly break the degeneracy in planetary internal structures would be to characterize the atmosphere of TOI-1452\,b.

\begin{deluxetable*}{cccccc}
\tablecaption{Chemical ratios by weight for the TOI-1452 system}
\tablehead{
\colhead{Chemical} & \colhead{TOI-1452} & \colhead{Sun$\dag$} & TOI-1452\,b & TOI-1452\,b & TOI-1452\,b \\[-0.1cm]
 Ratios & & & No assumptions & Stellar priors & Bare rock   \\
}

\startdata
Fe/Mg & $1.48^{+0.36}_{-0.29}$ & 1.83$\pm$0.25 & $3.4^{+6.3}_{-2.2}$ & $1.5^{+0.4}_{-0.4}$& $1.2^{+1.2}_{-0.6}$\\
Mg/Si & $0.86^{+0.37}_{-0.26}$ & 1.06$\pm$0.13 & 0.86$^*$ &  0.86$^*$&  0.86$^*$ \\
C/O & $0.48^{+0.11}_{-0.09}$ & 0.41$\pm$0.07 & ${\hat{}}$ & ${\hat{}}$ & ${\hat{}}$ \\
\hline
CMF & -- & -- & $0.30^{+0.20}_{-0.17}$ & $0.18^{+0.06}_{-0.06}$ & $0.19^{+0.18}_{-0.12}$ \\
WMF & -- & -- & $0.27^{+0.20}_{-0.15}$ & $0.22^{+0.21}_{-0.13}$ & -- \\
\enddata
\tablecomments{
$^{\dag}$ Photospheric abundance ratios from \citet{Asplund2009}.\\
$^{*}$ Mg/Si ratio of the planets are fixed to the star TOI-1452 ratio.\\
$\hat{}\ $  The interior models assume no carbon compounds. \\
}
\label{table:ratios}
\end{deluxetable*}

\subsection{Atmospheric characterization prospect} \label{sec:atmo_char}

\begin{figure}[t]
\centering
    \includegraphics[width=1\linewidth]{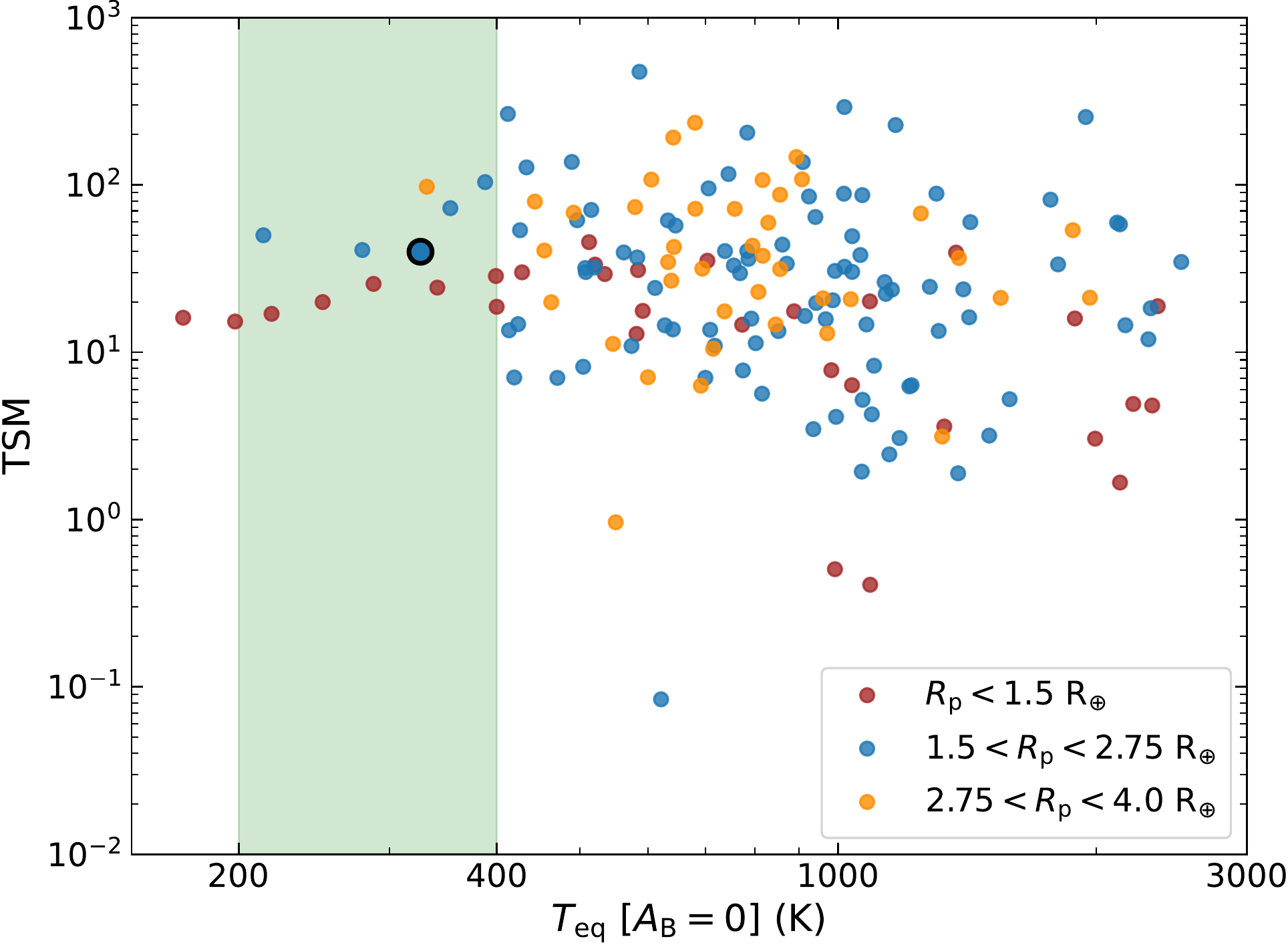}
  \caption{Transmission spectroscopy metric (TSM, \citealt{Kempton_2018}) as a function of planetary equilibrium temperature for small exoplanets ($R_{\rm p} < 4$\,R$_{\oplus}$) with well-established masses (error below 30\%). The green region corresponds to an arbitrary temperate temperature interval (200\,K $< T_{\rm eq} <$ 400\,K). TOI-1452\,b is among the best temperate targets for follow-up transit spectroscopy with JWST.}
  \label{fig:TSM}
\end{figure}

TOI-1452\,b is a prime target for follow-up transit spectroscopy with JWST. The system is located near Webb's Continuous Viewing Zone (CVZ), more precisely at a few degrees ($\sim$10$^{\circ}$) off the Northern CVZ, which means that it could be observed most of the year. Moreover, TOI-1452\,b is one of the few identified super-Earths in a temperate regime ($T_{\rm eq}$ between 200 and 400\,K) orbiting a relatively bright star amenable to transmission spectroscopy observations (see Figure~\ref{fig:TSM} and Table~\ref{table:temperate_planets}).
The expected strength of the atmospheric signal is characterized by the
Transmission Spectroscopy Metric (TSM, \citealt{Kempton_2018}), which is proportional to the host star's $J$ magnitude and the planet's atmospheric scale height.
Figure~\ref{fig:TSM} displays the TSM as a function of equilibrium temperature for known small exoplanets with available mass measurements. The sample is restricted to systems with well-determined masses (relative uncertainty $< 30$\%) since a constraint on surface gravity is essential to correctly interpret the transmission spectrum of an exoplanet \citep{Batalha_2019}. The temperate subset of Figure~\ref{fig:TSM} is detailed in Table~\ref{table:temperate_planets}. The TSM of TOI-1452\,b (39.9) is similar to well-known temperate systems such as LHS 1140\,b (50.0) and K2-18\,b (40.8), while being 60\% below the highest listed target in this subset, L231-32\,d (TOI-270\,d, 104.0). All seven host stars in Table~\ref{table:temperate_planets} have $T_{\rm eff}<4000$\,K (or average $T_{\rm eff}$ of 3225\,K), confirming the high interest of M dwarfs for planetary atmospheric characterization. Note that the high-value target L~98-59\,d (TSM above 200, \citealt{Cloutier_2019}; \citealt{Demangeon_2021}) was just barely excluded from Table~\ref{table:temperate_planets} due to its $T_{\rm eq}$ (409\,K) being slightly above our 400\,K cut. Our subset also excludes intriguing planets with plausible temperate environment, but deprived of mass measurement (or imprecise mass), such as TOI-700\,c (TSM = 79.7, \citealt{Gilbert_2020}), TOI-1266\,c (TSM = 48.8, \citealt{Demory_2020}, and K2-3\,c (TSM = 25.5, \citealt{Damasso_2018}). The TOI-1452 system is a unique target to explore the atmospheric properties of temperate planets within the radius valley.
This paper provides the first mass determination needed for the interpretation of future transmission spectra.

\begin{deluxetable*}{ccccccccccc}
\centering
\tablecaption{Transmission Spectroscopy Metric (TSM) for a subset of well-characterized small exoplanets in a temperate equilibrium temperature regime (200\,K $\leq T_{\rm eq} \leq$ 400\,K)}

\tablehead{\colhead{Planet} & \colhead{$P$} & \colhead{$M_{\rm p}$} & \colhead{$R_{\rm p}$} & \colhead{$T_{\rm eq}$\,$[A_{\rm B} = 0]$} & \colhead{$J$} & \colhead{$T_{\rm eff}$} & \colhead{$M_{\star}$} & \colhead{$R_{\star}$} & \colhead{TSM} & \colhead{Ref.}\\[-0.1cm]
& (days) & (M$_{\oplus}$) & (R$_{\oplus}$) & (K) & (mag) & (K) & (M$_{\odot}$) & (R$_{\odot}$) & &
}
\startdata
L231-32\,d & 11.380 & 4.78 & 2.133 & 388 & 9.10 & 3506 & 0.39 & 0.38 & 104.0 & (1) \\
TOI-1231\,b & 24.246 & 15.4 & 3.65 & 331 & 8.88 & 3553 & 0.48 & 0.48 & 97.6 & (2)\\
LTT~3780\,c & 12.252 & 8.6 & 2.30 & 353 & 9.01 & 3331 & 0.40 & 0.37 & 72.6 & (3)\\
LHS 1140\,b & 24.739 & 6.38 & 1.635 & 214 & 9.61 & 2988 & 0.19 & 0.21 & 50.0 & (4)\\
K2-18\,b & 32.940 & 8.63 & 2.610 & 279 & 9.76 & 3457 & 0.50 & 0.44 & 40.8 & (5)\\
\textbf{TOI-1452\,b} & 11.062 & 4.82 & 1.672 & 326 & 10.60 & 3185 & 0.25 & 0.28 & 39.9 & (6)\\
TRAPPIST-1\,b & 1.511 & 1.374 & 1.116 & 399 & 11.35 & 2566 & 0.09 & 0.12 & 28.6 & (7)\\
TRAPPIST-1\,d & 4.049 & 0.388 & 0.788 & 287 & 11.35 & 2566 & 0.09 & 0.12 & 25.6 & (7)\\
TRAPPIST-1\,c & 2.422 & 1.308 & 1.097 & 341 & 11.35 & 2566 & 0.09 & 0.12 & 24.3 & (7)\\
TRAPPIST-1\,e & 6.101 & 0.692 & 0.920 & 251 & 11.35 & 2566 & 0.09 & 0.12 & 19.9 & (7)\\
LHS 1140\,c & 3.777 & 1.76 & 1.169 & 400 & 9.61 & 2988 & 0.19 & 0.21 & 18.7 & (4)\\
TRAPPIST-1\,f & 9.208 & 1.039 & 1.045 & 218 & 11.35 & 2566 & 0.09 & 0.12 & 17.0 & (7)\\
\enddata
\tablerefs{(1) \cite{VanEylen_2021}. (2) \cite{Burt_2021}. (3) \cite{Cloutier_2020a}. (4) \cite{Lillo-Box_2020}. (5) \cite{Benneke_2019}. (6) This work. (7) \cite{Agol_2021}.}
\label{table:temperate_planets}
\end{deluxetable*}

\subsection{Implications for the emergence of the M dwarf radius valley} \label{sec:radius_valley}

Planets on either side of the radius valley differ by their composition, typically `rocky' for the smaller super-Earths, and `gaseous' for the larger mini-Neptunes. This transition occurs as a consequence of a varying envelope mass fraction: adding an H-He envelope up to a few percents of the total mass of a planet essentially doubles its observable radius (\citealt{Lopez_2014}; \citealt{Chen_2016}). Thermally-driven atmospheric escape processes such as photoevaporation (\citealt{Owen_2013}, \citeyear{Owen_2017}; \citealt{Lopez_2014}; \citealt{Lopez_2018}; \citealt{Wu_2019}) and core-powered mass loss (\citealt{Ginzburg_2018}; \citealt{Gupta_2019}, \citeyear{Gupta_2020}) have been proposed as radius valley emergence mechanisms. In these models, super-Earths and mini-Neptunes originate from the same population of planets that form with an extended H-He envelope around Earth-like core, with the population of rocky super-Earths emerging after losing their primordial atmospheres to hydrodynamic escape. Another possible scenario is to assemble rocky super-Earths at late times after most or all of the gas has been dissipated from the protoplanetary disk (\citealt{Lee_2014}; \citealt{Lopez_2018}; \citealt{Lee_2021}). The two classes of planets would form on different timescales, resulting in a bimodal distribution without relying on any subsequent atmospheric escape.

Each of the aforementioned mechanisms predicts that the rocky-to-gaseous transition ($R_{\rm valley}$) varies with parameters such as orbital period $P$ and stellar mass $M_{\star}$. Photoevaporation, core-powered mass loss, and gas-poor accretion models predict a negative slope in $R_{\rm p}$--$P$ space, respectively $R_{\rm valley} \propto P^{-0.25 \textendash 0.15}$ (\citealt{Owen_2017}; \citealt{Lopez_2018}; \citealt{Mordasini_2020}), $R_{\rm valley} \propto P^{-0.11}$ \citep{Gupta_2019}, and $R_{\rm valley} \propto P^{-0.08}$ \citep{Lee_2021}. Conversely, the formation of super-Earths strictly by the merging of planetary embryos in a gas-depleted environment, analogous to the formation of terrestrial planets in the Solar System, would produce a positive slope ($R_{\rm valley} \propto P^{0.11}$, \citealt{Lopez_2018}). One way to test the proposed models is to compare these predictions to the real population of exoplanets. 

From occurrence rate calculations of small close-in planets around Sun-like stars, \cite{Martinez_2019} measured a $d \log R_{\rm valley} / d \log P = -0.11 \pm 0.02$, consistent with thermally-driven mass loss and gas-poor formation. Using a similar methodology but for planets around low-mass stars with $T_{\rm eff} < 4700$\,K, \cite{Cloutier-Menou_2020} obtained a $d \log R_{\rm valley} / d \log P = 0.058 \pm 0.022$, where the positive sign suggests that the gas-depleted formation of super-Earths may be dominant around M dwarfs. These distinct slope measurements carve out regions in the $R_{\rm p}$--$P$ parameter space where the models make opposing predictions regarding the bulk composition of a planet (i.e., either rocky or gaseous). This framework to test radius valley emergence models around M dwarfs was introduced in \cite{Cloutier-Menou_2020} and has since been applied to a number of transiting planets (TOIs 776\,b; \citealt{Luque_2021}, 1235\,b; \citealt{Cloutier_2020b}, 1634\,b; \citealt{Cloutier_2021}, 1685\,b; \citealt{Bluhm_2021}).

\begin{figure}[t]
\centering
    \includegraphics[width=1\linewidth]{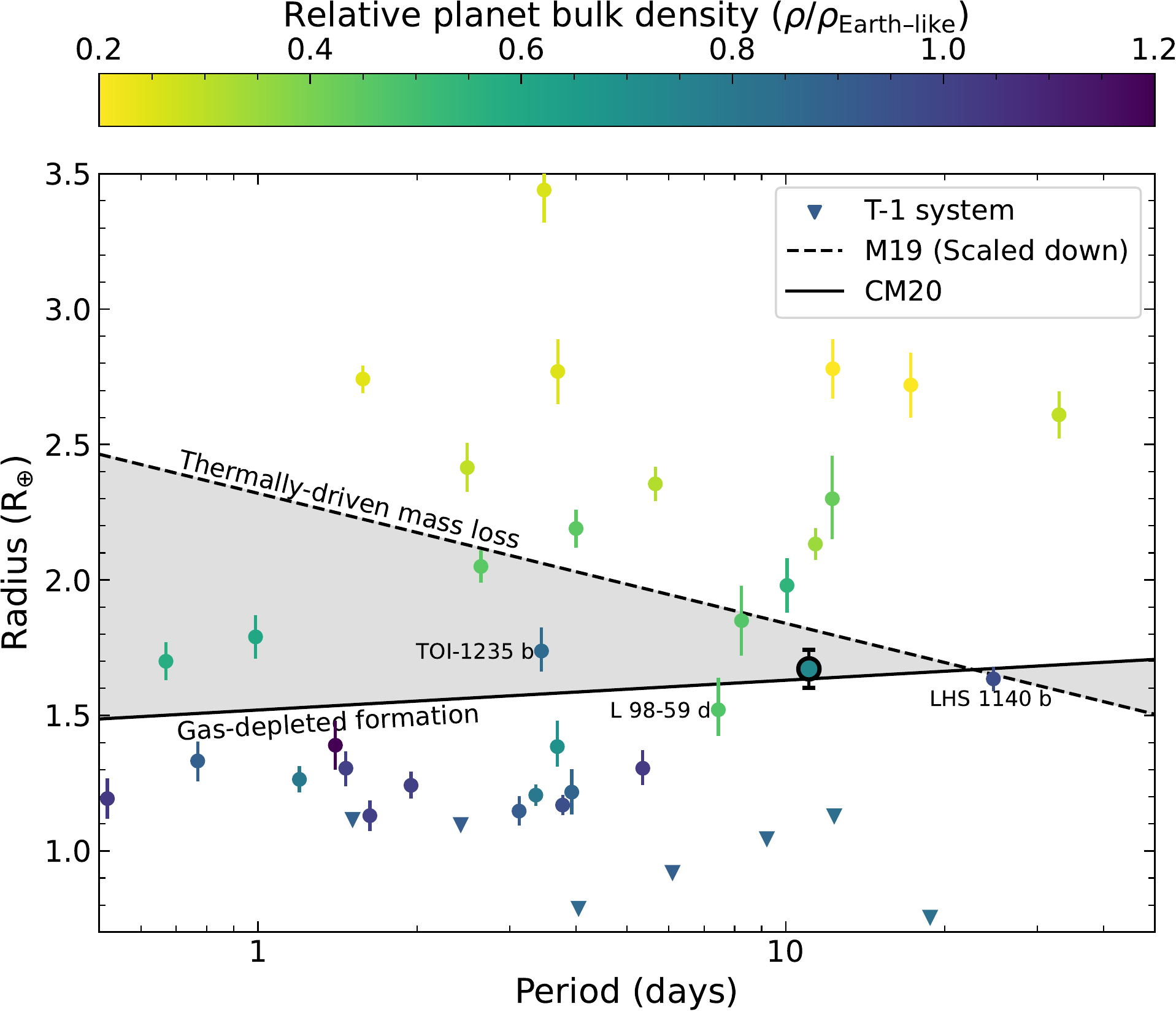}
  \caption{Period--radius diagram of exoplanets around low-mass stars ($T_{\rm eff} \leq 4000$\,K) with mass and radius known with precision better than 30\%. The TRAPPIST-1 (T-1) planets are represented by triangles. Empirical \textit{valley} for low-mass stars from \citealt{Cloutier-Menou_2020} (CM20), consistent with gas-depleted formation, and from \citealt{Martinez_2019} (M19), compatible with thermally-driven atmospheric mass loss and scaled down to represent the stellar mass population of CM20. TOI-1452\,b lies below M19 and has a small bulk density compared to a terrestrial planet of the same mass ($\rho / \rho_{\textrm{Earth-like}} \approx 0.76$). This is inconsistent with photoevaporation and core-powered mass loss predictions.}
  \label{fig:RP}
\end{figure}

Figure~\ref{fig:RP} presents the period--radius diagram for exoplanets around M-dwarf hosts ($T_{\rm eff} < 4000$\,K). 
Each planet is color-coded by its bulk density relative to the Earth-like structure model of \citealt{Valencia2007} (see Figure~\ref{fig:MR}). TOI-1452\,b sits on or slightly above the empirical \textit{valley} of \cite{Cloutier-Menou_2020}, while being considerably below the slope measured by \cite{Martinez_2019}, scaled down to match the median stellar mass of the \cite{Cloutier-Menou_2020} sample (using Equation 11 therein). The locus of TOI-1452\,b in Figure~\ref{fig:RP}, combined with our density estimate, are incompatible with the photoevaporation and core-powered mass loss models. However, the likely intermediate nature of TOI-1452\,b cannot strongly support the alternative gas-depleted formation scenario either as the dominant mechanism for the emergence of the M-dwarf radius valley. A volatile-rich interior for TOI-1452\,b could indicate a different formation pathway, e.g., one without significant gas accretion during the disk lifetime.

Figure~\ref{fig:RP} also highlights three other systems
presenting similarities with TOI-1452\,b, namely TOI-1235\,b \citep{Cloutier_2020b}, L~98-59\,d (\citealt{Cloutier_2019}; \citealt{Demangeon_2021}) and LHS 1140\,b (\citealt{Dittmann_2017}; \citealt{Lillo-Box_2020}). All four planets have a similar size, while spanning a large interval in periods. TOI-1235\,b ($P = 3.445$\,days, $R_{\rm p} =  1.738$\,R$_{\oplus}$) and LHS 1140\,b ($P = 24.737$\,days, $R_{\rm p} = 1.635$\,R$_{\oplus}$) have densities compatible with bona fide super-Earths; their position in Figure~\ref{fig:RP} indicates that they are probably examples of the largest terrestrial planets that can be assembled around M dwarfs without accreting a substantial hydrogen envelope. On the other hand, L~98-59\,d ($P = 7.451$\,days, $R_{\rm p} = 1.521$\,R$_{\oplus}$) is a likely water-rich ($\sim$30\%) planet that may be approaching, like TOI-1452\,b, the minimum size for volatile-rich objects around a low-mass star. These four systems constitute benchmarks for understanding the formation and evolution of planets within the radius valley.

\section{Summary \& Conclusion} \label{sec:conclusion}

This paper reports the discovery and characterization of the transiting temperate exoplanet TOI-1452\,b. A joint analysis of transit observations from TESS and other ground-based telescopes combined with radial velocity measurements from SPIRou and IRD, yields a mass of $4.82 \pm 1.30$\,M$_{\oplus}$ and a radius of $1.672 \pm 0.071$\,R$_{\oplus}$. These physical parameters are consistent with either a rocky world with a Fe/Mg ratio similar to the host star (Fe/Mg $=1.2^{+1.2}_{-0.6}$ by weight), a water-rich interior (either $22^{+21}_{-13} \%$ H$_2$O by weight, if stellar priors are assumed for the planetary refractory ratios, or $27^{+20}_{-15} \%$ H$_2$O if no assumptions are made) or a terrestrial planet surrounded by a $\lesssim $1\% H-He atmosphere. Orbiting its M4 host star ($T_{\rm eff} = 3185 \pm 50$\,K) every $11.06201 \pm 0.00002$\,days, the planet receives about twice as much radiation than the Earth ($S = 1.8 \pm 0.2$\,S$_{\oplus}$), corresponding to a blackbody temperature of $326\pm7$\,K. The results of our interior modeling and the fact that the planet receives modest irradiation make TOI-1452\,b a good candidate water world.

TOI-1452\,b is a prime target for upcoming atmospheric characterization efforts with JWST, featuring a high Transmission Spectroscopy Metric compared to other known temperate exoplanets. Transit spectroscopy observations with JWST should reveal the true nature of this intriguing exoplanet lying within the radius valley, whether this is a rocky world or one with a volatile envelope. Being observable with JWST most of the year, TOI-1452\,b is a unique system for studying exoplanets at the transition between super-Earths and mini-Neptunes.

\acknowledgments

We acknowledge the use of public TESS Alert data from pipelines at the TESS Science Office and at the TESS Science Processing Operations Center. This research has made use of the Exoplanet Follow-up Observation Program website, which is operated by the California Institute of Technology, under contract with the National Aeronautics and Space Administration under the Exoplanet Exploration Program. Resources supporting this work were provided by the NASA High-End Computing (HEC) Program through the NASA Advanced Supercomputing (NAS) Division at Ames Research Center for the production of the SPOC data products. This paper includes data collected by the TESS mission that are publicly available from the Mikulski Archive for Space Telescopes (MAST). \par Based on observations obtained at the Canada-France-Hawaii Telescope (CFHT) which is operated from the summit of Maunakea by the National Research Council of Canada, the Institut National des Sciences de l'Univers of the Centre National de la Recherche Scientique of France, and the University of Hawaii. The observations at the Canada-France-Hawaii Telescope were performed with care and respect from the summit of Maunakea which is a signicant cultural and historic site. \par Based on observations obtained at the Observatoire du Mont-Mégantic, financed by Université de Montréal, Université Laval, the Canada Economic Development program and the Ministère de l'Économie et de l'Innovation. \par This paper is also based on observations made with the MuSCAT3 instrument, developed by the Astrobiology Center and under financial supports by JSPS KAKENHI (JP18H05439) and JST PRESTO (JPMJPR1775), at Faulkes Telescope North on Maui, HI, operated by the Las Cumbres Observatory. \par This work is partly supported by the Natural Science and Engineering Research Council of Canada and the Institute for Research on Exoplanets through the Trottier Family Foundation \par This work is partly supported by MEXT/JSPS KAKENHI Grant Numbers JP22000005, JP15H02063, JP17H04574, JP18H05439, JP18H05442, JP19K14783, JP21H00035, JP21K13975, JP21K20376, JST CREST Grant Number JPMJCR1761, and the Astrobiology Center of National Institutes of Natural Sciences (NINS) (Grant Numbers AB031010, AB031014).  \par We thank Dr.\ Martin Turbet for the suggestions to improve the discussion section. \par We acknowledge very useful feedback and discussion from  Dr.\ Ansgar Reiners, regarding the importance to properly check the contamination from the nearby companion in the SPIRou and IRD spectra and its effect on the final radial velocities. \par JFD acknowledges funding from the European Research Council (ERC) under the H2020 research \& innovation programme (grant agreement \#740651 NewWorlds). \par This work has been carried out within the framework of the NCCR PlanetS supported by the Swiss National Science Foundation. \par PC thanks the LSSTC Data Science Fellowship Program, which is funded by LSSTC, NSF Cybertraining Grant \#1829740, the Brinson Foundation, and the Moore Foundation; her participation in the program has benefited this work. \par A.C. and X.D. acknowledges funding from the ANR of France under contract number ANR\-18\-CE31\-0019 (SPlaSH). This work is supported by the French National Research Agency in the framework of the \textit{Investissements d'Avenir} program (ANR-15-IDEX-02), through the funding of the ``Origin of Life" project of the Grenoble-Alpes University. \par F.K. acknowledge the ANR project [SPlaSH] from the French Agence Nationale de Recherche with reference ANR-18-CE31-0019-02 \par J.H.C.M. is supported in the form of a work contract funded by Fundação para a Ciência e Tecnologia (FCT) with the reference DL 57/2016/CP1364/CT0007; and also supported from FCT through national funds and by FEDER-Fundo Europeu de Desenvolvimento Regional through COMPETE2020- \textit{Programa Operacional Competitividade e Internacionalização} for these grants UIDB/04434/2020 \& UIDP/04434/2020, PTDC/FIS-AST/32113/2017 \& POCI-01-0145-FEDER-032113, PTDC/FIS-AST/28953/2017 \& POCI-01-0145-FEDER-028953, PTDC/FIS-AST/29942/2017. \par This project has received funding from the European Research Council (ERC) under the European Union's Horizon 2020 research and innovation programme under grant agreement No 716155 (SACCRED). \par KG acknowledges the partial support from the Ministry of Science and Higher Education of the RF (grant 075-15-2020-780). \par TV acknowledges funding from the  \textit{Fonds de Recherche du Qu\'ebec - Nature et Technologie} (FRQNT).

\facilities{TESS, OMM-PESTO, MuSCAT3, KeckII/NIRC2, CFHT/SPIRou, Subaru/IRD}

\software{\texttt{emcee} \citep{Foreman-Mackey_2013}; \texttt{Astropy} \citep{Astropy_2018}; \texttt{radvel} \citep{Fulton_2018}; \texttt{matplotlib} \citep{Hunter_2007}; \texttt{celerite} \citep{celerite1_2017}; \texttt{celerite2} (\citealt{celerite1_2017}; \citeyear{celerite2_2018}); \texttt{juliet} \citep{Espinoza_2019}; \texttt{batman} \citep{Kreidberg_2015}; \texttt{SciPy} \citep{Virtanen_2020}; \texttt{NumPy} \citep{Harris_2020}; \texttt{photutils} \citep{Larry_Bradley_2020}; \texttt{Tapir} \citep{Jensen_2013}; \texttt{AstroImageJ} \citep{Collins_2017}}; \texttt{TRILEGAL} \citep{Girardi_2012}.

\bibliography{TOI-1452.bib}{}

\begin{thebibliography}{}
\expandafter\ifx\csname natexlab\endcsname\relax\def\natexlab#1{#1}\fi
\providecommand{\url}[1]{\href{#1}{#1}}
\providecommand{\dodoi}[1]{doi:~\href{http://doi.org/#1}{\nolinkurl{#1}}}
\providecommand{\doeprint}[1]{\href{http://ascl.net/#1}{\nolinkurl{http://ascl.net/#1}}}
\providecommand{\doarXiv}[1]{\href{https://arxiv.org/abs/#1}{\nolinkurl{https://arxiv.org/abs/#1}}}

\bibitem[{{Agol} {et~al.}(2021){Agol}, {Dorn}, {Grimm}, {Turbet}, {Ducrot},
  {Delrez}, {Gillon}, {Demory}, {Burdanov}, {Barkaoui}, {Benkhaldoun},
  {Bolmont}, {Burgasser}, {Carey}, {de Wit}, {Fabrycky}, {Foreman-Mackey},
  {Haldemann}, {Hernandez}, {Ingalls}, {Jehin}, {Langford}, {Leconte},
  {Lederer}, {Luger}, {Malhotra}, {Meadows}, {Morris}, {Pozuelos}, {Queloz},
  {Raymond}, {Selsis}, {Sestovic}, {Triaud}, \& {Van Grootel}}]{Agol_2021}
{Agol}, E., {Dorn}, C., {Grimm}, S.~L., {et~al.} 2021, Planet. Sci. J., 2, 1,
  \dodoi{10.3847/PSJ/abd022}

\bibitem[{{Ahumada} {et~al.}(2020){Ahumada}, {Prieto}, {Almeida}, {Anders},
  {Anderson}, {Andrews}, {Anguiano}, {Arcodia}, {Armengaud}, {Aubert}, {Avila},
  {Avila-Reese}, {Badenes}, {Balland}, {Barger}, {Barrera-Ballesteros}, {Basu},
  {Bautista}, {Beaton}, {Beers}, {Benavides}, {Bender}, {Bernardi}, {Bershady},
  {Beutler}, {Bidin}, {Bird}, {Bizyaev}, {Blanc}, {Blanton}, {Boquien},
  {Borissova}, {Bovy}, {Brandt}, {Brinkmann}, {Brownstein}, {Bundy}, {Bureau},
  {Burgasser}, {Burtin}, {Cano-D{\'\i}az}, {Capasso}, {Cappellari}, {Carrera},
  {Chabanier}, {Chaplin}, {Chapman}, {Cherinka}, {Chiappini}, {Doohyun Choi},
  {Chojnowski}, {Chung}, {Clerc}, {Coffey}, {Comerford}, {Comparat}, {da
  Costa}, {Cousinou}, {Covey}, {Crane}, {Cunha}, {Ilha}, {Dai}, {Damsted},
  {Darling}, {Davidson}, {Davies}, {Dawson}, {De}, {de la Macorra}, {De Lee},
  {Queiroz}, {Deconto Machado}, {de la Torre}, {Dell'Agli}, {du Mas des
  Bourboux}, {Diamond-Stanic}, {Dillon}, {Donor}, {Drory}, {Duckworth},
  {Dwelly}, {Ebelke}, {Eftekharzadeh}, {Davis Eigenbrot}, {Elsworth},
  {Eracleous}, {Erfanianfar}, {Escoffier}, {Fan}, {Farr},
  {Fern{\'a}ndez-Trincado}, {Feuillet}, {Finoguenov}, {Fofie},
  {Fraser-McKelvie}, {Frinchaboy}, {Fromenteau}, {Fu}, {Galbany}, {Garcia},
  {Garc{\'\i}a-Hern{\'a}ndez}, {Oehmichen}, {Ge}, {Maia}, {Geisler}, {Gelfand},
  {Goddy}, {Gonzalez-Perez}, {Grabowski}, {Green}, {Grier}, {Guo}, {Guy},
  {Harding}, {Hasselquist}, {Hawken}, {Hayes}, {Hearty}, {Hekker}, {Hogg},
  {Holtzman}, {Horta}, {Hou}, {Hsieh}, {Huber}, {Hunt}, {Chitham}, {Imig},
  {Jaber}, {Angel}, {Johnson}, {Jones}, {J{\"o}nsson}, {Jullo}, {Kim},
  {Kinemuchi}, {Kirkpatrick}, {Kite}, {Klaene}, {Kneib}, {Kollmeier}, {Kong},
  {Kounkel}, {Krishnarao}, {Lacerna}, {Lan}, {Lane}, {Law}, {Le Goff}, {Leung},
  {Lewis}, {Li}, {Lian}, {Lin}, {Long}, {Longa-Pe{\~n}a}, {Lundgren}, {Lyke},
  {Ted Mackereth}, {MacLeod}, {Majewski}, {Manchado}, {Maraston}, {Martini},
  {Masseron}, {Masters}, {Mathur}, {McDermid}, {Merloni}, {Merrifield},
  {M{\'e}sz{\'a}ros}, {Miglio}, {Minniti}, {Minsley}, {Miyaji}, {Mohammad},
  {Mosser}, {Mueller}, {Muna}, {Mu{\~n}oz-Guti{\'e}rrez}, {Myers}, {Nadathur},
  {Nair}, {Nandra}, {do Nascimento}, {Nevin}, {Newman}, {Nidever}, {Nitschelm},
  {Noterdaeme}, {O'Connell}, {Olmstead}, {Oravetz}, {Oravetz}, {Osorio},
  {Pace}, {Padilla}, {Palanque-Delabrouille}, {Palicio}, {Pan}, {Pan},
  {Parker}, {Paviot}, {Peirani}, {Ram{\'r}ez}, {Penny}, {Percival},
  {Perez-Fournon}, {P{\'e}rez-R{\`a}fols}, {Petitjean}, {Pieri},
  {Pinsonneault}, {Poovelil}, {Povick}, {Prakash}, {Price-Whelan}, {Raddick},
  {Raichoor}, {Ray}, {Rembold}, {Rezaie}, {Riffel}, {Riffel}, {Rix}, {Robin},
  {Roman-Lopes}, {Rom{\'a}n-Z{\'u}{\~n}iga}, {Rose}, {Ross}, {Rossi},
  {Rowlands}, {Rubin}, {Salvato}, {S{\'a}nchez}, {S{\'a}nchez-Menguiano},
  {S{\'a}nchez-Gallego}, {Sayres}, {Schaefer}, {Schiavon}, {Schimoia},
  {Schlafly}, {Schlegel}, {Schneider}, {Schultheis}, {Schwope}, {Seo},
  {Serenelli}, {Shafieloo}, {Shamsi}, {Shao}, {Shen}, {Shetrone}, {Shirley},
  {Aguirre}, {Simon}, {Skrutskie}, {Slosar}, {Smethurst}, {Sobeck}, {Sodi},
  {Souto}, {Stark}, {Stassun}, {Steinmetz}, {Stello}, {Stermer},
  {Storchi-Bergmann}, {Streblyanska}, {Stringfellow}, {Stutz}, {Su{\'a}rez},
  {Sun}, {Taghizadeh-Popp}, {Talbot}, {Tayar}, {Thakar}, {Theriault}, {Thomas},
  {Thomas}, {Tinker}, {Tojeiro}, {Toledo}, {Tremonti}, {Troup}, {Tuttle},
  {Unda-Sanzana}, {Valentini}, {Vargas-Gonz{\'a}lez}, {Vargas-Maga{\~n}a},
  {V{\'a}zquez-Mata}, {Vivek}, {Wake}, {Wang}, {Weaver}, {Weijmans}, {Wild},
  {Wilson}, {Wilson}, {Wolthuis}, {Wood-Vasey}, {Yan}, {Yang}, {Y{\`e}che},
  {Zamora}, {Zarrouk}, {Zasowski}, {Zhang}, {Zhao}, {Zhao}, {Zheng}, {Zheng},
  {Zhu}, \& {Zou}}]{Ahumada_2020}
{Ahumada}, R., {Prieto}, C.~A., {Almeida}, A., {et~al.} 2020, \apjs, 249, 3,
  \dodoi{10.3847/1538-4365/ab929e}

\bibitem[{{Akeson} {et~al.}(2013){Akeson}, {Chen}, {Ciardi}, {Crane}, {Good},
  {Harbut}, {Jackson}, {Kane}, {Laity}, {Leifer}, {Lynn}, {McElroy}, {Papin},
  {Plavchan}, {Ram{\'\i}rez}, {Rey}, {von Braun}, {Wittman}, {Abajian}, {Ali},
  {Beichman}, {Beekley}, {Berriman}, {Berukoff}, {Bryden}, {Chan}, {Groom},
  {Lau}, {Payne}, {Regelson}, {Saucedo}, {Schmitz}, {Stauffer}, {Wyatt}, \&
  {Zhang}}]{Akeson_2013}
{Akeson}, R.~L., {Chen}, X., {Ciardi}, D., {et~al.} 2013, \pasp, 125, 989,
  \dodoi{10.1086/672273}

\bibitem[{{Allard} {et~al.}(2012{\natexlab{a}}){Allard}, {Homeier}, \&
  {Freytag}}]{allard2012models}
{Allard}, F., {Homeier}, D., \& {Freytag}, B. 2012{\natexlab{a}}, Philosophical
  Transactions of the Royal Society of London Series A, 370, 2765,
  \dodoi{10.1098/rsta.2011.0269}

\bibitem[{{Allard} {et~al.}(2013){Allard}, {Homeier}, {Freytag},
  {Schaffenberger}, \& {Rajpurohit}}]{Allard_2013}
{Allard}, F., {Homeier}, D., {Freytag}, B., {Schaffenberger}, W., \&
  {Rajpurohit}, A.~S. 2013, Memorie della Societa Astronomica Italiana
  Supplementi, 24, 128.
\newblock \doarXiv{1302.6559}

\bibitem[{{Allard} {et~al.}(2012{\natexlab{b}}){Allard}, {Homeier}, {Freytag},
  \& {Sharp}}]{Allard_2012}
{Allard}, F., {Homeier}, D., {Freytag}, B., \& {Sharp}, C.~M.
  2012{\natexlab{b}}, in EAS Publications Series, Vol.~57, EAS Publications
  Series, ed. C.~{Reyl{\'e}}, C.~{Charbonnel}, \& M.~{Schultheis}, 3--43,
  \dodoi{10.1051/eas/1257001}

\bibitem[{Artigau {et~al.}(2018)Artigau, Saint-Antoine, Lévesque, Vallée,
  Doyon, Hernandez, \& Moutou}]{Artigau_2018}
Artigau, E., Saint-Antoine, J., Lévesque, P.-L., {et~al.} 2018, High Energy,
  Optical, and Infrared Detectors for Astronomy VIII, 10709, 107091P,
  \dodoi{10.1117/12.2314475}

\bibitem[{{Artigau} {et~al.}(2014){Artigau}, {Astudillo-Defru}, {Delfosse},
  {Bouchy}, {Bonfils}, {Lovis}, {Pepe}, {Moutou}, {Donati}, {Doyon}, \&
  {Malo}}]{Artigau_2014}
{Artigau}, {\'E}., {Astudillo-Defru}, N., {Delfosse}, X., {et~al.} 2014, in
  Society of Photo-Optical Instrumentation Engineers (SPIE) Conference Series,
  Vol. 9149, Observatory Operations: Strategies, Processes, and Systems V, ed.
  A.~B. {Peck}, C.~R. {Benn}, \& R.~L. {Seaman}, 914905,
  \dodoi{10.1117/12.2056385}

\bibitem[{{Artigau} {et~al.}(2021){Artigau}, {H{\'e}brard}, {Cadieux},
  {Vandal}, {Cook}, {Doyon}, {Gagn{\'e}}, {Moutou}, {Martioli}, {Frasca},
  {Jahandar}, {Lafreni{\`e}re}, {Malo}, {Donati}, {Cort{\'e}s-Zuleta},
  {Boisse}, {Delfosse}, {Carmona}, {Fouqu{\'e}}, {Morin}, {Rowe}, {Marino},
  {Papini}, {Ciardi}, {Lund}, {Martins}, {Pelletier}, {Arnold}, {Bouchy},
  {Forveille}, {Santos}, {Bonfils}, {Figueira}, {Fausnaugh}, {Ricker},
  {Latham}, {Seager}, {Winn}, {Jenkins}, {Ting}, {Torres}, \& {Gomes da
  Silva}}]{Artigau_2021}
{Artigau}, {\'E}., {H{\'e}brard}, G., {Cadieux}, C., {et~al.} 2021, \aj, 162,
  144, \dodoi{10.3847/1538-3881/ac096d}

\bibitem[{{Artigau} {et~al.}(2022){Artigau}, Cadieux, Cook, Doyon, Vandal,
  Donati, Moutou, Delfosse, Fouqué, Martioli, Bouchy, Parsons, Carmona,
  Dumusque, Astudillo-Defru, Bonfils, \& Mignon}]{Artigau_2022}
{Artigau}, E., Cadieux, C., Cook, N.~J., {et~al.} 2022, Line-by-line velocity
  measurements, an outlier-resistant method for precision velocimetry.
\newblock \doarXiv{2207.13524}

\bibitem[{{Asplund} {et~al.}(2009){Asplund}, {Grevesse}, {Sauval}, \&
  {Scott}}]{Asplund2009}
{Asplund}, M., {Grevesse}, N., {Sauval}, A.~J., \& {Scott}, P. 2009, \araa, 47,
  481, \dodoi{10.1146/annurev.astro.46.060407.145222}

\bibitem[{{Astropy Collaboration} {et~al.}(2018){Astropy Collaboration},
  {Price-Whelan}, {Sip{\H{o}}cz}, {G{\"u}nther}, {Lim}, {Crawford}, {Conseil},
  {Shupe}, {Craig}, {Dencheva}, {Ginsburg}, {VanderPlas}, {Bradley},
  {P{\'e}rez-Su{\'a}rez}, {de Val-Borro}, {Aldcroft}, {Cruz}, {Robitaille},
  {Tollerud}, {Ardelean}, {Babej}, {Bach}, {Bachetti}, {Bakanov}, {Bamford},
  {Barentsen}, {Barmby}, {Baumbach}, {Berry}, {Biscani}, {Boquien}, {Bostroem},
  {Bouma}, {Brammer}, {Bray}, {Breytenbach}, {Buddelmeijer}, {Burke},
  {Calderone}, {Cano Rodr{\'\i}guez}, {Cara}, {Cardoso}, {Cheedella}, {Copin},
  {Corrales}, {Crichton}, {D'Avella}, {Deil}, {Depagne}, {Dietrich}, {Donath},
  {Droettboom}, {Earl}, {Erben}, {Fabbro}, {Ferreira}, {Finethy}, {Fox},
  {Garrison}, {Gibbons}, {Goldstein}, {Gommers}, {Greco}, {Greenfield},
  {Groener}, {Grollier}, {Hagen}, {Hirst}, {Homeier}, {Horton}, {Hosseinzadeh},
  {Hu}, {Hunkeler}, {Ivezi{\'c}}, {Jain}, {Jenness}, {Kanarek}, {Kendrew},
  {Kern}, {Kerzendorf}, {Khvalko}, {King}, {Kirkby}, {Kulkarni}, {Kumar},
  {Lee}, {Lenz}, {Littlefair}, {Ma}, {Macleod}, {Mastropietro}, {McCully},
  {Montagnac}, {Morris}, {Mueller}, {Mumford}, {Muna}, {Murphy}, {Nelson},
  {Nguyen}, {Ninan}, {N{\"o}the}, {Ogaz}, {Oh}, {Parejko}, {Parley}, {Pascual},
  {Patil}, {Patil}, {Plunkett}, {Prochaska}, {Rastogi}, {Reddy Janga},
  {Sabater}, {Sakurikar}, {Seifert}, {Sherbert}, {Sherwood-Taylor}, {Shih},
  {Sick}, {Silbiger}, {Singanamalla}, {Singer}, {Sladen}, {Sooley},
  {Sornarajah}, {Streicher}, {Teuben}, {Thomas}, {Tremblay}, {Turner},
  {Terr{\'o}n}, {van Kerkwijk}, {de la Vega}, {Watkins}, {Weaver}, {Whitmore},
  {Woillez}, {Zabalza}, \& {Astropy Contributors}}]{Astropy_2018}
{Astropy Collaboration}, {Price-Whelan}, A.~M., {Sip{\H{o}}cz}, B.~M., {et~al.}
  2018, \aj, 156, 123, \dodoi{10.3847/1538-3881/aabc4f}

\bibitem[{{Barnes} \& {Heller}(2013)}]{Barnes2013}
{Barnes}, R., \& {Heller}, R. 2013, Astrobiology, 13, 279,
  \dodoi{10.1089/ast.2012.0867}

\bibitem[{{Batalha} {et~al.}(2019){Batalha}, {Lewis}, {Fortney}, {Batalha},
  {Kempton}, {Lewis}, \& {Line}}]{Batalha_2019}
{Batalha}, N.~E., {Lewis}, T., {Fortney}, J.~J., {et~al.} 2019, \apjl, 885,
  L25, \dodoi{10.3847/2041-8213/ab4909}

\bibitem[{{Bean} {et~al.}(2018){Bean}, {Stevenson}, {Batalha},
  {Berta-Thompson}, {Kreidberg}, {Crouzet}, {Benneke}, {Line}, {Sing},
  {Wakeford}, {Knutson}, {Kempton}, {D{\'e}sert}, {Crossfield}, {Batalha}, {de
  Wit}, {Parmentier}, {Harrington}, {Moses}, {Lopez-Morales}, {Alam}, {Blecic},
  {Bruno}, {Carter}, {Chapman}, {Decin}, {Dragomir}, {Evans}, {Fortney},
  {Fraine}, {Gao}, {Garc{\'\i}a Mu{\~n}oz}, {Gibson}, {Goyal}, {Heng}, {Hu},
  {Kendrew}, {Kilpatrick}, {Krick}, {Lagage}, {Lendl}, {Louden}, {Madhusudhan},
  {Mandell}, {Mansfield}, {May}, {Morello}, {Morley}, {Nikolov}, {Redfield},
  {Roberts}, {Schlawin}, {Spake}, {Todorov}, {Tsiaras}, {Venot}, {Waalkes},
  {Wheatley}, {Zellem}, {Angerhausen}, {Barrado}, {Carone}, {Casewell},
  {Cubillos}, {Damiano}, {de Val-Borro}, {Drummond}, {Edwards}, {Endl},
  {Espinoza}, {France}, {Gizis}, {Greene}, {Henning}, {Hong}, {Ingalls}, {Iro},
  {Irwin}, {Kataria}, {Lahuis}, {Leconte}, {Lillo-Box}, {Lines}, {Lothringer},
  {Mancini}, {Marchis}, {Mayne}, {Palle}, {Rauscher}, {Roudier}, {Shkolnik},
  {Southworth}, {Swain}, {Taylor}, {Teske}, {Tinetti}, {Tremblin}, {Tucker},
  {van Boekel}, {Waldmann}, {Weaver}, \& {Zingales}}]{Bean_2018}
{Bean}, J.~L., {Stevenson}, K.~B., {Batalha}, N.~M., {et~al.} 2018, \pasp, 130,
  114402, \dodoi{10.1088/1538-3873/aadbf3}

\bibitem[{{Benneke} {et~al.}(2019){Benneke}, {Wong}, {Piaulet}, {Knutson},
  {Lothringer}, {Morley}, {Crossfield}, {Gao}, {Greene}, {Dressing},
  {Dragomir}, {Howard}, {McCullough}, {Kempton}, {Fortney}, \&
  {Fraine}}]{Benneke_2019}
{Benneke}, B., {Wong}, I., {Piaulet}, C., {et~al.} 2019, \apjl, 887, L14,
  \dodoi{10.3847/2041-8213/ab59dc}

\bibitem[{{Berger} {et~al.}(2020){Berger}, {Huber}, {Gaidos}, {van Saders}, \&
  {Weiss}}]{Berger_2020}
{Berger}, T.~A., {Huber}, D., {Gaidos}, E., {van Saders}, J.~L., \& {Weiss},
  L.~M. 2020, \aj, 160, 108, \dodoi{10.3847/1538-3881/aba18a}

\bibitem[{Bertaux {et~al.}(2014)Bertaux, Lallement, Ferron, Boonne, \&
  Bodichon}]{Bertaux_2014}
Bertaux, J.~L., Lallement, R., Ferron, S., Boonne, C., \& Bodichon, R. 2014,
  Astronomy \& Astrophysics, 564, A46, \dodoi{10.1051/0004-6361/201322383}

\bibitem[{{Bluhm} {et~al.}(2021){Bluhm}, {Pall{\'e}}, {Molaverdikhani},
  {Kemmer}, {Hatzes}, {Kossakowski}, {Stock}, {Caballero}, {Lillo-Box},
  {B{\'e}jar}, {Soto}, {Amado}, {Brown}, {Cadieux}, {Cloutier}, {Collins},
  {Collins}, {Cort{\'e}s-Contreras}, {Doyon}, {Dreizler}, {Espinoza}, {Fukui},
  {Gonz{\'a}lez-{\'A}lvarez}, {Henning}, {Horne}, {Jeffers}, {Jenkins},
  {Jensen}, {Kaminski}, {Kielkopf}, {Kusakabe}, {K{\"u}rster},
  {Lafreni{\`e}re}, {Luque}, {Murgas}, {Montes}, {Morales}, {Narita},
  {Passegger}, {Quirrenbach}, {Sch{\"o}fer}, {Reffert}, {Reiners}, {Ribas},
  {Ricker}, {Seager}, {Schweitzer}, {Schwarz}, {Tamura}, {Trifonov},
  {Vanderspek}, {Winn}, {Zechmeister}, \& {Zapatero Osorio}}]{Bluhm_2021}
{Bluhm}, P., {Pall{\'e}}, E., {Molaverdikhani}, K., {et~al.} 2021, \aap, 650,
  A78, \dodoi{10.1051/0004-6361/202140688}

\bibitem[{{Bolmont} {et~al.}(2017){Bolmont}, {Selsis}, {Owen}, {Ribas},
  {Raymond}, {Leconte}, \& {Gillon}}]{Bolmont2017}
{Bolmont}, E., {Selsis}, F., {Owen}, J.~E., {et~al.} 2017, \mnras, 464, 3728,
  \dodoi{10.1093/mnras/stw2578}

\bibitem[{{Boucher} {et~al.}(2021){Boucher}, {Darveau-Bernier}, {Pelletier},
  {Lafreni{\`e}re}, {Artigau}, {Cook}, {Allart}, {Radica}, {Doyon}, {Benneke},
  {Arnold}, {Bonfils}, {Bourrier}, {Cloutier}, {Gomes da Silva}, {Deibert},
  {Delfosse}, {Donati}, {Ehrenreich}, {Figueira}, {Forveille}, {Fouqu{\'e}},
  {Gagn{\'e}}, {Gaidos}, {H{\'e}brard}, {Jayawardhana}, {Klein}, {Lovis},
  {Martins}, {Martioli}, {Moutou}, \& {Santos}}]{Boucher_2021}
{Boucher}, A., {Darveau-Bernier}, A., {Pelletier}, S., {et~al.} 2021, \aj, 162,
  233, \dodoi{10.3847/1538-3881/ac1f8e}

\bibitem[{{Bouchy} {et~al.}(2001){Bouchy}, {Pepe}, \& {Queloz}}]{Bouchy_2001}
{Bouchy}, F., {Pepe}, F., \& {Queloz}, D. 2001, \aap, 374, 733,
  \dodoi{10.1051/0004-6361:20010730}

\bibitem[{Bradley {et~al.}(2020)Bradley, Sip{\H o}cz, Robitaille, Tollerud,
  Vin{\'{\i}}cius, Deil, Barbary, Wilson, Busko, G{\"u}nther, Cara, Conseil,
  Bostroem, Droettboom, Bray, Bratholm, Lim, Barentsen, Craig, Pascual, Perren,
  Greco, Donath, de~Val-Borro, Kerzendorf, Bach, Weaver, D'Eugenio, Souchereau,
  \& Ferreira}]{Larry_Bradley_2020}
Bradley, L., Sip{\H o}cz, B., Robitaille, T., {et~al.} 2020, astropy/photutils:
  1.0.0, 1.0.0,  Zenodo, \dodoi{10.5281/zenodo.4044744}

\bibitem[{{Burt} {et~al.}(2021){Burt}, {Dragomir}, {Molli{\`e}re},
  {Youngblood}, {Garc{\'\i}a Mu{\~n}oz}, {McCann}, {Kreidberg}, {Huang},
  {Collins}, {Eastman}, {Abe}, {Almenara}, {Crossfield}, {Ziegler},
  {Rodriguez}, {Mamajek}, {Stassun}, {Halverson}, {Villanueva}, {Butler},
  {Wang}, {Schwarz}, {Ricker}, {Vanderspek}, {Latham}, {Seager}, {Winn},
  {Jenkins}, {Agabi}, {Bonfils}, {Ciardi}, {Cointepas}, {Crane}, {Crouzet},
  {Dransfield}, {Feng}, {Furlan}, {Guillot}, {Gupta}, {Howell}, {Jensen},
  {Law}, {Mann}, {Marie-Sainte}, {Matson}, {Matthews}, {M{\'e}karnia},
  {Pepper}, {Scott}, {Shectman}, {Schlieder}, {Schmider}, {Stevens}, {Teske},
  {Triaud}, {Charbonneau}, {Berta-Thompson}, {Burke}, {Daylan}, {Barclay},
  {Wohler}, \& {Brasseur}}]{Burt_2021}
{Burt}, J.~A., {Dragomir}, D., {Molli{\`e}re}, P., {et~al.} 2021, \aj, 162, 87,
  \dodoi{10.3847/1538-3881/ac0432}

\bibitem[{{Chabrier}(2001)}]{Chabrier_2001}
{Chabrier}, G. 2001, \apj, 554, 1274, \dodoi{10.1086/321401}

\bibitem[{{Chambers} {et~al.}(2016){Chambers}, {Magnier}, {Metcalfe},
  {Flewelling}, {Huber}, {Waters}, {Denneau}, {Draper}, {Farrow}, {Finkbeiner},
  {Holmberg}, {Koppenhoefer}, {Price}, {Rest}, {Saglia}, {Schlafly}, {Smartt},
  {Sweeney}, {Wainscoat}, {Burgett}, {Chastel}, {Grav}, {Heasley}, {Hodapp},
  {Jedicke}, {Kaiser}, {Kudritzki}, {Luppino}, {Lupton}, {Monet}, {Morgan},
  {Onaka}, {Shiao}, {Stubbs}, {Tonry}, {White}, {Ba{\~n}ados}, {Bell},
  {Bender}, {Bernard}, {Boegner}, {Boffi}, {Botticella}, {Calamida},
  {Casertano}, {Chen}, {Chen}, {Cole}, {Deacon}, {Frenk}, {Fitzsimmons},
  {Gezari}, {Gibbs}, {Goessl}, {Goggia}, {Gourgue}, {Goldman}, {Grant},
  {Grebel}, {Hambly}, {Hasinger}, {Heavens}, {Heckman}, {Henderson}, {Henning},
  {Holman}, {Hopp}, {Ip}, {Isani}, {Jackson}, {Keyes}, {Koekemoer}, {Kotak},
  {Le}, {Liska}, {Long}, {Lucey}, {Liu}, {Martin}, {Masci}, {McLean}, {Mindel},
  {Misra}, {Morganson}, {Murphy}, {Obaika}, {Narayan}, {Nieto-Santisteban},
  {Norberg}, {Peacock}, {Pier}, {Postman}, {Primak}, {Rae}, {Rai}, {Riess},
  {Riffeser}, {Rix}, {R{\"o}ser}, {Russel}, {Rutz}, {Schilbach}, {Schultz},
  {Scolnic}, {Strolger}, {Szalay}, {Seitz}, {Small}, {Smith}, {Soderblom},
  {Taylor}, {Thomson}, {Taylor}, {Thakar}, {Thiel}, {Thilker}, {Unger},
  {Urata}, {Valenti}, {Wagner}, {Walder}, {Walter}, {Watters}, {Werner},
  {Wood-Vasey}, \& {Wyse}}]{Chambers_2016}
{Chambers}, K.~C., {Magnier}, E.~A., {Metcalfe}, N., {et~al.} 2016, arXiv
  e-prints, arXiv:1612.05560.
\newblock \doarXiv{1612.05560}

\bibitem[{{Chen} \& {Rogers}(2016)}]{Chen_2016}
{Chen}, H., \& {Rogers}, L.~A. 2016, \apj, 831, 180,
  \dodoi{10.3847/0004-637X/831/2/180}

\bibitem[{{Cloutier} \& {Menou}(2020)}]{Cloutier-Menou_2020}
{Cloutier}, R., \& {Menou}, K. 2020, \aj, 159, 211,
  \dodoi{10.3847/1538-3881/ab8237}

\bibitem[{{Cloutier} {et~al.}(2019){Cloutier}, {Astudillo-Defru}, {Bonfils},
  {Jenkins}, {Berdi{\~n}as}, {Ricker}, {Vanderspek}, {Latham}, {Seager},
  {Winn}, {Jenkins}, {Almenara}, {Bouchy}, {Delfosse}, {D{\'\i}az},
  {D{\'\i}az}, {Doyon}, {Figueira}, {Forveille}, {Kurtovic}, {Lovis}, {Mayor},
  {Menou}, {Morgan}, {Morris}, {Muirhead}, {Murgas}, {Pepe}, {Santos},
  {S{\'e}gransan}, {Smith}, {Tenenbaum}, {Torres}, {Udry}, {Vezie}, \&
  {Villasenor}}]{Cloutier_2019}
{Cloutier}, R., {Astudillo-Defru}, N., {Bonfils}, X., {et~al.} 2019, \aap, 629,
  A111, \dodoi{10.1051/0004-6361/201935957}

\bibitem[{{Cloutier} {et~al.}(2020{\natexlab{a}}){Cloutier}, {Eastman},
  {Rodriguez}, {Astudillo-Defru}, {Bonfils}, {Mortier}, {Watson}, {Stalport},
  {Pinamonti}, {Lienhard}, {Harutyunyan}, {Damasso}, {Latham}, {Collins},
  {Massey}, {Irwin}, {Winters}, {Charbonneau}, {Ziegler}, {Matthews},
  {Crossfield}, {Kreidberg}, {Quinn}, {Ricker}, {Vanderspek}, {Seager}, {Winn},
  {Jenkins}, {Vezie}, {Udry}, {Twicken}, {Tenenbaum}, {Sozzetti},
  {S{\'e}gransan}, {Schlieder}, {Sasselov}, {Santos}, {Rice}, {Rackham},
  {Poretti}, {Piotto}, {Phillips}, {Pepe}, {Molinari}, {Mignon}, {Micela},
  {Melo}, {de Medeiros}, {Mayor}, {Matson}, {Martinez Fiorenzano}, {Mann},
  {Magazz{\'u}}, {Lovis}, {L{\'o}pez-Morales}, {Lopez}, {Lissauer},
  {L{\'e}pine}, {Law}, {Kielkopf}, {Johnson}, {Jensen}, {Howell}, {Gonzales},
  {Ghedina}, {Forveille}, {Figueira}, {Dumusque}, {Dressing}, {Doyon},
  {D{\'\i}az}, {Fabrizio}, {Delfosse}, {Cosentino}, {Conti}, {Collins},
  {Cameron}, {Ciardi}, {Caldwell}, {Burke}, {Buchhave}, {Brice{\~n}o}, {Boyd},
  {Bouchy}, {Beichman}, {Artigau}, \& {Almenara}}]{Cloutier_2020a}
{Cloutier}, R., {Eastman}, J.~D., {Rodriguez}, J.~E., {et~al.}
  2020{\natexlab{a}}, \aj, 160, 3, \dodoi{10.3847/1538-3881/ab91c2}

\bibitem[{{Cloutier} {et~al.}(2020{\natexlab{b}}){Cloutier}, {Rodriguez},
  {Irwin}, {Charbonneau}, {Stassun}, {Mortier}, {Latham}, {Isaacson}, {Howard},
  {Udry}, {Wilson}, {Watson}, {Pinamonti}, {Lienhard}, {Giacobbe}, {Guerra},
  {Collins}, {Beiryla}, {Esquerdo}, {Matthews}, {Matson}, {Howell}, {Furlan},
  {Crossfield}, {Winters}, {Nava}, {Ment}, {Lopez}, {Ricker}, {Vanderspek},
  {Seager}, {Jenkins}, {Ting}, {Tenenbaum}, {Sozzetti}, {Sha}, {S{\'e}gransan},
  {Schlieder}, {Sasselov}, {Roy}, {Robertson}, {Rice}, {Poretti}, {Piotto},
  {Phillips}, {Pepper}, {Pepe}, {Molinari}, {Mocnik}, {Micela}, {Mayor},
  {Martinez Fiorenzano}, {Mallia}, {Lubin}, {Lovis}, {L{\'o}pez-Morales},
  {Kosiarek}, {Kielkopf}, {Kane}, {Jensen}, {Isopi}, {Huber}, {Hill},
  {Harutyunyan}, {Gonzales}, {Giacalone}, {Ghedina}, {Ercolino}, {Dumusque},
  {Dressing}, {Damasso}, {Dalba}, {Cosentino}, {Conti}, {Col{\'o}n}, {Collins},
  {Cameron}, {Ciardi}, {Christiansen}, {Chontos}, {Cecconi}, {Caldwell},
  {Burke}, {Buchhave}, {Beichman}, {Behmard}, {Beard}, \& {Akana
  Murphy}}]{Cloutier_2020b}
{Cloutier}, R., {Rodriguez}, J.~E., {Irwin}, J., {et~al.} 2020{\natexlab{b}},
  \aj, 160, 22, \dodoi{10.3847/1538-3881/ab9534}

\bibitem[{{Cloutier} {et~al.}(2021){Cloutier}, {Charbonneau}, {Stassun},
  {Murgas}, {Mortier}, {Massey}, {Lissauer}, {Latham}, {Irwin}, {Haywood},
  {Guerra}, {Girardin}, {Giacalone}, {Bosch-Cabot}, {Bieryla}, {Winn},
  {Watson}, {Vanderspek}, {Udry}, {Tamura}, {Sozzetti}, {Shporer},
  {S{\'e}gransan}, {Seager}, {Savel}, {Sasselov}, {Rose}, {Ricker}, {Rice},
  {Quintana}, {Quinn}, {Piotto}, {Phillips}, {Pepe}, {Pedani}, {Parviainen},
  {Palle}, {Narita}, {Molinari}, {Micela}, {McDermott}, {Mayor}, {Matson},
  {Martinez Fiorenzano}, {Lovis}, {L{\'o}pez-Morales}, {Kusakabe}, {Jensen},
  {Jenkins}, {Huang}, {Howell}, {Harutyunyan}, {F{\H{u}}r{\'e}sz}, {Fukui},
  {Esquerdo}, {Esparza-Borges}, {Dumusque}, {Dressing}, {Fabrizio}, {Collins},
  {Cameron}, {Christiansen}, {Cecconi}, {Buchhave}, {Boschin}, \&
  {Andreuzzi}}]{Cloutier_2021}
{Cloutier}, R., {Charbonneau}, D., {Stassun}, K.~G., {et~al.} 2021, \aj, 162,
  79, \dodoi{10.3847/1538-3881/ac0157}

\bibitem[{Collins {et~al.}(2017)Collins, Kielkopf, Stassun, \&
  Hessman}]{Collins_2017}
Collins, K.~A., Kielkopf, J.~F., Stassun, K.~G., \& Hessman, F.~V. 2017, The
  Astronomical Journal, 153, 77, \dodoi{10.3847/1538-3881/153/2/77}

\bibitem[{{Courcol} {et~al.}(2015){Courcol}, {Bouchy}, {Pepe}, {Santerne},
  {Delfosse}, {Arnold}, {Astudillo-Defru}, {Boisse}, {Bonfils}, {Borgniet},
  {Bourrier}, {Cabrera}, {Deleuil}, {Demangeon}, {D{\'\i}az}, {Ehrenreich},
  {Forveille}, {H{\'e}brard}, {Lagrange}, {Montagnier}, {Moutou}, {Rey},
  {Santos}, {S{\'e}gransan}, {Udry}, \& {Wilson}}]{Courcol_2015}
{Courcol}, B., {Bouchy}, F., {Pepe}, F., {et~al.} 2015, \aap, 581, A38,
  \dodoi{10.1051/0004-6361/201526329}

\bibitem[{{Crossfield} {et~al.}(2019){Crossfield}, {Waalkes}, {Newton},
  {Narita}, {Muirhead}, {Ment}, {Matthews}, {Kraus}, {Kostov}, {Kosiarek},
  {Kane}, {Isaacson}, {Halverson}, {Gonzales}, {Everett}, {Dragomir},
  {Collins}, {Chontos}, {Berardo}, {Winters}, {Winn}, {Scott}, {Rojas-Ayala},
  {Rizzuto}, {Petigura}, {Peterson}, {Mocnik}, {Mikal-Evans}, {Mehrle},
  {Matson}, {Kuzuhara}, {Irwin}, {Huber}, {Huang}, {Howell}, {Howard},
  {Hirano}, {Fulton}, {Dupuy}, {Dressing}, {Dalba}, {Charbonneau}, {Burt},
  {Berta-Thompson}, {Benneke}, {Watanabe}, {Twicken}, {Tamura}, {Schlieder},
  {Seager}, {Rose}, {Ricker}, {Quintana}, {L{\'e}pine}, {Latham}, {Kotani},
  {Jenkins}, {Hori}, {Colon}, \& {Caldwell}}]{Crossfield_2019}
{Crossfield}, I. J.~M., {Waalkes}, W., {Newton}, E.~R., {et~al.} 2019, \apjl,
  883, L16, \dodoi{10.3847/2041-8213/ab3d30}

\bibitem[{{Cushing} {et~al.}(2005){Cushing}, {Rayner}, \&
  {Vacca}}]{Cushing_2005}
{Cushing}, M.~C., {Rayner}, J.~T., \& {Vacca}, W.~D. 2005, \apj, 623, 1115,
  \dodoi{10.1086/428040}

\bibitem[{{Damasso} {et~al.}(2018){Damasso}, {Bonomo}, {Astudillo-Defru},
  {Bonfils}, {Malavolta}, {Sozzetti}, {Lopez}, {Zeng}, {Haywood}, {Irwin},
  {Mortier}, {Vanderburg}, {Maldonado}, {Lanza}, {Affer}, {Almenara},
  {Benatti}, {Biazzo}, {Bignamini}, {Borsa}, {Bouchy}, {Buchhave}, {Cameron},
  {Carleo}, {Charbonneau}, {Claudi}, {Cosentino}, {Covino}, {Delfosse},
  {Desidera}, {Di Fabrizio}, {Dressing}, {Esposito}, {Fares}, {Figueira},
  {Fiorenzano}, {Forveille}, {Giacobbe}, {Gonz{\'a}lez-{\'A}lvarez}, {Gratton},
  {Harutyunyan}, {Johnson}, {Latham}, {Leto}, {Lopez-Morales}, {Lovis},
  {Maggio}, {Mancini}, {Masiero}, {Mayor}, {Micela}, {Molinari}, {Motalebi},
  {Murgas}, {Nascimbeni}, {Pagano}, {Pepe}, {Phillips}, {Piotto}, {Poretti},
  {Rainer}, {Rice}, {Santos}, {Sasselov}, {Scandariato}, {S{\'e}gransan},
  {Smareglia}, {Udry}, {Watson}, \& {W{\"u}nsche}}]{Damasso_2018}
{Damasso}, M., {Bonomo}, A.~S., {Astudillo-Defru}, N., {et~al.} 2018, \aap,
  615, A69, \dodoi{10.1051/0004-6361/201732459}

\bibitem[{David {et~al.}(2021)David, Contardo, Sandoval, Angus, Lu, Bedell,
  Curtis, Foreman-Mackey, Fulton, Grunblatt, \& et~al.}]{David_2021}
David, T.~J., Contardo, G., Sandoval, A., {et~al.} 2021, The Astronomical
  Journal, 161, 265, \dodoi{10.3847/1538-3881/abf439}

\bibitem[{{Demangeon} {et~al.}(2021){Demangeon}, {Zapatero Osorio}, {Alibert},
  {Barros}, {Adibekyan}, {Tabernero}, {Antoniadis-Karnavas}, {Camacho},
  {Su{\'a}rez Mascare{\~n}o}, {Oshagh}, {Micela}, {Sousa}, {Lovis}, {Pepe},
  {Rebolo}, {Cristiani}, {Santos}, {Allart}, {Allende Prieto}, {Bossini},
  {Bouchy}, {Cabral}, {Damasso}, {Di Marcantonio}, {D'Odorico}, {Ehrenreich},
  {Faria}, {Figueira}, {G{\'e}nova Santos}, {Haldemann}, {Hara}, {Gonz{\'a}lez
  Hern{\'a}ndez}, {Lavie}, {Lillo-Box}, {Lo Curto}, {Martins}, {M{\'e}gevand},
  {Mehner}, {Molaro}, {Nunes}, {Pall{\'e}}, {Pasquini}, {Poretti}, {Sozzetti},
  \& {Udry}}]{Demangeon_2021}
{Demangeon}, O.~D.~S., {Zapatero Osorio}, M.~R., {Alibert}, Y., {et~al.} 2021,
  \aap, 653, A41, \dodoi{10.1051/0004-6361/202140728}

\bibitem[{{Demory} {et~al.}(2020){Demory}, {Pozuelos}, {G{\'o}mez Maqueo Chew},
  {Sabin}, {Petrucci}, {Schroffenegger}, {Grimm}, {Sestovic}, {Gillon},
  {McCormac}, {Barkaoui}, {Benz}, {Bieryla}, {Bouchy}, {Burdanov}, {Collins},
  {de Wit}, {Dressing}, {Garcia}, {Giacalone}, {Guerra}, {Haldemann}, {Heng},
  {Jehin}, {Jofr{\'e}}, {Kane}, {Lillo-Box}, {Maign{\'e}}, {Mordasini},
  {Morris}, {Niraula}, {Queloz}, {Rackham}, {Savel}, {Soubkiou}, {Srdoc},
  {Stassun}, {Triaud}, {Zambelli}, {Ricker}, {Latham}, {Seager}, {Winn},
  {Jenkins}, {Calvario-Vel{\'a}squez}, {Franco Herrera}, {Colorado}, {Cadena
  Zepeda}, {Figueroa}, {Watson}, {Lugo-Ibarra}, {Carigi}, {Guisa}, {Herrera},
  {Sierra D{\'\i}az}, {Su{\'a}rez}, {Barrado}, {Batalha}, {Benkhaldoun},
  {Chontos}, {Dai}, {Essack}, {Ghachoui}, {Huang}, {Huber}, {Isaacson},
  {Lissauer}, {Morales-Calder{\'o}n}, {Robertson}, {Roy}, {Twicken},
  {Vanderburg}, \& {Weiss}}]{Demory_2020}
{Demory}, B.~O., {Pozuelos}, F.~J., {G{\'o}mez Maqueo Chew}, Y., {et~al.} 2020,
  \aap, 642, A49, \dodoi{10.1051/0004-6361/202038616}

\bibitem[{{Dittmann} {et~al.}(2017){Dittmann}, {Irwin}, {Charbonneau},
  {Bonfils}, {Astudillo-Defru}, {Haywood}, {Berta-Thompson}, {Newton},
  {Rodriguez}, {Winters}, {Tan}, {Almenara}, {Bouchy}, {Delfosse}, {Forveille},
  {Lovis}, {Murgas}, {Pepe}, {Santos}, {Udry}, {W{\"u}nsche}, {Esquerdo},
  {Latham}, \& {Dressing}}]{Dittmann_2017}
{Dittmann}, J.~A., {Irwin}, J.~M., {Charbonneau}, D., {et~al.} 2017, \nat, 544,
  333, \dodoi{10.1038/nature22055}

\bibitem[{{Donati} {et~al.}(1997){Donati}, {Semel}, {Carter}, {Rees}, \&
  {Collier Cameron}}]{Donati_1997}
{Donati}, J.~F., {Semel}, M., {Carter}, B.~D., {Rees}, D.~E., \& {Collier
  Cameron}, A. 1997, \mnras, 291, 658, \dodoi{10.1093/mnras/291.4.658}

\bibitem[{{Donati} {et~al.}(2018){Donati}, {Kouach}, {Lacombe}, {Baratchart},
  {Doyon}, {Delfosse}, {Artigau}, {Moutou}, {H{\'e}brard}, {Bouchy}, {Bouvier},
  {Alencar}, {Saddlemyer}, {Par{\`e}s}, {Rabou}, {Micheau}, {Dolon}, {Barrick},
  {Hernandez}, {Wang}, {Reshetov}, {Striebig}, {Challita}, {Carmona},
  {Tibault}, {Martioli}, {Figueira}, {Boisse}, \& {Pepe}}]{Donati_2018}
{Donati}, J.-F., {Kouach}, D., {Lacombe}, M., {et~al.} 2018, {SPIRou: A NIR
  Spectropolarimeter/High-Precision Velocimeter for the CFHT}, ed. H.~J. {Deeg}
  \& J.~A. {Belmonte}, 107, \dodoi{10.1007/978-3-319-55333-7\_107}

\bibitem[{{Donati} {et~al.}(2020){Donati}, {Kouach}, {Moutou}, {Doyon},
  {Delfosse}, {Artigau}, {Baratchart}, {Lacombe}, {Barrick}, {H{\'e}brard},
  {Bouchy}, {Saddlemyer}, {Par{\`e}s}, {Rabou}, {Micheau}, {Dolon}, {Reshetov},
  {Challita}, {Carmona}, {Striebig}, {Thibault}, {Martioli}, {Cook},
  {Fouqu{\'e}}, {Vermeulen}, {Wang}, {Arnold}, {Pepe}, {Boisse}, {Figueira},
  {Bouvier}, {Ray}, {Feugeade}, {Morin}, {Alencar}, {Hobson}, {Castilho},
  {Udry}, {Santos}, {Hernandez}, {Benedict}, {Vall{\'e}e}, {Gallou}, {Dupieux},
  {Larrieu}, {Perruchot}, {Sottile}, {Moreau}, {Usher}, {Baril}, {Wildi},
  {Chazelas}, {Malo}, {Bonfils}, {Loop}, {Kerley}, {Wevers}, {Dunn}, {Pazder},
  {Macdonald}, {Dubois}, {Carri{\'e}}, {Valentin}, {Henault}, {Yan}, \&
  {Steinmetz}}]{Donati_2020}
{Donati}, J.~F., {Kouach}, D., {Moutou}, C., {et~al.} 2020, \mnras, 498, 5684,
  \dodoi{10.1093/mnras/staa2569}

\bibitem[{{Dreizler} {et~al.}(2020){Dreizler}, {Crossfield}, {Kossakowski},
  {Plavchan}, {Jeffers}, {Kemmer}, {Luque}, {Espinoza}, {Pall{\'e}}, {Stassun},
  {Matthews}, {Cale}, {Caballero}, {Schlecker}, {Lillo-Box}, {Zechmeister},
  {Lalitha}, {Reiners}, {Soubkiou}, {Bitsch}, {Zapatero Osorio}, {Chaturvedi},
  {Hatzes}, {Ricker}, {Vanderspek}, {Latham}, {Seager}, {Winn}, {Jenkins},
  {Aceituno}, {Amado}, {Barkaoui}, {Barbieri}, {Batalha}, {Bauer}, {Benneke},
  {Benkhaldoun}, {Beichman}, {Berberian}, {Burt}, {Butler}, {Caldwell},
  {Chintada}, {Chontos}, {Christiansen}, {Ciardi}, {Cifuentes}, {Collins},
  {Collins}, {Combs}, {Cort{\'e}s-Contreras}, {Crane}, {Daylan}, {Dragomir},
  {Esparza-Borges}, {Evans}, {Feng}, {Flowers}, {Fukui}, {Fulton}, {Furlan},
  {Gaidos}, {Geneser}, {Giacalone}, {Gillon}, {Gonzales}, {Gorjian}, {Hellier},
  {Hidalgo}, {Howard}, {Howell}, {Huber}, {Isaacson}, {Jehin}, {Jensen},
  {Kaminski}, {Kane}, {Kawauchi}, {Kielkopf}, {Klahr}, {Kosiarek}, {Kreidberg},
  {K{\"u}rster}, {Lafarga}, {Livingston}, {Louie}, {Mann}, {Madrigal-Aguado},
  {Matson}, {Mocnik}, {Morales}, {Muirhead}, {Murgas}, {Nandakumar}, {Narita},
  {Nowak}, {Oshagh}, {Parviainen}, {Passegger}, {Pollacco}, {Pozuelos},
  {Quirrenbach}, {Reefe}, {Ribas}, {Robertson}, {Rodr{\'\i}guez-L{\'o}pez},
  {Rose}, {Roy}, {Schweitzer}, {Schlieder}, {Shectman}, {Tanner},
  {{\c{S}}enavc{\i}}, {Teske}, {Twicken}, {Villasenor}, {Wang}, {Weiss},
  {Wittrock}, {Y{\i}lmaz}, \& {Zohrabi}}]{Dreizler_2020}
{Dreizler}, S., {Crossfield}, I.~J.~M., {Kossakowski}, D., {et~al.} 2020, \aap,
  644, A127, \dodoi{10.1051/0004-6361/202038016}

\bibitem[{{Dressing} \& {Charbonneau}(2015)}]{Dressing_2015}
{Dressing}, C.~D., \& {Charbonneau}, D. 2015, \apj, 807, 45,
  \dodoi{10.1088/0004-637X/807/1/45}

\bibitem[{{El-Badry} {et~al.}(2021){El-Badry}, {Rix}, \&
  {Heintz}}]{El-Badry_2021}
{El-Badry}, K., {Rix}, H.-W., \& {Heintz}, T.~M. 2021, \mnras, 506, 2269,
  \dodoi{10.1093/mnras/stab323}

\bibitem[{{Espinoza}(2018)}]{Espinoza_2018}
{Espinoza}, N. 2018, Research Notes of the American Astronomical Society, 2,
  209, \dodoi{10.3847/2515-5172/aaef38}

\bibitem[{{Espinoza} {et~al.}(2019){Espinoza}, {Kossakowski}, \&
  {Brahm}}]{Espinoza_2019}
{Espinoza}, N., {Kossakowski}, D., \& {Brahm}, R. 2019, \mnras, 490, 2262,
  \dodoi{10.1093/mnras/stz2688}

\bibitem[{{Foreman-Mackey}(2018)}]{celerite2_2018}
{Foreman-Mackey}, D. 2018, Research Notes of the American Astronomical Society,
  2, 31, \dodoi{10.3847/2515-5172/aaaf6c}

\bibitem[{{Foreman-Mackey} {et~al.}(2017){Foreman-Mackey}, {Agol},
  {Ambikasaran}, \& {Angus}}]{celerite1_2017}
{Foreman-Mackey}, D., {Agol}, E., {Ambikasaran}, S., \& {Angus}, R. 2017, \aj,
  154, 220, \dodoi{10.3847/1538-3881/aa9332}

\bibitem[{{Foreman-Mackey} {et~al.}(2013){Foreman-Mackey}, {Hogg}, {Lang}, \&
  {Goodman}}]{Foreman-Mackey_2013}
{Foreman-Mackey}, D., {Hogg}, D.~W., {Lang}, D., \& {Goodman}, J. 2013, \pasp,
  125, 306, \dodoi{10.1086/670067}

\bibitem[{{Foreman-Mackey} {et~al.}(2019){Foreman-Mackey}, {Farr}, {Sinha},
  {Archibald}, {Hogg}, {Sanders}, {Zuntz}, {Williams}, {Nelson}, {de
  Val-Borro}, {Erhardt}, {Pashchenko}, \& {Pla}}]{Foreman-Mackey_2019}
{Foreman-Mackey}, D., {Farr}, W., {Sinha}, M., {et~al.} 2019, The Journal of
  Open Source Software, 4, 1864, \dodoi{10.21105/joss.01864}

\bibitem[{{Fressin} {et~al.}(2013){Fressin}, {Torres}, {Charbonneau}, {Bryson},
  {Christiansen}, {Dressing}, {Jenkins}, {Walkowicz}, \&
  {Batalha}}]{Fressin_2013}
{Fressin}, F., {Torres}, G., {Charbonneau}, D., {et~al.} 2013, \apj, 766, 81,
  \dodoi{10.1088/0004-637X/766/2/81}

\bibitem[{{Fukui} {et~al.}(2011){Fukui}, {Narita}, {Tristram}, {Sumi}, {Abe},
  {Itow}, {Sullivan}, {Bond}, {Hirano}, {Tamura}, {Bennett}, {Furusawa},
  {Hayashi}, {Hearnshaw}, {Hosaka}, {Kamiya}, {Kobara}, {Korpela}, {Kilmartin},
  {Lin}, {Ling}, {Makita}, {Masuda}, {Matsubara}, {Miyake}, {Muraki}, {Nagaya},
  {Nishimoto}, {Ohnishi}, {Omori}, {Perrott}, {Rattenbury}, {Saito}, {Skuljan},
  {Suzuki}, {Sweatman}, \& {Wada}}]{2011PASJ...63..287F}
{Fukui}, A., {Narita}, N., {Tristram}, P.~J., {et~al.} 2011, \pasj, 63, 287,
  \dodoi{10.1093/pasj/63.1.287}

\bibitem[{{Fulton} \& {Petigura}(2018)}]{Fulton-Petigura_2018}
{Fulton}, B.~J., \& {Petigura}, E.~A. 2018, \aj, 156, 264,
  \dodoi{10.3847/1538-3881/aae828}

\bibitem[{{Fulton} {et~al.}(2018){Fulton}, {Petigura}, {Blunt}, \&
  {Sinukoff}}]{Fulton_2018}
{Fulton}, B.~J., {Petigura}, E.~A., {Blunt}, S., \& {Sinukoff}, E. 2018, \pasp,
  130, 044504, \dodoi{10.1088/1538-3873/aaaaa8}

\bibitem[{{Fulton} {et~al.}(2017){Fulton}, {Petigura}, {Howard}, {Isaacson},
  {Marcy}, {Cargile}, {Hebb}, {Weiss}, {Johnson}, {Morton}, {Sinukoff},
  {Crossfield}, \& {Hirsch}}]{Fulton_2017}
{Fulton}, B.~J., {Petigura}, E.~A., {Howard}, A.~W., {et~al.} 2017, \aj, 154,
  109, \dodoi{10.3847/1538-3881/aa80eb}

\bibitem[{{Gagn{\'e}} {et~al.}(2018){Gagn{\'e}}, {Mamajek}, {Malo}, {Riedel},
  {Rodriguez}, {Lafreni{\`e}re}, {Faherty}, {Roy-Loubier}, {Pueyo}, {Robin}, \&
  {Doyon}}]{Gagne_2018}
{Gagn{\'e}}, J., {Mamajek}, E.~E., {Malo}, L., {et~al.} 2018, \apj, 856, 23,
  \dodoi{10.3847/1538-4357/aaae09}

\bibitem[{{Gaia Collaboration} {et~al.}(2021){Gaia Collaboration}, {Brown},
  {Vallenari}, {Prusti}, {de Bruijne}, {Babusiaux}, {Biermann}, {Creevey},
  {Evans}, {Eyer}, {Hutton}, {Jansen}, {Jordi}, {Klioner}, {Lammers},
  {Lindegren}, {Luri}, {Mignard}, {Panem}, {Pourbaix}, {Randich}, {Sartoretti},
  {Soubiran}, {Walton}, {Arenou}, {Bailer-Jones}, {Bastian}, {Cropper},
  {Drimmel}, {Katz}, {Lattanzi}, {van Leeuwen}, {Bakker}, {Cacciari},
  {Casta{\~n}eda}, {De Angeli}, {Ducourant}, {Fabricius}, {Fouesneau},
  {Fr{\'e}mat}, {Guerra}, {Guerrier}, {Guiraud}, {Jean-Antoine Piccolo},
  {Masana}, {Messineo}, {Mowlavi}, {Nicolas}, {Nienartowicz}, {Pailler},
  {Panuzzo}, {Riclet}, {Roux}, {Seabroke}, {Sordo}, {Tanga}, {Th{\'e}venin},
  {Gracia-Abril}, {Portell}, {Teyssier}, {Altmann}, {Andrae}, {Bellas-Velidis},
  {Benson}, {Berthier}, {Blomme}, {Brugaletta}, {Burgess}, {Busso}, {Carry},
  {Cellino}, {Cheek}, {Clementini}, {Damerdji}, {Davidson}, {Delchambre},
  {Dell'Oro}, {Fern{\'a}ndez-Hern{\'a}ndez}, {Galluccio}, {Garc{\'\i}a-Lario},
  {Garcia-Reinaldos}, {Gonz{\'a}lez-N{\'u}{\~n}ez}, {Gosset}, {Haigron},
  {Halbwachs}, {Hambly}, {Harrison}, {Hatzidimitriou}, {Heiter},
  {Hern{\'a}ndez}, {Hestroffer}, {Hodgkin}, {Holl}, {Jan{\ss}en}, {Jevardat de
  Fombelle}, {Jordan}, {Krone-Martins}, {Lanzafame}, {L{\"o}ffler}, {Lorca},
  {Manteiga}, {Marchal}, {Marrese}, {Moitinho}, {Mora}, {Muinonen}, {Osborne},
  {Pancino}, {Pauwels}, {Petit}, {Recio-Blanco}, {Richards}, {Riello},
  {Rimoldini}, {Robin}, {Roegiers}, {Rybizki}, {Sarro}, {Siopis}, {Smith},
  {Sozzetti}, {Ulla}, {Utrilla}, {van Leeuwen}, {van Reeven}, {Abbas}, {Abreu
  Aramburu}, {Accart}, {Aerts}, {Aguado}, {Ajaj}, {Altavilla}, {{\'A}lvarez},
  {{\'A}lvarez Cid-Fuentes}, {Alves}, {Anderson}, {Anglada Varela}, {Antoja},
  {Audard}, {Baines}, {Baker}, {Balaguer-N{\'u}{\~n}ez}, {Balbinot}, {Balog},
  {Barache}, {Barbato}, {Barros}, {Barstow}, {Bartolom{\'e}}, {Bassilana},
  {Bauchet}, {Baudesson-Stella}, {Becciani}, {Bellazzini}, {Bernet}, {Bertone},
  {Bianchi}, {Blanco-Cuaresma}, {Boch}, {Bombrun}, {Bossini}, {Bouquillon},
  {Bragaglia}, {Bramante}, {Breedt}, {Bressan}, {Brouillet}, {Bucciarelli},
  {Burlacu}, {Busonero}, {Butkevich}, {Buzzi}, {Caffau}, {Cancelliere},
  {C{\'a}novas}, {Cantat-Gaudin}, {Carballo}, {Carlucci}, {Carnerero},
  {Carrasco}, {Casamiquela}, {Castellani}, {Castro-Ginard}, {Castro Sampol},
  {Chaoul}, {Charlot}, {Chemin}, {Chiavassa}, {Cioni}, {Comoretto}, {Cooper},
  {Cornez}, {Cowell}, {Crifo}, {Crosta}, {Crowley}, {Dafonte}, {Dapergolas},
  {David}, {David}, {de Laverny}, {De Luise}, {De March}, {De Ridder}, {de
  Souza}, {de Teodoro}, {de Torres}, {del Peloso}, {del Pozo}, {Delbo},
  {Delgado}, {Delgado}, {Delisle}, {Di Matteo}, {Diakite}, {Diener},
  {Distefano}, {Dolding}, {Eappachen}, {Edvardsson}, {Enke}, {Esquej}, {Fabre},
  {Fabrizio}, {Faigler}, {Fedorets}, {Fernique}, {Fienga}, {Figueras},
  {Fouron}, {Fragkoudi}, {Fraile}, {Franke}, {Gai}, {Garabato},
  {Garcia-Gutierrez}, {Garc{\'\i}a-Torres}, {Garofalo}, {Gavras}, {Gerlach},
  {Geyer}, {Giacobbe}, {Gilmore}, {Girona}, {Giuffrida}, {Gomel}, {Gomez},
  {Gonzalez-Santamaria}, {Gonz{\'a}lez-Vidal}, {Granvik},
  {Guti{\'e}rrez-S{\'a}nchez}, {Guy}, {Hauser}, {Haywood}, {Helmi}, {Hidalgo},
  {Hilger}, {H{\l}adczuk}, {Hobbs}, {Holland}, {Huckle}, {Jasniewicz},
  {Jonker}, {Juaristi Campillo}, {Julbe}, {Karbevska}, {Kervella}, {Khanna},
  {Kochoska}, {Kontizas}, {Kordopatis}, {Korn}, {Kostrzewa-Rutkowska},
  {Kruszy{\'n}ska}, {Lambert}, {Lanza}, {Lasne}, {Le Campion}, {Le Fustec},
  {Lebreton}, {Lebzelter}, {Leccia}, {Leclerc}, {Lecoeur-Taibi}, {Liao},
  {Licata}, {Lindstr{\o}m}, {Lister}, {Livanou}, {Lobel}, {Madrero Pardo},
  {Managau}, {Mann}, {Marchant}, {Marconi}, {Marcos Santos}, {Marinoni},
  {Marocco}, {Marshall}, {Martin Polo}, {Mart{\'\i}n-Fleitas}, {Masip},
  {Massari}, {Mastrobuono-Battisti}, {Mazeh}, {McMillan}, {Messina},
  {Michalik}, {Millar}, {Mints}, {Molina}, {Molinaro}, {Moln{\'a}r},
  {Montegriffo}, {Mor}, {Morbidelli}, {Morel}, {Morris}, {Mulone}, {Munoz},
  {Muraveva}, {Murphy}, {Musella}, {Noval}, {Ord{\'e}novic}, {Orr{\`u}},
  {Osinde}, {Pagani}, {Pagano}, {Palaversa}, {Palicio}, {Panahi}, {Pawlak},
  {Pe{\~n}alosa Esteller}, {Penttil{\"a}}, {Piersimoni}, {Pineau}, {Plachy},
  {Plum}, {Poggio}, {Poretti}, {Poujoulet}, {Pr{\v{s}}a}, {Pulone}, {Racero},
  {Ragaini}, {Rainer}, {Raiteri}, {Rambaux}, {Ramos}, {Ramos-Lerate}, {Re
  Fiorentin}, {Regibo}, {Reyl{\'e}}, {Ripepi}, {Riva}, {Rixon}, {Robichon},
  {Robin}, {Roelens}, {Rohrbasser}, {Romero-G{\'o}mez}, {Rowell}, {Royer},
  {Rybicki}, {Sadowski}, {Sagrist{\`a} Sell{\'e}s}, {Sahlmann}, {Salgado},
  {Salguero}, {Samaras}, {Sanchez Gimenez}, {Sanna}, {Santove{\~n}a},
  {Sarasso}, {Schultheis}, {Sciacca}, {Segol}, {Segovia}, {S{\'e}gransan},
  {Semeux}, {Shahaf}, {Siddiqui}, {Siebert}, {Siltala}, {Slezak}, {Smart},
  {Solano}, {Solitro}, {Souami}, {Souchay}, {Spagna}, {Spoto}, {Steele},
  {Steidelm{\"u}ller}, {Stephenson}, {S{\"u}veges}, {Szabados}, {Szegedi-Elek},
  {Taris}, {Tauran}, {Taylor}, {Teixeira}, {Thuillot}, {Tonello}, {Torra},
  {Torra}, {Turon}, {Unger}, {Vaillant}, {van Dillen}, {Vanel}, {Vecchiato},
  {Viala}, {Vicente}, {Voutsinas}, {Weiler}, {Wevers}, {Wyrzykowski}, {Yoldas},
  {Yvard}, {Zhao}, {Zorec}, {Zucker}, {Zurbach}, \&
  {Zwitter}}]{Gaia_Collaboration_2021}
{Gaia Collaboration}, {Brown}, A.~G.~A., {Vallenari}, A., {et~al.} 2021, \aap,
  649, A1, \dodoi{10.1051/0004-6361/202039657}

\bibitem[{{Gilbert} {et~al.}(2020){Gilbert}, {Barclay}, {Schlieder},
  {Quintana}, {Hord}, {Kostov}, {Lopez}, {Rowe}, {Hoffman}, {Walkowicz},
  {Silverstein}, {Rodriguez}, {Vanderburg}, {Suissa}, {Airapetian}, {Clement},
  {Raymond}, {Mann}, {Kruse}, {Lissauer}, {Col{\'o}n}, {Kopparapu},
  {Kreidberg}, {Zieba}, {Collins}, {Quinn}, {Howell}, {Ziegler}, {Vrijmoet},
  {Adams}, {Arney}, {Boyd}, {Brande}, {Burke}, {Cacciapuoti}, {Chance},
  {Christiansen}, {Covone}, {Daylan}, {Dineen}, {Dressing}, {Essack},
  {Fauchez}, {Galgano}, {Howe}, {Kaltenegger}, {Kane}, {Lam}, {Lee}, {Lewis},
  {Logsdon}, {Mandell}, {Monsue}, {Mullally}, {Mullally}, {Paudel},
  {Pidhorodetska}, {Plavchan}, {Reyes}, {Rinehart}, {Rojas-Ayala}, {Smith},
  {Stassun}, {Tenenbaum}, {Vega}, {Villanueva}, {Wolf}, {Youngblood}, {Ricker},
  {Vanderspek}, {Latham}, {Seager}, {Winn}, {Jenkins}, {Bakos}, {Brice{\~n}o},
  {Ciardi}, {Cloutier}, {Conti}, {Couperus}, {Di Sora}, {Eisner}, {Everett},
  {Gan}, {Hartman}, {Henry}, {Isopi}, {Jao}, {Jensen}, {Law}, {Mallia},
  {Matson}, {Shappee}, {Le Wood}, \& {Winters}}]{Gilbert_2020}
{Gilbert}, E.~A., {Barclay}, T., {Schlieder}, J.~E., {et~al.} 2020, \aj, 160,
  116, \dodoi{10.3847/1538-3881/aba4b2}

\bibitem[{{Ginzburg} {et~al.}(2018){Ginzburg}, {Schlichting}, \&
  {Sari}}]{Ginzburg_2018}
{Ginzburg}, S., {Schlichting}, H.~E., \& {Sari}, R. 2018, \mnras, 476, 759,
  \dodoi{10.1093/mnras/sty290}

\bibitem[{{Girardi} {et~al.}(2012){Girardi}, {Barbieri}, {Groenewegen},
  {Marigo}, {Bressan}, {Rocha-Pinto}, {Santiago}, {Camargo}, \& {da
  Costa}}]{Girardi_2012}
{Girardi}, L., {Barbieri}, M., {Groenewegen}, M. A.~T., {et~al.} 2012, in
  Astrophysics and Space Science Proceedings, Vol.~26, Red Giants as Probes of
  the Structure and Evolution of the Milky Way, 165,
  \dodoi{10.1007/978-3-642-18418-5\_17}

\bibitem[{{Guerrero} {et~al.}(2021){Guerrero}, {Seager}, {Huang}, {Vanderburg},
  {Garcia Soto}, {Mireles}, {Hesse}, {Fong}, {Glidden}, {Shporer}, {Latham},
  {Collins}, {Quinn}, {Burt}, {Dragomir}, {Crossfield}, {Vanderspek},
  {Fausnaugh}, {Burke}, {Ricker}, {Daylan}, {Essack}, {G{\"u}nther}, {Osborn},
  {Pepper}, {Rowden}, {Sha}, {Villanueva}, {Yahalomi}, {Yu}, {Ballard},
  {Batalha}, {Berardo}, {Chontos}, {Dittmann}, {Esquerdo}, {Mikal-Evans},
  {Jayaraman}, {Krishnamurthy}, {Louie}, {Mehrle}, {Niraula}, {Rackham},
  {Rodriguez}, {Rowden}, {Sousa-Silva}, {Watanabe}, {Wong}, {Zhan},
  {Zivanovic}, {Christiansen}, {Ciardi}, {Swain}, {Lund}, {Mullally},
  {Fleming}, {Rodriguez}, {Boyd}, {Quintana}, {Barclay}, {Col{\'o}n},
  {Rinehart}, {Schlieder}, {Clampin}, {Jenkins}, {Twicken}, {Caldwell},
  {Coughlin}, {Henze}, {Lissauer}, {Morris}, {Rose}, {Smith}, {Tenenbaum},
  {Ting}, {Wohler}, {Bakos}, {Bean}, {Berta-Thompson}, {Bieryla}, {Bouma},
  {Buchhave}, {Butler}, {Charbonneau}, {Doty}, {Ge}, {Holman}, {Howard},
  {Kaltenegger}, {Kane}, {Kjeldsen}, {Kreidberg}, {Lin}, {Minsky}, {Narita},
  {Paegert}, {P{\'a}l}, {Palle}, {Sasselov}, {Spencer}, {Sozzetti}, {Stassun},
  {Torres}, {Udry}, \& {Winn}}]{Guerrero_2021}
{Guerrero}, N.~M., {Seager}, S., {Huang}, C.~X., {et~al.} 2021, \apjs, 254, 39,
  \dodoi{10.3847/1538-4365/abefe1}

\bibitem[{{Guillot} \& {Morel}(1995)}]{Guillot1995}
{Guillot}, T., \& {Morel}, P. 1995, \aaps, 109, 109

\bibitem[{{G{\"u}nther} {et~al.}(2019){G{\"u}nther}, {Pozuelos}, {Dittmann},
  {Dragomir}, {Kane}, {Daylan}, {Feinstein}, {Huang}, {Morton}, {Bonfanti},
  {Bouma}, {Burt}, {Collins}, {Lissauer}, {Matthews}, {Montet}, {Vanderburg},
  {Wang}, {Winters}, {Ricker}, {Vanderspek}, {Latham}, {Seager}, {Winn},
  {Jenkins}, {Armstrong}, {Barkaoui}, {Batalha}, {Bean}, {Caldwell}, {Ciardi},
  {Collins}, {Crossfield}, {Fausnaugh}, {Furesz}, {Gan}, {Gillon}, {Guerrero},
  {Horne}, {Howell}, {Ireland}, {Isopi}, {Jehin}, {Kielkopf}, {Lepine},
  {Mallia}, {Matson}, {Myers}, {Palle}, {Quinn}, {Relles}, {Rojas-Ayala},
  {Schlieder}, {Sefako}, {Shporer}, {Su{\'a}rez}, {Tan}, {Ting}, {Twicken}, \&
  {Waite}}]{Gunther_2019}
{G{\"u}nther}, M.~N., {Pozuelos}, F.~J., {Dittmann}, J.~A., {et~al.} 2019,
  Nature Astronomy, 3, 1099, \dodoi{10.1038/s41550-019-0845-5}

\bibitem[{{Gupta} \& {Schlichting}(2019)}]{Gupta_2019}
{Gupta}, A., \& {Schlichting}, H.~E. 2019, \mnras, 487, 24,
  \dodoi{10.1093/mnras/stz1230}

\bibitem[{{Gupta} \& {Schlichting}(2020)}]{Gupta_2020}
---. 2020, \mnras, 493, 792, \dodoi{10.1093/mnras/staa315}

\bibitem[{{Gustafsson} {et~al.}(2008){Gustafsson}, {Edvardsson}, {Eriksson},
  {J{\o}rgensen}, {Nordlund}, \& {Plez}}]{Gustafsson_2008}
{Gustafsson}, B., {Edvardsson}, B., {Eriksson}, K., {et~al.} 2008, \aap, 486,
  951, \dodoi{10.1051/0004-6361:200809724}

\bibitem[{{Hardegree-Ullman} {et~al.}(2020){Hardegree-Ullman}, {Zink},
  {Christiansen}, {Dressing}, {Ciardi}, \& {Schlieder}}]{Hardegree-Ullman_2020}
{Hardegree-Ullman}, K.~K., {Zink}, J.~K., {Christiansen}, J.~L., {et~al.} 2020,
  \apjs, 247, 28, \dodoi{10.3847/1538-4365/ab7230}

\bibitem[{{Harris} {et~al.}(2020){Harris}, {Millman}, {van der Walt},
  {Gommers}, {Virtanen}, {Cournapeau}, {Wieser}, {Taylor}, {Berg}, {Smith},
  {Kern}, {Picus}, {Hoyer}, {van Kerkwijk}, {Brett}, {Haldane}, {del R{\'\i}o},
  {Wiebe}, {Peterson}, {G{\'e}rard-Marchant}, {Sheppard}, {Reddy}, {Weckesser},
  {Abbasi}, {Gohlke}, \& {Oliphant}}]{Harris_2020}
{Harris}, C.~R., {Millman}, K.~J., {van der Walt}, S.~J., {et~al.} 2020, \nat,
  585, 357, \dodoi{10.1038/s41586-020-2649-2}

\bibitem[{{Hauschildt} {et~al.}(1999){Hauschildt}, {Allard}, \&
  {Baron}}]{Hauschildt_1999}
{Hauschildt}, P.~H., {Allard}, F., \& {Baron}, E. 1999, \apj, 512, 377,
  \dodoi{10.1086/306745}

\bibitem[{{Hemley} {et~al.}(1987){Hemley}, {Jephcoat}, {Mao}, {Zha}, {Finger},
  \& {Cox}}]{Hemley1987}
{Hemley}, R.~J., {Jephcoat}, A.~P., {Mao}, H.~K., {et~al.} 1987, \nat, 330,
  737, \dodoi{10.1038/330737a0}

\bibitem[{{Higson} {et~al.}(2019){Higson}, {Handley}, {Hobson}, \&
  {Lasenby}}]{Higson_2019}
{Higson}, E., {Handley}, W., {Hobson}, M., \& {Lasenby}, A. 2019, Statistics
  and Computing, 29, 891, \dodoi{10.1007/s11222-018-9844-0}

\bibitem[{{Hirano} {et~al.}(2020){Hirano}, {Kuzuhara}, {Kotani}, {Omiya},
  {Kudo}, {Harakawa}, {Vievard}, {Kurokawa}, {Nishikawa}, {Tamura}, {Hodapp},
  {Ishizuka}, {Jacobson}, {Konishi}, {Serizawa}, {Ueda}, {Gaidos}, \&
  {Sato}}]{2020PASJ...72...93H}
{Hirano}, T., {Kuzuhara}, M., {Kotani}, T., {et~al.} 2020, \pasj, 72, 93,
  \dodoi{10.1093/pasj/psaa085}

\bibitem[{{Hobson} {et~al.}(2021){Hobson}, {Bouchy}, {Cook}, {Artigau},
  {Moutou}, {Boisse}, {Lovis}, {Carmona}, {Delfosse}, {Donati}, \& {SPIRou
  Team}}]{Hobson_2021}
{Hobson}, M.~J., {Bouchy}, F., {Cook}, N.~J., {et~al.} 2021, \aap, 648, A48,
  \dodoi{10.1051/0004-6361/202038413}

\bibitem[{Horne(1986)}]{Horne_1986}
Horne, K. 1986, Publications of the Astronomical Society of the Pacific, 98,
  609, \dodoi{10.1086/131801}

\bibitem[{{Howard} {et~al.}(2010){Howard}, {Marcy}, {Johnson}, {Fischer},
  {Wright}, {Isaacson}, {Valenti}, {Anderson}, {Lin}, \& {Ida}}]{Howard_2010}
{Howard}, A.~W., {Marcy}, G.~W., {Johnson}, J.~A., {et~al.} 2010, Science, 330,
  653, \dodoi{10.1126/science.1194854}

\bibitem[{{Howard} {et~al.}(2012){Howard}, {Marcy}, {Bryson}, {Jenkins},
  {Rowe}, {Batalha}, {Borucki}, {Koch}, {Dunham}, {Gautier}, {Van Cleve},
  {Cochran}, {Latham}, {Lissauer}, {Torres}, {Brown}, {Gilliland}, {Buchhave},
  {Caldwell}, {Christensen-Dalsgaard}, {Ciardi}, {Fressin}, {Haas}, {Howell},
  {Kjeldsen}, {Seager}, {Rogers}, {Sasselov}, {Steffen}, {Basri},
  {Charbonneau}, {Christiansen}, {Clarke}, {Dupree}, {Fabrycky}, {Fischer},
  {Ford}, {Fortney}, {Tarter}, {Girouard}, {Holman}, {Johnson}, {Klaus},
  {Machalek}, {Moorhead}, {Morehead}, {Ragozzine}, {Tenenbaum}, {Twicken},
  {Quinn}, {Isaacson}, {Shporer}, {Lucas}, {Walkowicz}, {Welsh}, {Boss},
  {Devore}, {Gould}, {Smith}, {Morris}, {Prsa}, {Morton}, {Still}, {Thompson},
  {Mullally}, {Endl}, \& {MacQueen}}]{Howard_2012}
{Howard}, A.~W., {Marcy}, G.~W., {Bryson}, S.~T., {et~al.} 2012, \apjs, 201,
  15, \dodoi{10.1088/0067-0049/201/2/15}

\bibitem[{{Hunter}(2007)}]{Hunter_2007}
{Hunter}, J.~D. 2007, Computing in Science and Engineering, 9, 90,
  \dodoi{10.1109/MCSE.2007.55}

\bibitem[{{Husser} {et~al.}(2013){Husser}, {Wende-von Berg}, {Dreizler},
  {Homeier}, {Reiners}, {Barman}, \& {Hauschildt}}]{husser2013new}
{Husser}, T.~O., {Wende-von Berg}, S., {Dreizler}, S., {et~al.} 2013, \aap,
  553, A6, \dodoi{10.1051/0004-6361/201219058}

\bibitem[{{Jenkins}(2002)}]{Jenkins_2002}
{Jenkins}, J.~M. 2002, \apj, 575, 493, \dodoi{10.1086/341136}

\bibitem[{{Jenkins} {et~al.}(2020){Jenkins}, {Tenenbaum}, {Seader}, {Burke},
  {McCauliff}, {Smith}, {Twicken}, \& {Chandrasekaran}}]{Jenkins_2020}
{Jenkins}, J.~M., {Tenenbaum}, P., {Seader}, S., {et~al.} 2020, {Kepler Data
  Processing Handbook: Transiting Planet Search}, Kepler Science Document
  KSCI-19081-003

\bibitem[{{Jenkins} {et~al.}(2010){Jenkins}, {Chandrasekaran}, {McCauliff},
  {Caldwell}, {Tenenbaum}, {Li}, {Klaus}, {Cote}, \& {Middour}}]{Jenkins_2010}
{Jenkins}, J.~M., {Chandrasekaran}, H., {McCauliff}, S.~D., {et~al.} 2010, in
  Society of Photo-Optical Instrumentation Engineers (SPIE) Conference Series,
  Vol. 7740, Software and Cyberinfrastructure for Astronomy, ed. N.~M.
  {Radziwill} \& A.~{Bridger}, 77400D, \dodoi{10.1117/12.856764}

\bibitem[{{Jenkins} {et~al.}(2016){Jenkins}, {Twicken}, {McCauliff},
  {Campbell}, {Sanderfer}, {Lung}, {Mansouri-Samani}, {Girouard}, {Tenenbaum},
  {Klaus}, {Smith}, {Caldwell}, {Chacon}, {Henze}, {Heiges}, {Latham},
  {Morgan}, {Swade}, {Rinehart}, \& {Vanderspek}}]{Jenkins_2016}
{Jenkins}, J.~M., {Twicken}, J.~D., {McCauliff}, S., {et~al.} 2016, in Society
  of Photo-Optical Instrumentation Engineers (SPIE) Conference Series, Vol.
  9913, Software and Cyberinfrastructure for Astronomy IV, ed. G.~{Chiozzi} \&
  J.~C. {Guzman}, 99133E, \dodoi{10.1117/12.2233418}

\bibitem[{{Jensen}(2013)}]{Jensen_2013}
{Jensen}, E. 2013, {Tapir: A web interface for transit/eclipse observability}.
\newblock \doeprint{1306.007}

\bibitem[{{Kempton} {et~al.}(2018){Kempton}, {Bean}, {Louie}, {Deming}, {Koll},
  {Mansfield}, {Christiansen}, {L{\'o}pez-Morales}, {Swain}, {Zellem},
  {Ballard}, {Barclay}, {Barstow}, {Batalha}, {Beatty}, {Berta-Thompson},
  {Birkby}, {Buchhave}, {Charbonneau}, {Cowan}, {Crossfield}, {de Val-Borro},
  {Doyon}, {Dragomir}, {Gaidos}, {Heng}, {Hu}, {Kane}, {Kreidberg}, {Mallonn},
  {Morley}, {Narita}, {Nascimbeni}, {Pall{\'e}}, {Quintana}, {Rauscher},
  {Seager}, {Shkolnik}, {Sing}, {Sozzetti}, {Stassun}, {Valenti}, \& {von
  Essen}}]{Kempton_2018}
{Kempton}, E. M.~R., {Bean}, J.~L., {Louie}, D.~R., {et~al.} 2018, \pasp, 130,
  114401, \dodoi{10.1088/1538-3873/aadf6f}

\bibitem[{{Kiman} {et~al.}(2019){Kiman}, {Schmidt}, {Angus}, {Cruz}, {Faherty},
  \& {Rice}}]{Kiman_2019}
{Kiman}, R., {Schmidt}, S.~J., {Angus}, R., {et~al.} 2019, \aj, 157, 231,
  \dodoi{10.3847/1538-3881/ab1753}

\bibitem[{{Kipping}(2013)}]{Kipping_2013}
{Kipping}, D.~M. 2013, \mnras, 435, 2152, \dodoi{10.1093/mnras/stt1435}

\bibitem[{{Kotani} {et~al.}(2018){Kotani}, {Tamura}, {Nishikawa}, {Ueda},
  {Kuzuhara}, {Omiya}, {Hashimoto}, {Ishizuka}, {Hirano}, {Suto}, {Kurokawa},
  {Kokubo}, {Mori}, {Tanaka}, {Kashiwagi}, {Konishi}, {Kudo}, {Sato},
  {Jacobson}, {Hodapp}, {Hall}, {Aoki}, {Usuda}, {Nishiyama}, {Nakajima},
  {Ikeda}, {Yamamuro}, {Morino}, {Baba}, {Hosokawa}, {Ishikawa}, {Narita},
  {Kokubo}, {Hayano}, {Izumiura}, {Kambe}, {Kusakabe}, {Kwon}, {Ikoma}, {Hori},
  {Genda}, {Fukui}, {Fujii}, {Kawahara}, {Olivier}, {Jovanovic}, {Harakawa},
  {Hayashi}, {Hidai}, {Machida}, {Matsuo}, {Nagata}, {Ogihara}, {Takami},
  {Takato}, {Terada}, \& {Oh}}]{2018SPIE10702E..11K}
{Kotani}, T., {Tamura}, M., {Nishikawa}, J., {et~al.} 2018, in Society of
  Photo-Optical Instrumentation Engineers (SPIE) Conference Series, Vol. 10702,
  Ground-based and Airborne Instrumentation for Astronomy VII, ed. C.~J.
  {Evans}, L.~{Simard}, \& H.~{Takami}, 1070211, \dodoi{10.1117/12.2311836}

\bibitem[{{Kreidberg}(2015)}]{Kreidberg_2015}
{Kreidberg}, L. 2015, \pasp, 127, 1161, \dodoi{10.1086/683602}

\bibitem[{{Lee} {et~al.}(2014){Lee}, {Chiang}, \& {Ormel}}]{Lee_2014}
{Lee}, E.~J., {Chiang}, E., \& {Ormel}, C.~W. 2014, \apj, 797, 95,
  \dodoi{10.1088/0004-637X/797/2/95}

\bibitem[{{Lee} \& {Connors}(2021)}]{Lee_2021}
{Lee}, E.~J., \& {Connors}, N.~J. 2021, \apj, 908, 32,
  \dodoi{10.3847/1538-4357/abd6c7}

\bibitem[{{Li} {et~al.}(2019){Li}, {Tenenbaum}, {Twicken}, {Burke}, {Jenkins},
  {Quintana}, {Rowe}, \& {Seader}}]{Li_2019}
{Li}, J., {Tenenbaum}, P., {Twicken}, J.~D., {et~al.} 2019, \pasp, 131, 024506,
  \dodoi{10.1088/1538-3873/aaf44d}

\bibitem[{{Lillo-Box} {et~al.}(2014){Lillo-Box}, {Barrado}, \&
  {Bouy}}]{Lillo-Box_2014}
{Lillo-Box}, J., {Barrado}, D., \& {Bouy}, H. 2014, \aap, 566, A103,
  \dodoi{10.1051/0004-6361/201423497}

\bibitem[{{Lillo-Box} {et~al.}(2020){Lillo-Box}, {Figueira}, {Leleu},
  {Acu{\~n}a}, {Faria}, {Hara}, {Santos}, {Correia}, {Robutel}, {Deleuil},
  {Barrado}, {Sousa}, {Bonfils}, {Mousis}, {Almenara}, {Astudillo-Defru},
  {Marcq}, {Udry}, {Lovis}, \& {Pepe}}]{Lillo-Box_2020}
{Lillo-Box}, J., {Figueira}, P., {Leleu}, A., {et~al.} 2020, \aap, 642, A121,
  \dodoi{10.1051/0004-6361/202038922}

\bibitem[{{Lopez} \& {Fortney}(2014)}]{Lopez_2014}
{Lopez}, E.~D., \& {Fortney}, J.~J. 2014, \apj, 792, 1,
  \dodoi{10.1088/0004-637X/792/1/1}

\bibitem[{{Lopez} \& {Rice}(2018)}]{Lopez_2018}
{Lopez}, E.~D., \& {Rice}, K. 2018, \mnras, 479, 5303,
  \dodoi{10.1093/mnras/sty1707}

\bibitem[{{Luque} {et~al.}(2021){Luque}, {Serrano}, {Molaverdikhani}, {Nixon},
  {Livingston}, {Guenther}, {Pall{\'e}}, {Madhusudhan}, {Nowak}, {Korth},
  {Cochran}, {Hirano}, {Chaturvedi}, {Goffo}, {Albrecht}, {Barrag{\'a}n},
  {Brice{\~n}o}, {Cabrera}, {Charbonneau}, {Cloutier}, {Collins}, {Collins},
  {Col{\'o}n}, {Crossfield}, {Csizmadia}, {Dai}, {Deeg}, {Esposito},
  {Fridlund}, {Gandolfi}, {Georgieva}, {Glidden}, {Goeke}, {Grziwa}, {Hatzes},
  {Henze}, {Howell}, {Irwin}, {Jenkins}, {Jensen}, {K{\'a}bath}, {Kidwell},
  {Kielkopf}, {Knudstrup}, {Lam}, {Latham}, {Lissauer}, {Mann}, {Matthews},
  {Mireles}, {Narita}, {Paegert}, {Persson}, {Redfield}, {Ricker}, {Rodler},
  {Schlieder}, {Scott}, {Seager}, {{\v{S}}ubjak}, {Tan}, {Ting}, {Vanderspek},
  {Van Eylen}, {Winn}, \& {Ziegler}}]{Luque_2021}
{Luque}, R., {Serrano}, L.~M., {Molaverdikhani}, K., {et~al.} 2021, \aap, 645,
  A41, \dodoi{10.1051/0004-6361/202039455}

\bibitem[{{Majewski} {et~al.}(2016){Majewski}, {APOGEE Team}, \& {APOGEE-2
  Team}}]{Majewski_2016}
{Majewski}, S.~R., {APOGEE Team}, \& {APOGEE-2 Team}. 2016, Astronomische
  Nachrichten, 337, 863, \dodoi{10.1002/asna.201612387}

\bibitem[{{Mann} {et~al.}(2015){Mann}, {Feiden}, {Gaidos}, {Boyajian}, \& {von
  Braun}}]{Mann_2015}
{Mann}, A.~W., {Feiden}, G.~A., {Gaidos}, E., {Boyajian}, T., \& {von Braun},
  K. 2015, \apj, 804, 64, \dodoi{10.1088/0004-637X/804/1/64}

\bibitem[{{Mann} {et~al.}(2013){Mann}, {Gaidos}, \& {Ansdell}}]{Mann2013}
{Mann}, A.~W., {Gaidos}, E., \& {Ansdell}, M. 2013, \apj, 779, 188,
  \dodoi{10.1088/0004-637X/779/2/188}

\bibitem[{{Mann} {et~al.}(2019){Mann}, {Dupuy}, {Kraus}, {Gaidos}, {Ansdell},
  {Ireland}, {Rizzuto}, {Hung}, {Dittmann}, {Factor}, {Feiden}, {Martinez},
  {Ru{\'\i}z-Rodr{\'\i}guez}, \& {Thao}}]{Mann_2019}
{Mann}, A.~W., {Dupuy}, T., {Kraus}, A.~L., {et~al.} 2019, \apj, 871, 63,
  \dodoi{10.3847/1538-4357/aaf3bc}

\bibitem[{{Martinez} {et~al.}(2019){Martinez}, {Cunha}, {Ghezzi}, \&
  {Smith}}]{Martinez_2019}
{Martinez}, C.~F., {Cunha}, K., {Ghezzi}, L., \& {Smith}, V.~V. 2019, \apj,
  875, 29, \dodoi{10.3847/1538-4357/ab0d93}

\bibitem[{{Martioli} {et~al.}(2020){Martioli}, {H{\'e}brard}, {Moutou},
  {Donati}, {Artigau}, {Cale}, {Cook}, {Dalal}, {Delfosse}, {Forveille},
  {Gaidos}, {Plavchan}, {Berberian}, {Carmona}, {Cloutier}, {Doyon},
  {Fouqu{\'e}}, {Klein}, {Lecavelier des Etangs}, {Manset}, {Morin}, {Tanner},
  {Teske}, \& {Wang}}]{Martioli_2020}
{Martioli}, E., {H{\'e}brard}, G., {Moutou}, C., {et~al.} 2020, \aap, 641, L1,
  \dodoi{10.1051/0004-6361/202038695}

\bibitem[{{Martioli} {et~al.}(2022){Martioli}, {H{\'e}brard}, {Fouqu{\'e}},
  {Artigau}, {Donati}, {Cadieux}, {Bellotti}, {Lecavelier des Etangs}, {Doyon},
  {do Nascimento}, {Arnold}, {Carmona}, {Cook}, {Cortes-Zuleta}, {de Almeida},
  {Delfosse}, {Folsom}, {K{\"o}nig}, {Moutou}, {Ould-Elhkim}, {Petit},
  {Stassun}, {Vidotto}, {Vandal}, {Benneke}, {Boisse}, {Bonfils}, {Boyd},
  {Brasseur}, {Charbonneau}, {Cloutier}, {Collins}, {Cristofari}, {Crossfield},
  {D{\'\i}az}, {Fausnaugh}, {Figueira}, {Forveille}, {Furlan}, {Girardin},
  {Gnilka}, {Gomes da Silva}, {Gu}, {Guerra}, {Howell}, {Hussain}, {Jenkins},
  {Kiefer}, {Latham}, {Matson}, {Matthews}, {Morin}, {Naves}, {Ricker},
  {Seager}, {Takami}, {Twicken}, {Vanderburg}, {Vanderspek}, \&
  {Winn}}]{Martioli_2022}
{Martioli}, E., {H{\'e}brard}, G., {Fouqu{\'e}}, P., {et~al.} 2022, \aap, 660,
  A86, \dodoi{10.1051/0004-6361/202142540}

\bibitem[{{Mayo} {et~al.}(2018){Mayo}, {Vanderburg}, {Latham}, {Bieryla},
  {Morton}, {Buchhave}, {Dressing}, {Beichman}, {Berlind}, {Calkins}, {Ciardi},
  {Crossfield}, {Esquerdo}, {Everett}, {Gonzales}, {Hirsch}, {Horch}, {Howard},
  {Howell}, {Livingston}, {Patel}, {Petigura}, {Schlieder}, {Scott}, {Schumer},
  {Sinukoff}, {Teske}, \& {Winters}}]{Mayo_2018}
{Mayo}, A.~W., {Vanderburg}, A., {Latham}, D.~W., {et~al.} 2018, \aj, 155, 136,
  \dodoi{10.3847/1538-3881/aaadff}

\bibitem[{McCully {et~al.}(2018)McCully, Turner, Volgenau, Harbeck, Valenti,
  Riba, Bachelet, Snyder, Kurczynski, Norbury, \&
  Street}]{curtis_mccully_2018_1257560}
McCully, C., Turner, M., Volgenau, N., {et~al.} 2018, LCOGT/banzai: Initial
  Release, 0.9.4,  Zenodo, \dodoi{10.5281/zenodo.1257560}

\bibitem[{{McDonald} {et~al.}(2019){McDonald}, {Kreidberg}, \&
  {Lopez}}]{McDonald_2019}
{McDonald}, G.~D., {Kreidberg}, L., \& {Lopez}, E. 2019, \apj, 876, 22,
  \dodoi{10.3847/1538-4357/ab1095}

\bibitem[{{McQuillan} {et~al.}(2013){McQuillan}, {Aigrain}, \&
  {Mazeh}}]{McQuillan_2013}
{McQuillan}, A., {Aigrain}, S., \& {Mazeh}, T. 2013, \mnras, 432, 1203,
  \dodoi{10.1093/mnras/stt536}

\bibitem[{Micheau {et~al.}(2018)Micheau, Challita, Gallou, Striebig, Kouach,
  Donati, Lacombe, Parès, Belot, Baratchart, Dubois, Barrick, Bouchy, \&
  Pepe}]{Micheau_2018}
Micheau, Y., Challita, Z., Gallou, G., {et~al.} 2018, in Ground-based and
  {Airborne} {Instrumentation} for {Astronomy} {VII}, ed. H.~Takami, C.~J.
  Evans, \& L.~Simard (Austin, United States: SPIE), 210,
  \dodoi{10.1117/12.2305937}

\bibitem[{{Mordasini}(2020)}]{Mordasini_2020}
{Mordasini}, C. 2020, \aap, 638, A52, \dodoi{10.1051/0004-6361/201935541}

\bibitem[{{Moutou} {et~al.}(2020){Moutou}, {Dalal}, {Donati}, {Martioli},
  {Folsom}, {Artigau}, {Boisse}, {Bouchy}, {Carmona}, {Cook}, {Delfosse},
  {Doyon}, {Fouqu{\'e}}, {Gaisn{\'e}}, {H{\'e}brard}, {Hobson}, {Klein},
  {Lecavelier des Etangs}, \& {Morin}}]{Moutou_2020}
{Moutou}, C., {Dalal}, S., {Donati}, J.~F., {et~al.} 2020, \aap, 642, A72,
  \dodoi{10.1051/0004-6361/202038108}

\bibitem[{{Mugrauer} \& {Michel}(2020)}]{Mugrauer_2020}
{Mugrauer}, M., \& {Michel}, K.-U. 2020, Astronomische Nachrichten, 341, 996,
  \dodoi{10.1002/asna.202013825}

\bibitem[{Muirhead {et~al.}(2018)Muirhead, Dressing, Mann, Rojas-Ayala,
  L{\'{e}}pine, Paegert, Lee, \& Oelkers}]{Muirhead_2018}
Muirhead, P.~S., Dressing, C.~D., Mann, A.~W., {et~al.} 2018, The Astronomical
  Journal, 155, 180, \dodoi{10.3847/1538-3881/aab710}

\bibitem[{{Narita} {et~al.}(2020){Narita}, {Fukui}, {Yamamuro}, {Harbeck},
  {Bowman}, {Elphick}, {Nation}, {Armstrong}, {Han}, {Abe}, {Ikoma}, {Isogai},
  {Kawauchi}, {Kurita}, {Kusakabe}, {de Leon}, {Livingston}, {Mori},
  {Nishiumi}, {Tamura}, {Watanabe}, {Volgenau}, {Heinrich-Josties}, {Foale},
  {Daily}, {McCully}, {Kirby}, {Smith}, {Haworth}, {Conway},
  {Storrie-Lombardi}, {Rosing}, {Chatelain}, {Bachelet}, {Johnson}, \&
  {Rabus}}]{2020SPIE11447E..5KN}
{Narita}, N., {Fukui}, A., {Yamamuro}, T., {et~al.} 2020, in Society of
  Photo-Optical Instrumentation Engineers (SPIE) Conference Series, Vol. 11447,
  Society of Photo-Optical Instrumentation Engineers (SPIE) Conference Series,
  114475K, \dodoi{10.1117/12.2559947}

\bibitem[{{Newton} {et~al.}(2018){Newton}, {Mondrik}, {Irwin}, {Winters}, \&
  {Charbonneau}}]{Newton_2018}
{Newton}, E.~R., {Mondrik}, N., {Irwin}, J., {Winters}, J.~G., \&
  {Charbonneau}, D. 2018, \aj, 156, 217, \dodoi{10.3847/1538-3881/aad73b}

\bibitem[{{Owen} \& {Murray-Clay}(2018)}]{Owen_2018}
{Owen}, J.~E., \& {Murray-Clay}, R. 2018, \mnras, 480, 2206,
  \dodoi{10.1093/mnras/sty1943}

\bibitem[{{Owen} \& {Wu}(2013)}]{Owen_2013}
{Owen}, J.~E., \& {Wu}, Y. 2013, \apj, 775, 105,
  \dodoi{10.1088/0004-637X/775/2/105}

\bibitem[{{Owen} \& {Wu}(2017)}]{Owen_2017}
---. 2017, \apj, 847, 29, \dodoi{10.3847/1538-4357/aa890a}

\bibitem[{{Pecaut} \& {Mamajek}(2013)}]{Pecaut_Mamajek_2013}
{Pecaut}, M.~J., \& {Mamajek}, E.~E. 2013, \apjs, 208, 9,
  \dodoi{10.1088/0067-0049/208/1/9}

\bibitem[{{Petigura} {et~al.}(2018){Petigura}, {Marcy}, {Winn}, {Weiss},
  {Fulton}, {Howard}, {Sinukoff}, {Isaacson}, {Morton}, \&
  {Johnson}}]{Petigura_2018}
{Petigura}, E.~A., {Marcy}, G.~W., {Winn}, J.~N., {et~al.} 2018, \aj, 155, 89,
  \dodoi{10.3847/1538-3881/aaa54c}

\bibitem[{{Piskunov} {et~al.}(1995){Piskunov}, {Kupka}, {Ryabchikova}, {Weiss},
  \& {Jeffery}}]{Piskunov_1995}
{Piskunov}, N.~E., {Kupka}, F., {Ryabchikova}, T.~A., {Weiss}, W.~W., \&
  {Jeffery}, C.~S. 1995, \aaps, 112, 525

\bibitem[{{Plotnykov} \& Valencia(2020)}]{Plotnykov2020}
{Plotnykov}, A., \& Valencia, D. 2020, MNRAS, 499, 932

\bibitem[{Ralchenko {et~al.}(2010)Ralchenko, Kramida, \&
  Reader}]{Ralchenko_2010}
Ralchenko, Y., Kramida, A., \& Reader, J. 2010, Gaithersburg, MD

\bibitem[{{Rayner} {et~al.}(2009){Rayner}, {Cushing}, \& {Vacca}}]{Rayner_2009}
{Rayner}, J.~T., {Cushing}, M.~C., \& {Vacca}, W.~D. 2009, \apjs, 185, 289,
  \dodoi{10.1088/0067-0049/185/2/289}

\bibitem[{{Reyl{\'e}} {et~al.}(2021){Reyl{\'e}}, {Jardine}, {Fouqu{\'e}},
  {Caballero}, {Smart}, \& {Sozzetti}}]{Reyle_2021}
{Reyl{\'e}}, C., {Jardine}, K., {Fouqu{\'e}}, P., {et~al.} 2021, \aap, 650,
  A201, \dodoi{10.1051/0004-6361/202140985}

\bibitem[{{Ricker} {et~al.}(2015){Ricker}, {Winn}, {Vanderspek}, {Latham},
  {Bakos}, {Bean}, {Berta-Thompson}, {Brown}, {Buchhave}, {Butler}, {Butler},
  {Chaplin}, {Charbonneau}, {Christensen-Dalsgaard}, {Clampin}, {Deming},
  {Doty}, {De Lee}, {Dressing}, {Dunham}, {Endl}, {Fressin}, {Ge}, {Henning},
  {Holman}, {Howard}, {Ida}, {Jenkins}, {Jernigan}, {Johnson}, {Kaltenegger},
  {Kawai}, {Kjeldsen}, {Laughlin}, {Levine}, {Lin}, {Lissauer}, {MacQueen},
  {Marcy}, {McCullough}, {Morton}, {Narita}, {Paegert}, {Palle}, {Pepe},
  {Pepper}, {Quirrenbach}, {Rinehart}, {Sasselov}, {Sato}, {Seager},
  {Sozzetti}, {Stassun}, {Sullivan}, {Szentgyorgyi}, {Torres}, {Udry}, \&
  {Villasenor}}]{Ricker_2015}
{Ricker}, G.~R., {Winn}, J.~N., {Vanderspek}, R., {et~al.} 2015, Journal of
  Astronomical Telescopes, Instruments, and Systems, 1, 014003,
  \dodoi{10.1117/1.JATIS.1.1.014003}

\bibitem[{{Rogers} {et~al.}(2021){Rogers}, {Gupta}, {Owen}, \&
  {Schlichting}}]{Rogers_2021}
{Rogers}, J.~G., {Gupta}, A., {Owen}, J.~E., \& {Schlichting}, H.~E. 2021,
  \mnras, 508, 5886, \dodoi{10.1093/mnras/stab2897}

\bibitem[{{Scora} {et~al.}(2020){Scora}, {Valencia}, {Morbidelli}, \&
  {Jacobson}}]{Scora2020}
{Scora}, J., {Valencia}, D., {Morbidelli}, A., \& {Jacobson}, S. 2020, \mnras,
  493, 4910, \dodoi{10.1093/mnras/staa568}

\bibitem[{Skilling(2006)}]{Skilling_2006}
Skilling, J. 2006, Bayesian Analysis, 1, 833 , \dodoi{10.1214/06-BA127}

\bibitem[{{Skrutskie} {et~al.}(2006){Skrutskie}, {Cutri}, {Stiening},
  {Weinberg}, {Schneider}, {Carpenter}, {Beichman}, {Capps}, {Chester},
  {Elias}, {Huchra}, {Liebert}, {Lonsdale}, {Monet}, {Price}, {Seitzer},
  {Jarrett}, {Kirkpatrick}, {Gizis}, {Howard}, {Evans}, {Fowler}, {Fullmer},
  {Hurt}, {Light}, {Kopan}, {Marsh}, {McCallon}, {Tam}, {Van Dyk}, \&
  {Wheelock}}]{Skrutskie_2006}
{Skrutskie}, M.~F., {Cutri}, R.~M., {Stiening}, R., {et~al.} 2006, \aj, 131,
  1163, \dodoi{10.1086/498708}

\bibitem[{{Smith} {et~al.}(2012){Smith}, {Stumpe}, {Van Cleve}, {Jenkins},
  {Barclay}, {Fanelli}, {Girouard}, {Kolodziejczak}, {McCauliff}, {Morris}, \&
  {Twicken}}]{Smith_2012}
{Smith}, J.~C., {Stumpe}, M.~C., {Van Cleve}, J.~E., {et~al.} 2012, \pasp, 124,
  1000, \dodoi{10.1086/667697}

\bibitem[{Sokal(1997)}]{Sokal_1997}
Sokal, A. 1997, Monte Carlo Methods in Statistical Mechanics: Foundations and
  New Algorithms, ed. C.~DeWitt-Morette, P.~Cartier, \& A.~Folacci (Boston, MA:
  Springer US), 131--192, \dodoi{10.1007/978-1-4899-0319-8_6}

\bibitem[{{Speagle}(2020)}]{Speagle_2020}
{Speagle}, J.~S. 2020, \mnras, 493, 3132, \dodoi{10.1093/mnras/staa278}

\bibitem[{{Stassun} {et~al.}(2017){Stassun}, {Collins}, \&
  {Gaudi}}]{Stassun_2017}
{Stassun}, K.~G., {Collins}, K.~A., \& {Gaudi}, B.~S. 2017, \aj, 153, 136,
  \dodoi{10.3847/1538-3881/aa5df3}

\bibitem[{{Stassun} {et~al.}(2018{\natexlab{a}}){Stassun}, {Corsaro}, {Pepper},
  \& {Gaudi}}]{Stassun_2018a}
{Stassun}, K.~G., {Corsaro}, E., {Pepper}, J.~A., \& {Gaudi}, B.~S.
  2018{\natexlab{a}}, \aj, 155, 22, \dodoi{10.3847/1538-3881/aa998a}

\bibitem[{{Stassun} \& {Torres}(2016)}]{Stassun_2016}
{Stassun}, K.~G., \& {Torres}, G. 2016, \aj, 152, 180,
  \dodoi{10.3847/0004-6256/152/6/180}

\bibitem[{{Stassun} \& {Torres}(2021)}]{Stassun_2021}
---. 2021, \apjl, 907, L33, \dodoi{10.3847/2041-8213/abdaad}

\bibitem[{{Stassun} {et~al.}(2018{\natexlab{b}}){Stassun}, {Oelkers}, {Pepper},
  {Paegert}, {De Lee}, {Torres}, {Latham}, {Charpinet}, {Dressing}, {Huber},
  {Kane}, {L{\'e}pine}, {Mann}, {Muirhead}, {Rojas-Ayala}, {Silvotti},
  {Fleming}, {Levine}, \& {Plavchan}}]{Stassun_2018b}
{Stassun}, K.~G., {Oelkers}, R.~J., {Pepper}, J., {et~al.} 2018{\natexlab{b}},
  \aj, 156, 102, \dodoi{10.3847/1538-3881/aad050}

\bibitem[{{Stassun} {et~al.}(2019){Stassun}, {Oelkers}, {Paegert}, {Torres},
  {Pepper}, {De Lee}, {Collins}, {Latham}, {Muirhead}, {Chittidi},
  {Rojas-Ayala}, {Fleming}, {Rose}, {Tenenbaum}, {Ting}, {Kane}, {Barclay},
  {Bean}, {Brassuer}, {Charbonneau}, {Ge}, {Lissauer}, {Mann}, {McLean},
  {Mullally}, {Narita}, {Plavchan}, {Ricker}, {Sasselov}, {Seager}, {Sharma},
  {Shiao}, {Sozzetti}, {Stello}, {Vanderspek}, {Wallace}, \&
  {Winn}}]{Stassun_2019}
{Stassun}, K.~G., {Oelkers}, R.~J., {Paegert}, M., {et~al.} 2019, \aj, 158,
  138, \dodoi{10.3847/1538-3881/ab3467}

\bibitem[{{Stetson}(1987)}]{Stetson_1987}
{Stetson}, P.~B. 1987, \pasp, 99, 191, \dodoi{10.1086/131977}

\bibitem[{{Stewart} \& {Ahrens}(2005)}]{Stewart_2005}
{Stewart}, S.~T., \& {Ahrens}, T.~J. 2005, Journal of Geophysical Research
  (Planets), 110, E03005, \dodoi{10.1029/2004JE002305}

\bibitem[{{Stumpe} {et~al.}(2014){Stumpe}, {Smith}, {Catanzarite}, {Van Cleve},
  {Jenkins}, {Twicken}, \& {Girouard}}]{Stumpe_2014}
{Stumpe}, M.~C., {Smith}, J.~C., {Catanzarite}, J.~H., {et~al.} 2014, \pasp,
  126, 100, \dodoi{10.1086/674989}

\bibitem[{{Stumpe} {et~al.}(2012){Stumpe}, {Smith}, {Van Cleve}, {Twicken},
  {Barclay}, {Fanelli}, {Girouard}, {Jenkins}, {Kolodziejczak}, {McCauliff}, \&
  {Morris}}]{Stumpe_2012}
{Stumpe}, M.~C., {Smith}, J.~C., {Van Cleve}, J.~E., {et~al.} 2012, \pasp, 124,
  985, \dodoi{10.1086/667698}

\bibitem[{{Sullivan} {et~al.}(2015){Sullivan}, {Winn}, {Berta-Thompson},
  {Charbonneau}, {Deming}, {Dressing}, {Latham}, {Levine}, {McCullough},
  {Morton}, {Ricker}, {Vanderspek}, \& {Woods}}]{Sullivan_2015}
{Sullivan}, P.~W., {Winn}, J.~N., {Berta-Thompson}, Z.~K., {et~al.} 2015, \apj,
  809, 77, \dodoi{10.1088/0004-637X/809/1/77}

\bibitem[{{Tamura} {et~al.}(2012){Tamura}, {Suto}, {Nishikawa}, {Kotani},
  {Sato}, {Aoki}, {Usuda}, {Kurokawa}, {Kashiwagi}, {Nishiyama}, {Ikeda},
  {Hall}, {Hodapp}, {Hashimoto}, {Morino}, {Inoue}, {Mizuno}, {Washizaki},
  {Tanaka}, {Suzuki}, {Kwon}, {Suenaga}, {Oh}, {Narita}, {Kokubo}, {Hayano},
  {Izumiura}, {Kambe}, {Kudo}, {Kusakabe}, {Ikoma}, {Hori}, {Omiya}, {Genda},
  {Fukui}, {Fujii}, {Guyon}, {Harakawa}, {Hayashi}, {Hidai}, {Hirano},
  {Kuzuhara}, {Machida}, {Matsuo}, {Nagata}, {Ohnuki}, {Ogihara}, {Oshino},
  {Suzuki}, {Takami}, {Takato}, {Takahashi}, {Tachinami}, \&
  {Terada}}]{2012SPIE.8446E..1TT}
{Tamura}, M., {Suto}, H., {Nishikawa}, J., {et~al.} 2012, in Society of
  Photo-Optical Instrumentation Engineers (SPIE) Conference Series, Vol. 8446,
  Ground-based and Airborne Instrumentation for Astronomy IV, ed. I.~S.
  {McLean}, S.~K. {Ramsay}, \& H.~{Takami}, 84461T, \dodoi{10.1117/12.925885}

\bibitem[{{Trotta}(2008)}]{Trotta_2008}
{Trotta}, R. 2008, Contemporary Physics, 49, 71,
  \dodoi{10.1080/00107510802066753}

\bibitem[{{Twicken} {et~al.}(2018){Twicken}, {Catanzarite}, {Clarke},
  {Girouard}, {Jenkins}, {Klaus}, {Li}, {McCauliff}, {Seader}, {Tenenbaum},
  {Wohler}, {Bryson}, {Burke}, {Caldwell}, {Haas}, {Henze}, \&
  {Sanderfer}}]{Twicken_2018}
{Twicken}, J.~D., {Catanzarite}, J.~H., {Clarke}, B.~D., {et~al.} 2018, \pasp,
  130, 064502, \dodoi{10.1088/1538-3873/aab694}

\bibitem[{{Valencia} {et~al.}(2013){Valencia}, {Guillot}, {Parmentier}, \&
  {Freedman}}]{Valencia2013}
{Valencia}, D., {Guillot}, T., {Parmentier}, V., \& {Freedman}, R.~S. 2013,
  \apj, 775, 10, \dodoi{10.1088/0004-637X/775/1/10}

\bibitem[{{Valencia} {et~al.}(2007){Valencia}, {Sasselov}, \&
  {O'Connell}}]{Valencia2007}
{Valencia}, D., {Sasselov}, D.~D., \& {O'Connell}, R.~J. 2007, \apj, 656, 545,
  \dodoi{10.1086/509800}

\bibitem[{{Van Eylen} {et~al.}(2018){Van Eylen}, {Agentoft}, {Lundkvist},
  {Kjeldsen}, {Owen}, {Fulton}, {Petigura}, \& {Snellen}}]{VanEylen_2018}
{Van Eylen}, V., {Agentoft}, C., {Lundkvist}, M.~S., {et~al.} 2018, \mnras,
  479, 4786, \dodoi{10.1093/mnras/sty1783}

\bibitem[{{Van Eylen} {et~al.}(2021){Van Eylen}, {Astudillo-Defru}, {Bonfils},
  {Livingston}, {Hirano}, {Luque}, {Lam}, {Justesen}, {Winn}, {Gandolfi},
  {Nowak}, {Palle}, {Albrecht}, {Dai}, {Campos Estrada}, {Owen},
  {Foreman-Mackey}, {Fridlund}, {Korth}, {Mathur}, {Forveille}, {Mikal-Evans},
  {Osborne}, {Ho}, {Almenara}, {Artigau}, {Barrag{\'a}n}, {Barros}, {Bouchy},
  {Cabrera}, {Caldwell}, {Charbonneau}, {Chaturvedi}, {Cochran}, {Csizmadia},
  {Damasso}, {Delfosse}, {De Medeiros}, {D{\'\i}az}, {Doyon}, {Esposito},
  {F{\H{u}}r{\'e}sz}, {Figueira}, {Georgieva}, {Goffo}, {Grziwa}, {Guenther},
  {Hatzes}, {Jenkins}, {Kabath}, {Knudstrup}, {Latham}, {Lavie}, {Lovis},
  {Mennickent}, {Mullally}, {Murgas}, {Narita}, {Pepe}, {Persson}, {Redfield},
  {Ricker}, {Santos}, {Seager}, {Serrano}, {Smith}, {Mascare{\~n}o}, {Subjak},
  {Twicken}, {Udry}, {Vanderspek}, \& {Zapatero Osorio}}]{VanEylen_2021}
{Van Eylen}, V., {Astudillo-Defru}, N., {Bonfils}, X., {et~al.} 2021, \mnras,
  507, 2154, \dodoi{10.1093/mnras/stab2143}

\bibitem[{{Virtanen} {et~al.}(2020){Virtanen}, {Gommers}, {Oliphant},
  {Haberland}, {Reddy}, {Cournapeau}, {Burovski}, {Peterson}, {Weckesser},
  {Bright}, {van der Walt}, {Brett}, {Wilson}, {Millman}, {Mayorov}, {Nelson},
  {Jones}, {Kern}, {Larson}, {Carey}, {Polat}, {Feng}, {Moore}, {VanderPlas},
  {Laxalde}, {Perktold}, {Cimrman}, {Henriksen}, {Quintero}, {Harris},
  {Archibald}, {Ribeiro}, {Pedregosa}, {van Mulbregt}, \& {SciPy 1. 0
  Contributors}}]{Virtanen_2020}
{Virtanen}, P., {Gommers}, R., {Oliphant}, T.~E., {et~al.} 2020, Nature
  Methods, 17, 261, \dodoi{10.1038/s41592-019-0686-2}

\bibitem[{{Winters} {et~al.}(2019){Winters}, {Medina}, {Irwin}, {Charbonneau},
  {Astudillo-Defru}, {Horch}, {Eastman}, {Vrijmoet}, {Henry}, {Diamond-Lowe},
  {Winston}, {Barclay}, {Bonfils}, {Ricker}, {Vanderspek}, {Latham}, {Seager},
  {Winn}, {Jenkins}, {Udry}, {Twicken}, {Teske}, {Tenenbaum}, {Pepe}, {Murgas},
  {Muirhead}, {Mink}, {Lovis}, {Levine}, {L{\'e}pine}, {Jao}, {Henze},
  {Fur{\'e}sz}, {Forveille}, {Figueira}, {Esquerdo}, {Dressing}, {D{\'\i}az},
  {Delfosse}, {Burke}, {Bouchy}, {Berlind}, \& {Almenara}}]{Winters_2019}
{Winters}, J.~G., {Medina}, A.~A., {Irwin}, J.~M., {et~al.} 2019, \aj, 158,
  152, \dodoi{10.3847/1538-3881/ab364d}

\bibitem[{{Wright} {et~al.}(2010){Wright}, {Eisenhardt}, {Mainzer}, {Ressler},
  {Cutri}, {Jarrett}, {Kirkpatrick}, {Padgett}, {McMillan}, {Skrutskie},
  {Stanford}, {Cohen}, {Walker}, {Mather}, {Leisawitz}, {Gautier}, {McLean},
  {Benford}, {Lonsdale}, {Blain}, {Mendez}, {Irace}, {Duval}, {Liu}, {Royer},
  {Heinrichsen}, {Howard}, {Shannon}, {Kendall}, {Walsh}, {Larsen}, {Cardon},
  {Schick}, {Schwalm}, {Abid}, {Fabinsky}, {Naes}, \& {Tsai}}]{Wright_2010}
{Wright}, E.~L., {Eisenhardt}, P. R.~M., {Mainzer}, A.~K., {et~al.} 2010, \aj,
  140, 1868, \dodoi{10.1088/0004-6256/140/6/1868}

\bibitem[{{Wu}(2019)}]{Wu_2019}
{Wu}, Y. 2019, \apj, 874, 91, \dodoi{10.3847/1538-4357/ab06f8}

\bibitem[{{Zacharias} {et~al.}(2013){Zacharias}, {Finch}, {Girard}, {Henden},
  {Bartlett}, {Monet}, \& {Zacharias}}]{Zacharias_2013}
{Zacharias}, N., {Finch}, C.~T., {Girard}, T.~M., {et~al.} 2013, \aj, 145, 44,
  \dodoi{10.1088/0004-6256/145/2/44}

\bibitem[{Zechmeister {et~al.}(2018)Zechmeister, Reiners, Amado, Azzaro, Bauer,
  B{\'e}jar, Caballero, Guenther, Hagen, Jeffers, Kaminski, Kürster,
  Launhardt, Montes, Morales, Quirrenbach, Reffert, Ribas, Seifert, Tal-Or, \&
  Wolthoff}]{Zechmeister_2018}
Zechmeister, M., Reiners, A., Amado, P.~J., {et~al.} 2018, Astronomy and
  Astrophysics, 609, A12, \dodoi{10.1051/0004-6361/201731483}

\end{thebibliography}
\bibliographystyle{aasjournal.bst}

\includeAffiliations
\allauthors

\appendix
\counterwithin{table}{section}
\counterwithin{figure}{section}

\section{TESS Light Curve}

We present the TESS multi-sector \texttt{PDCSAP} light curve in Figure~\ref{fig:TESS_lc_complete}, with the exception of sectors 14 and 21, which were previously shown in Figure~\ref{fig:TESS_lc_phase}.

\begin{figure*}[h!]
\minipage{0.49\textwidth}
  \centering
  \includegraphics[width=1\linewidth]{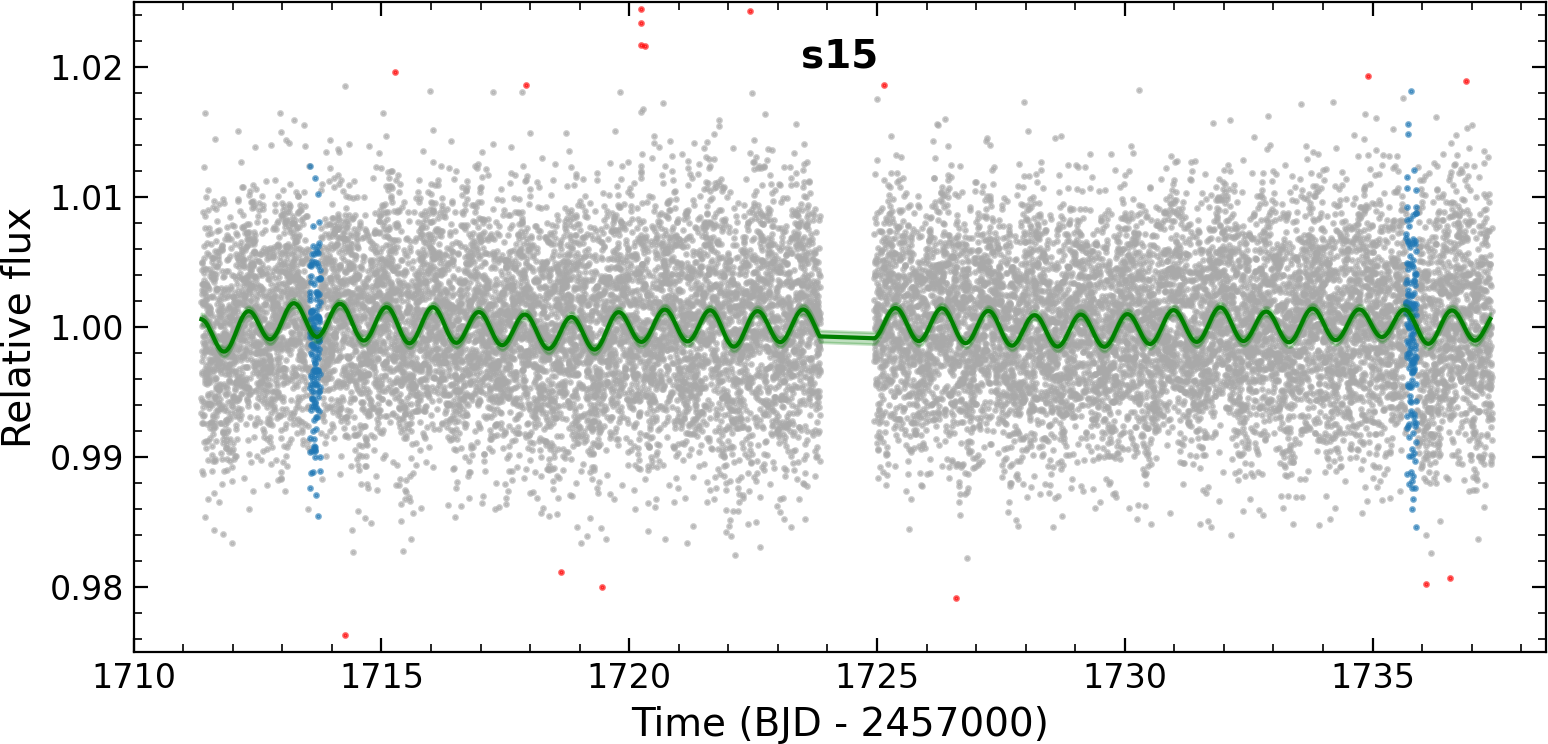}
  \endminipage\hfill
  \minipage{0.49\textwidth}
  \centering
  \includegraphics[width=1\linewidth]{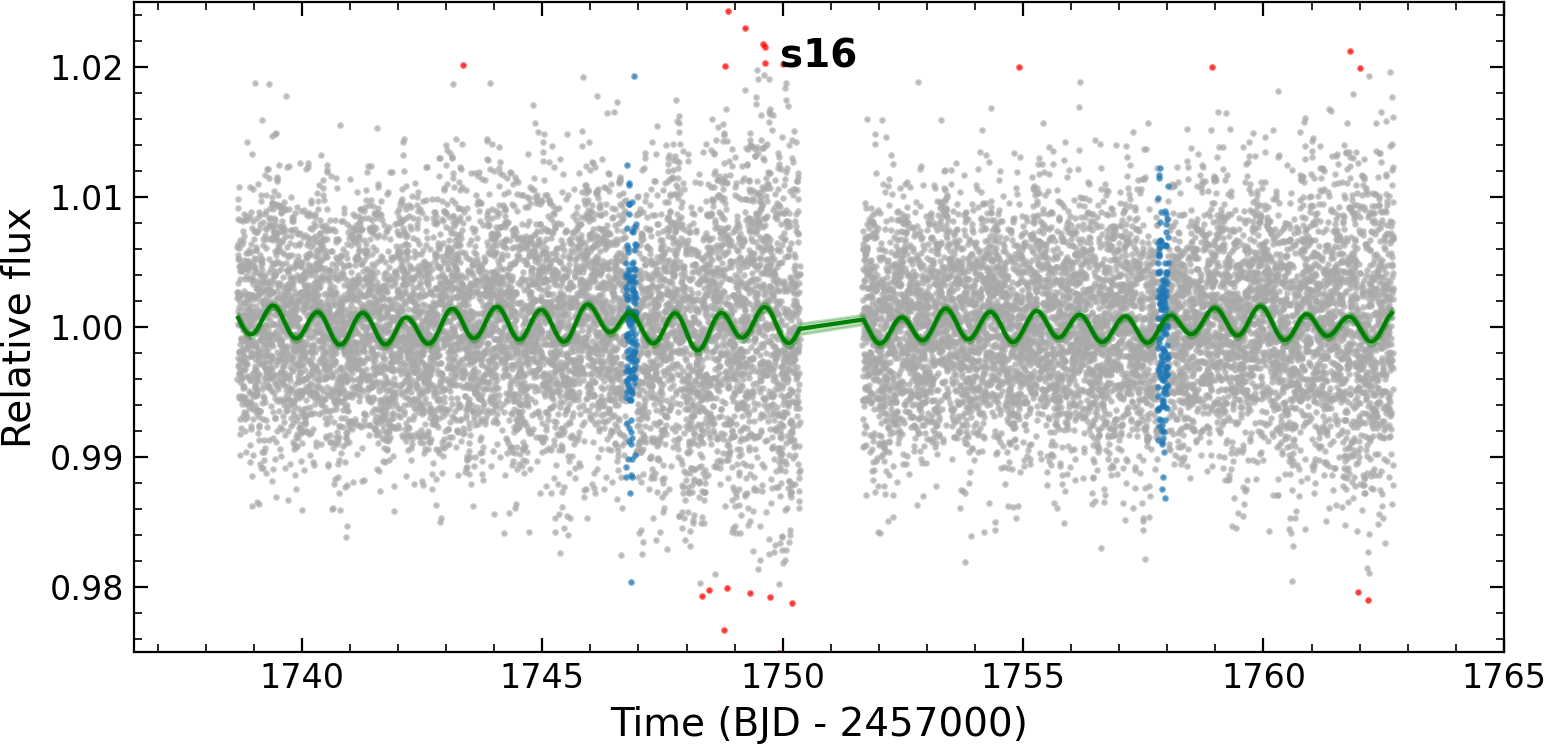}
  \endminipage\hfill\\[0.25cm]
  
  \minipage{0.49\textwidth}
  \centering
  \includegraphics[width=1\linewidth]{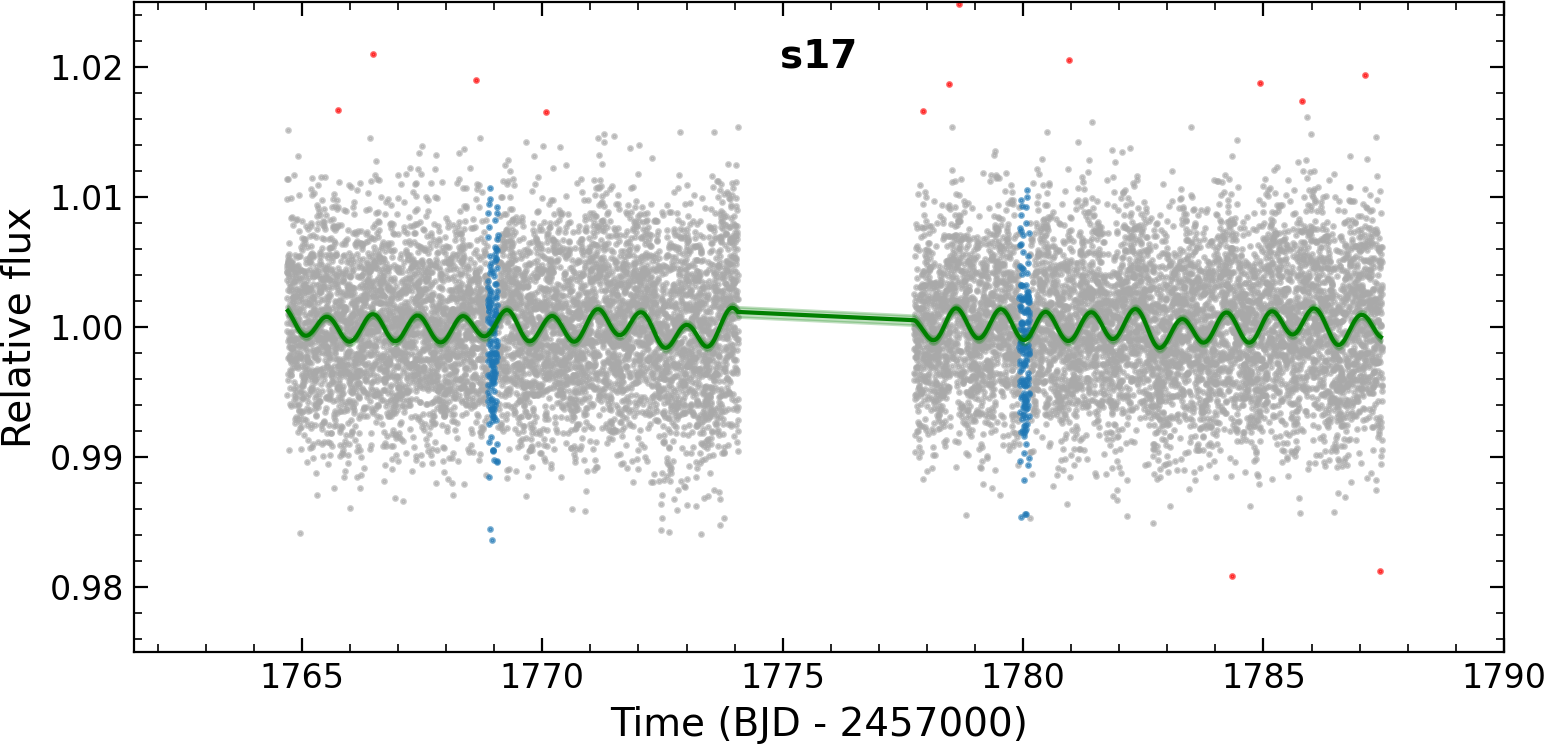}
  \endminipage\hfill
  \minipage{0.49\textwidth}
  \centering
  \includegraphics[width=1\linewidth]{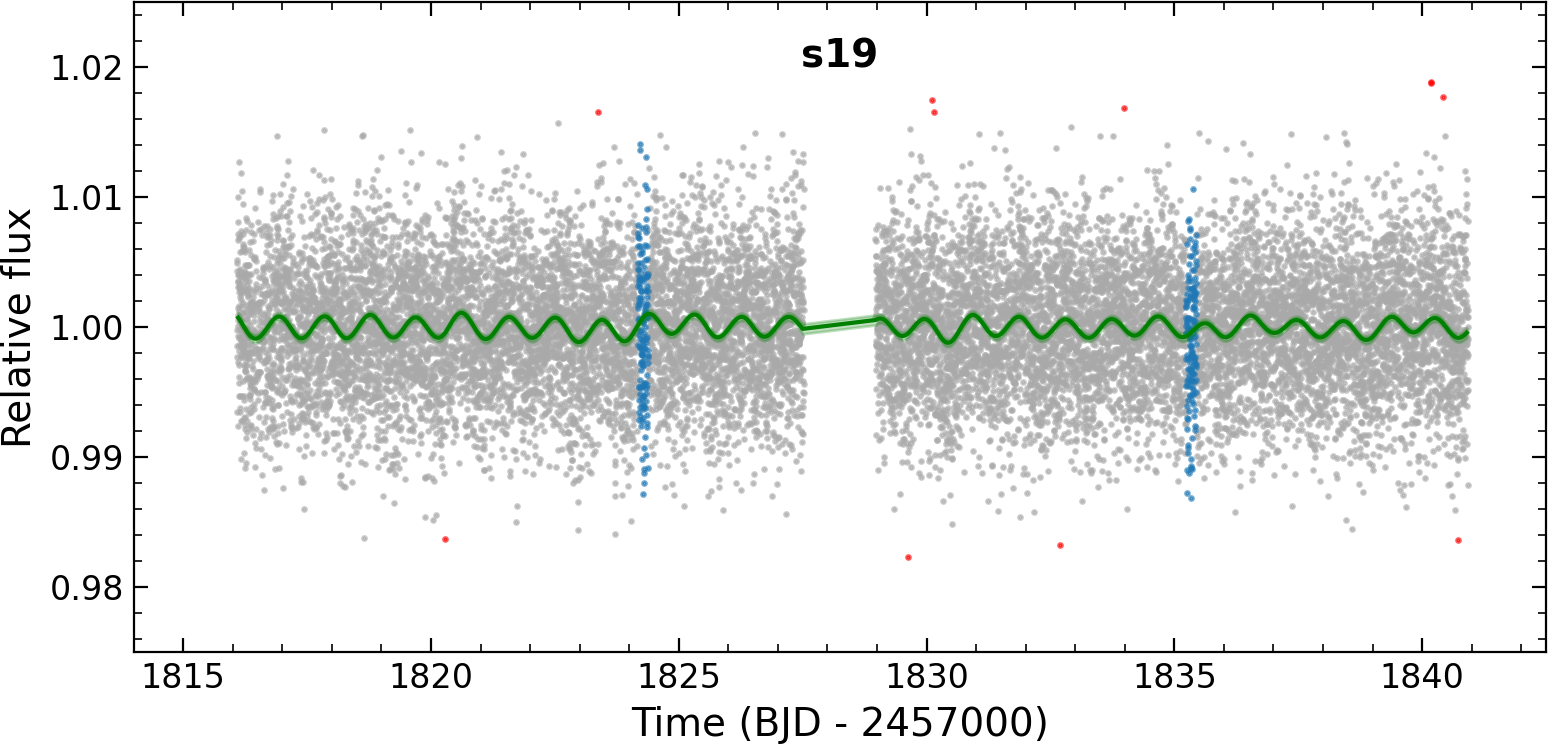}
  \endminipage\hfill\\[0.25cm]
  
  \minipage{0.49\textwidth}
  \centering
  \includegraphics[width=1\linewidth]{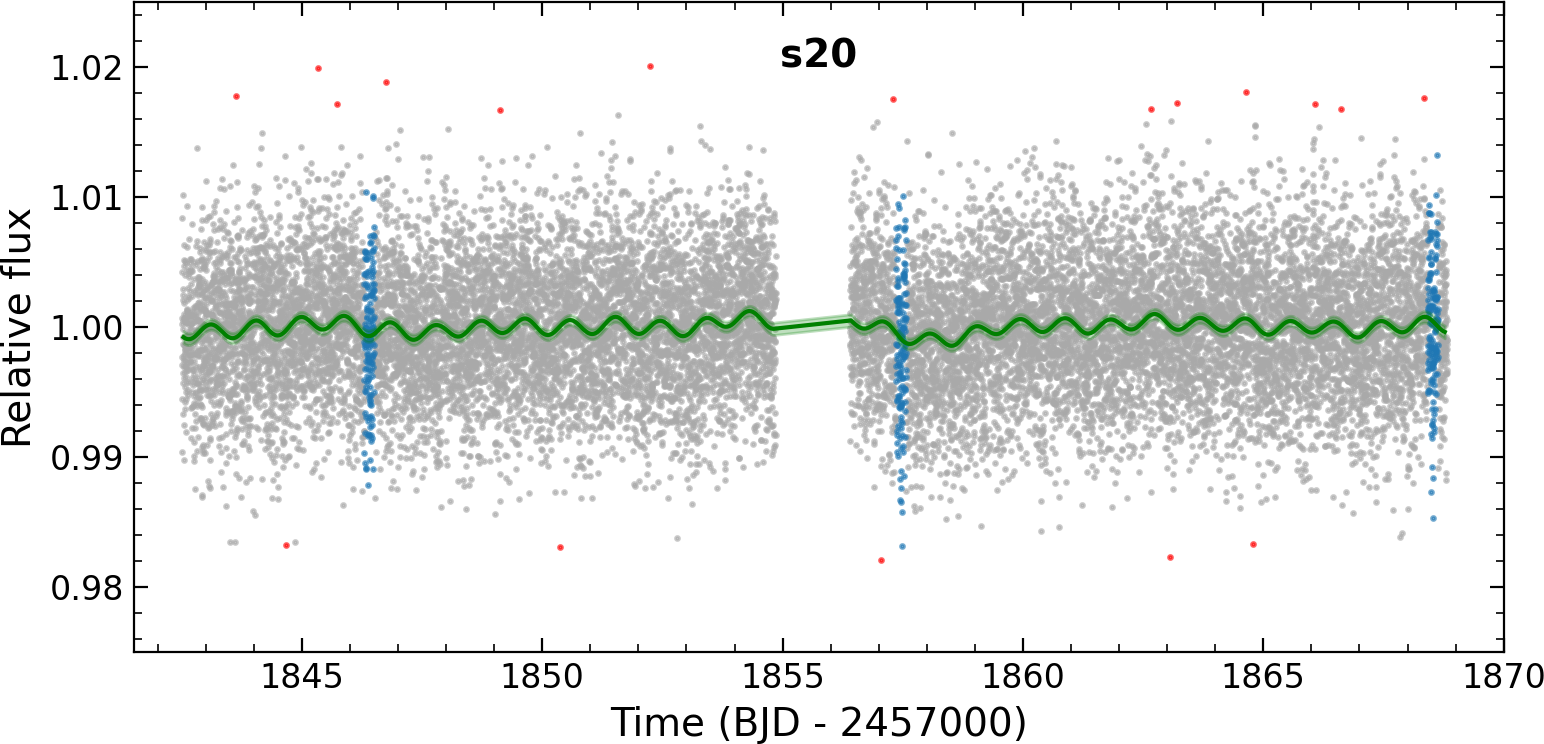}
  \endminipage\hfill
  \minipage{0.49\textwidth}
  \centering
  \includegraphics[width=1\linewidth]{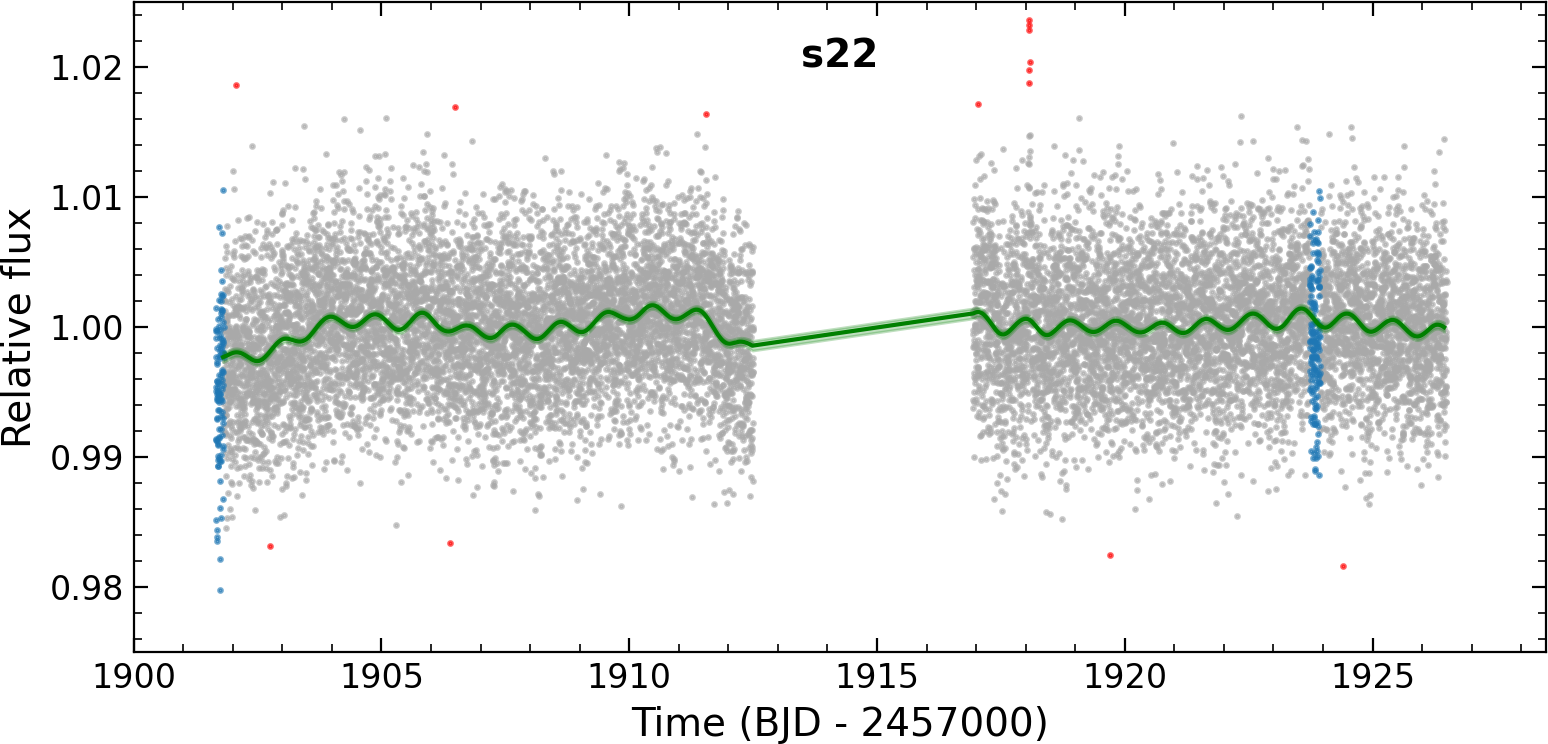}
  \endminipage\hfill\\[0.25cm]
  
  \minipage{0.49\textwidth}
  \centering
  \includegraphics[width=1\linewidth]{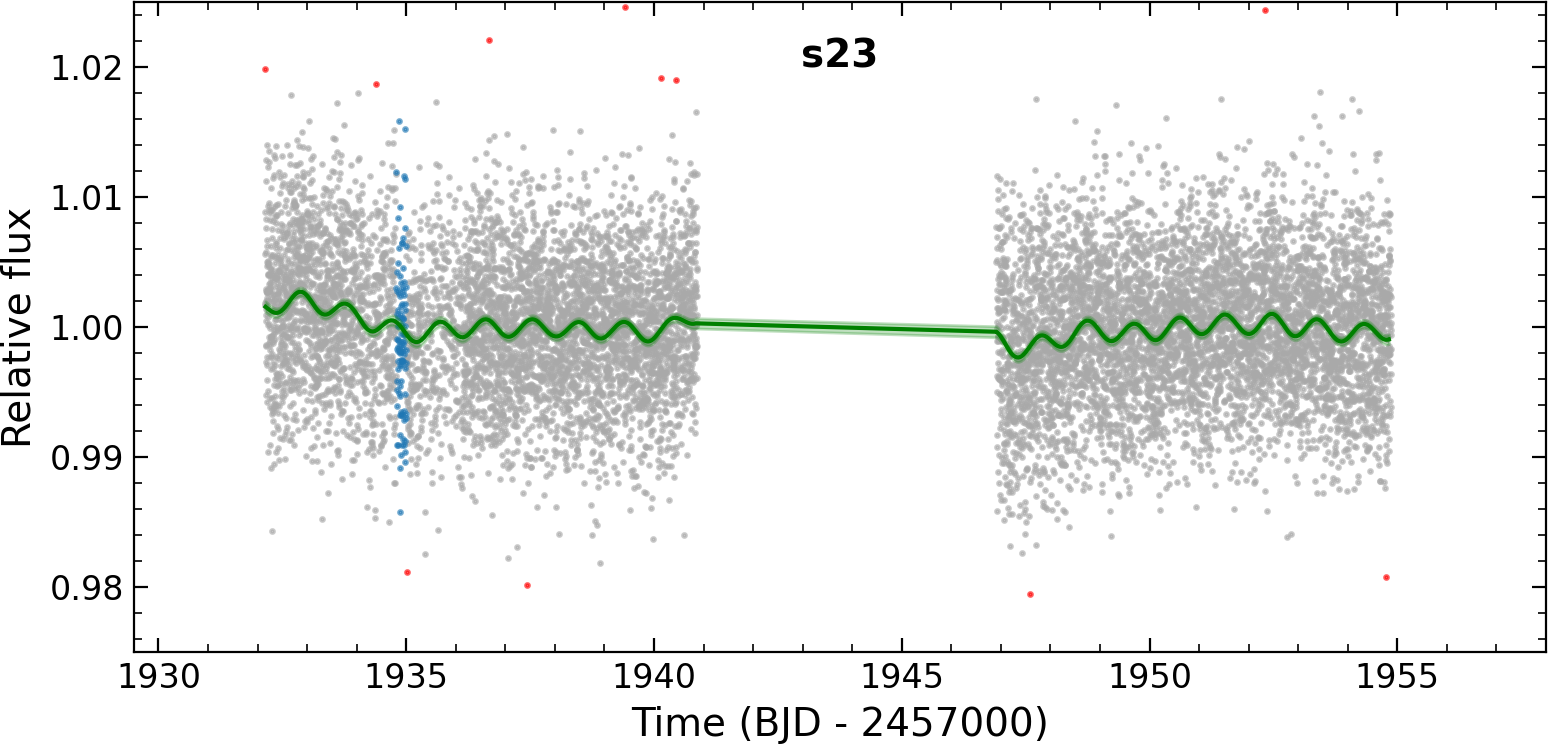}
  \endminipage\hfill
  \minipage{0.49\textwidth}
  \centering
  \includegraphics[width=1\linewidth]{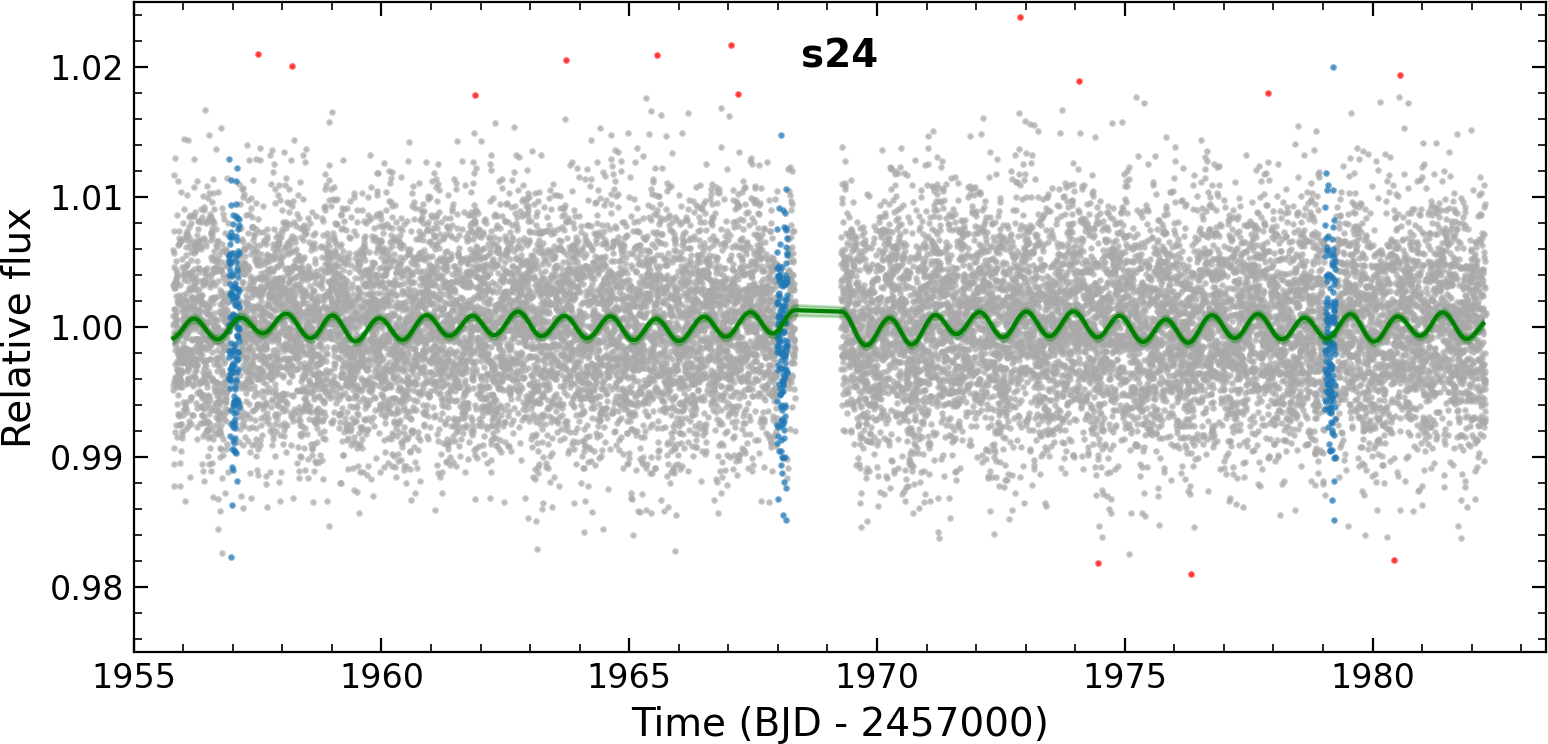}
  \endminipage\hfill\\[0.25cm]
  
  \minipage{0.49\textwidth}
  \centering
  \includegraphics[width=1\linewidth]{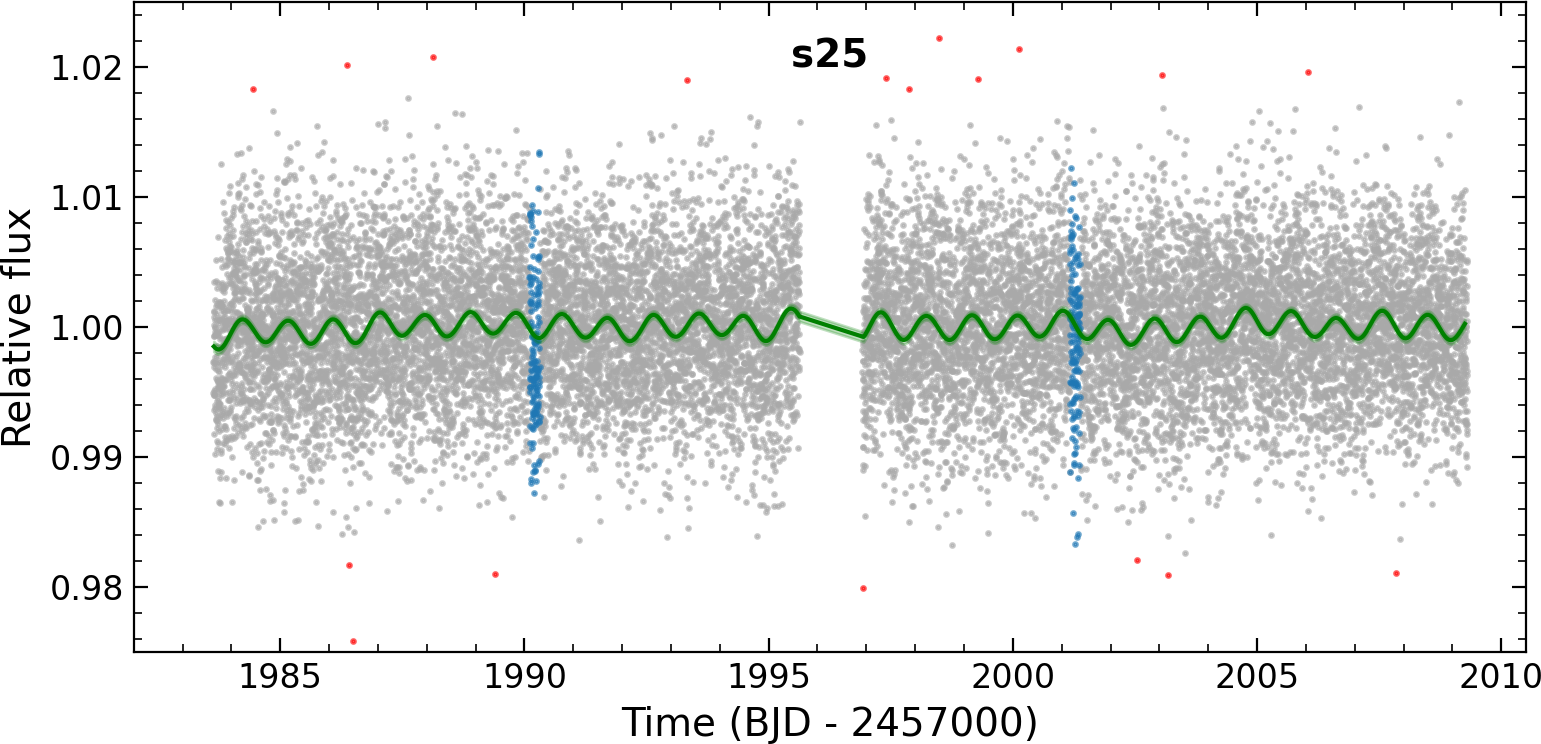}
  \endminipage\hfill
  \minipage{0.49\textwidth}
  \centering
  \includegraphics[width=1\linewidth]{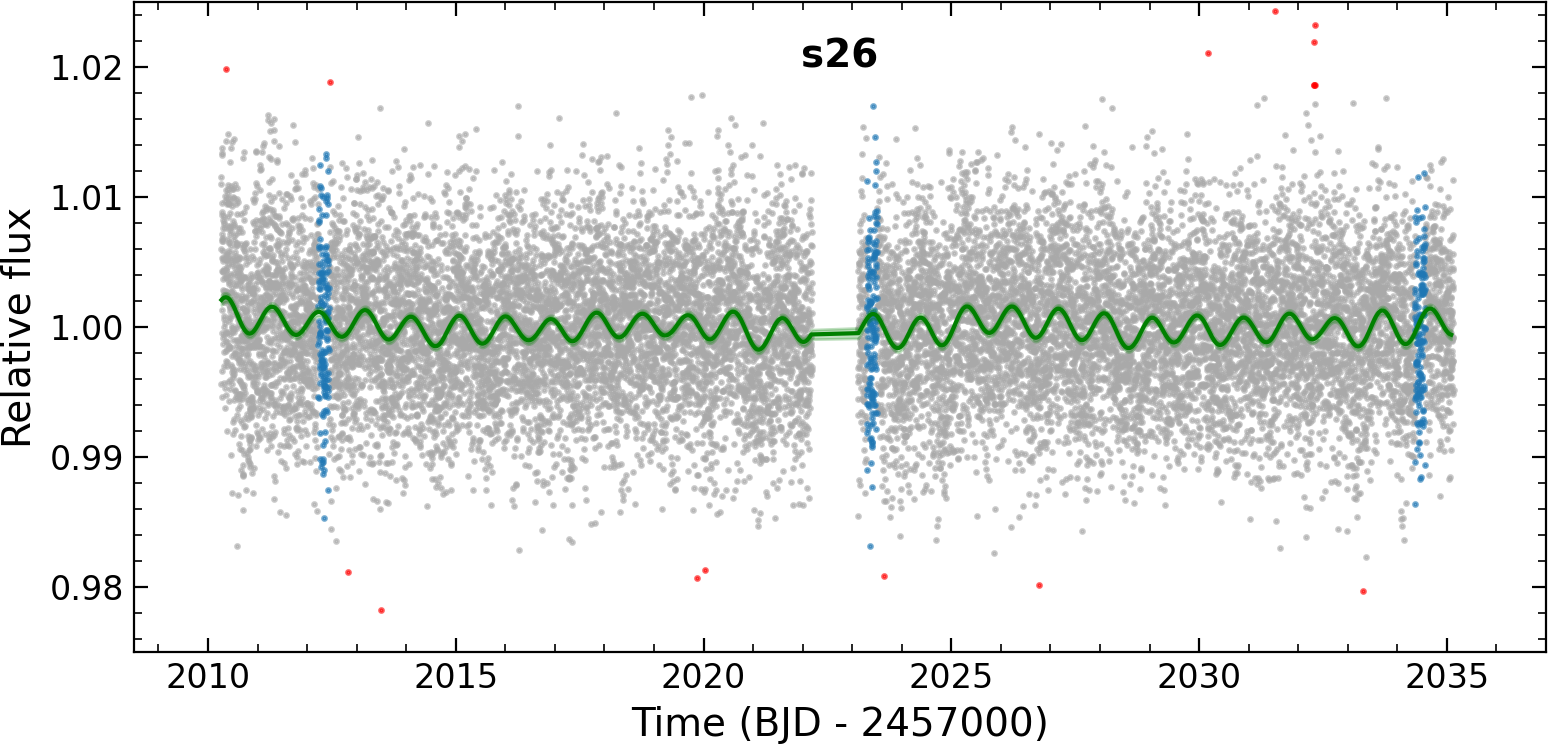}
  \endminipage\hfill\\[0.25cm]
  \caption{\textit{continue on the next page...}}
\end{figure*}
\setcounter{figure}{0}
\begin{figure*}[h!] 
  \minipage{0.49\textwidth}
  \centering
  \includegraphics[width=1\linewidth]{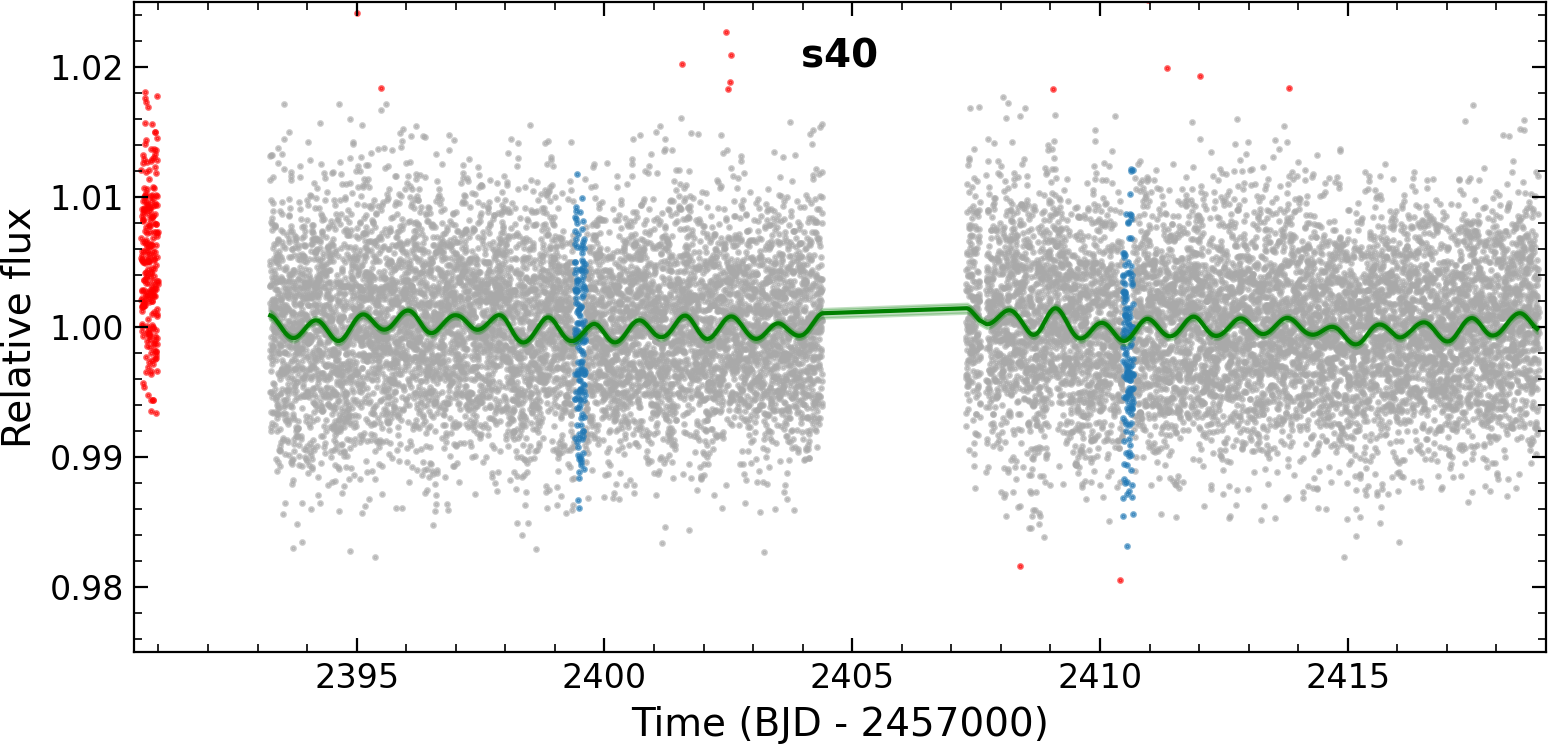}
  \endminipage\hfill
  \minipage{0.49\textwidth}
  \centering
  \includegraphics[width=1\linewidth]{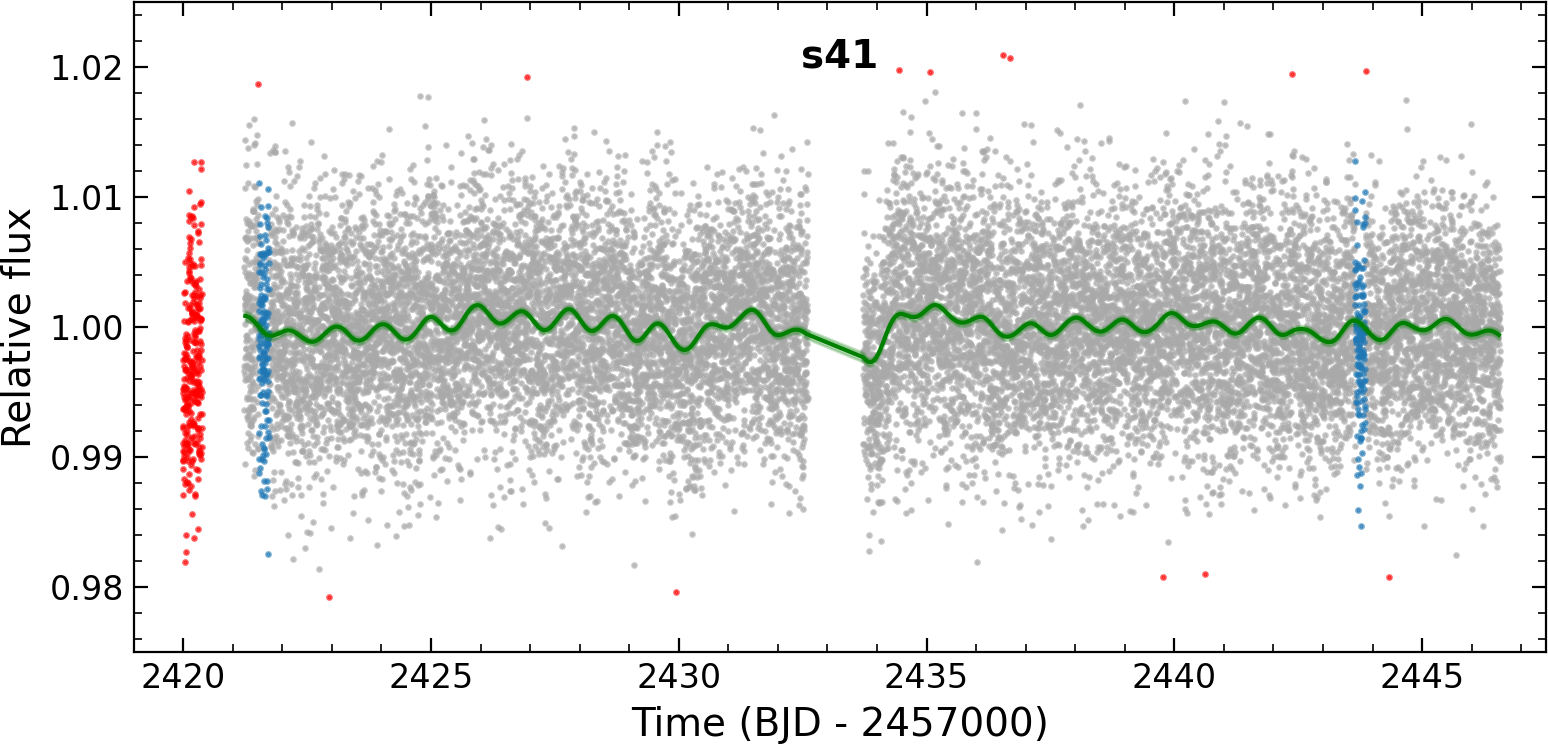}
  \endminipage\hfill\\[0.25cm]
  
    \minipage{0.49\textwidth}
  \centering
  \includegraphics[width=1\linewidth]{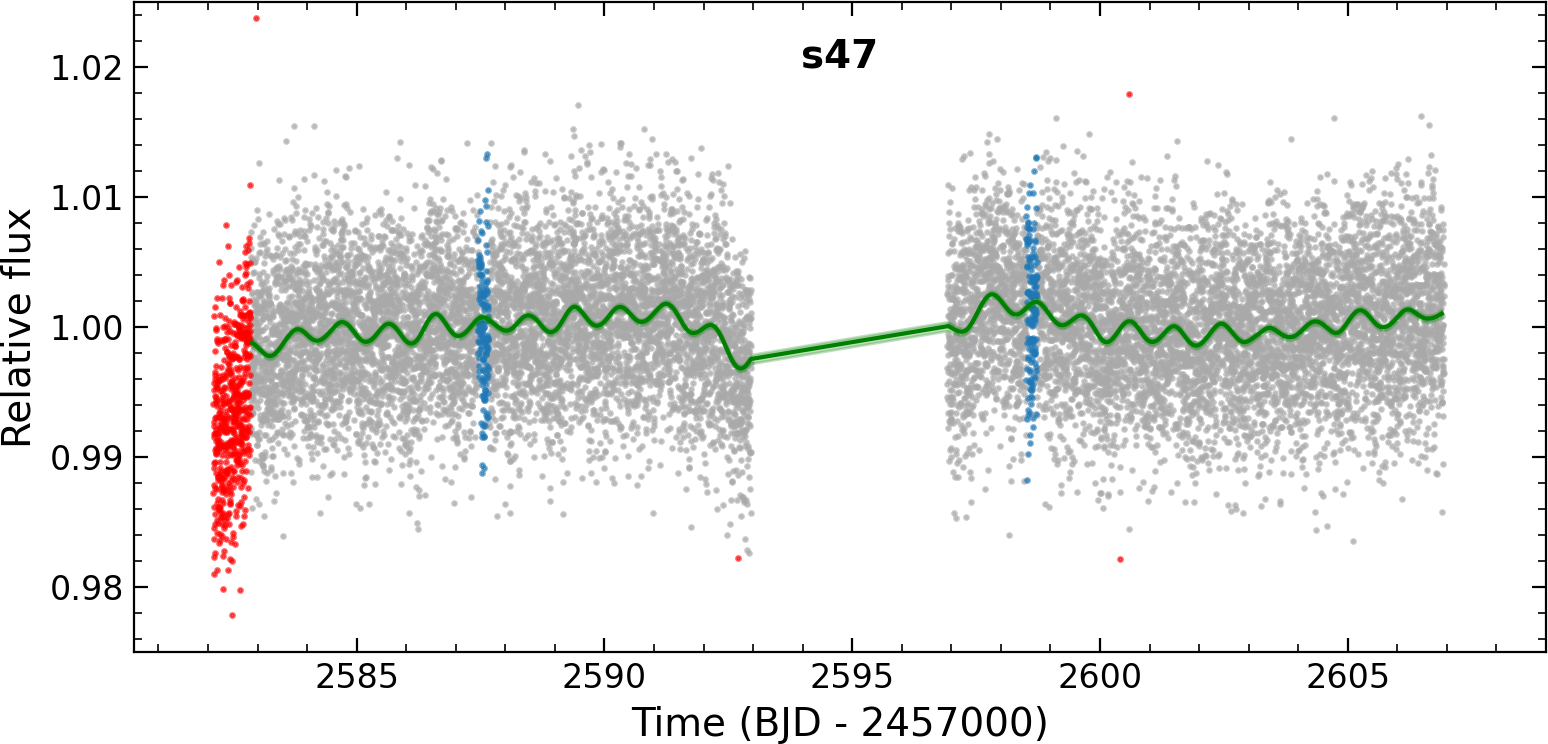}
  \endminipage\hfill
  \caption{Normalized \texttt{PDCSAP} light curve of TOI-1452 from sectors 15--26 (except 18), 40--41, and 47. Sectors 14 and 21 were previously shown in Fig.\ \ref{fig:TESS_lc_phase}. The blue data points highlight the epochs of the detected transits, while the red data points are outliers/stellar flares rejected by sigma clipping (3.5\,$\sigma$ clip) or manually (e.g., in sectors 40, 41, or 47). The gaps in the light curve coincide with data downlink when TESS is close to perigee. The full light curve was modeled with a quasi-periodic Gaussian Process (green curve, details in Sect.\ \ref{sec:tess_analysis}).}
\label{fig:TESS_lc_complete}
\end{figure*}

\section{Supplementary material regarding the joint transit-RV fit}

In this appendix, we summarize the RV component of the joint transit-RV models introduced in Section~\ref{sec:jointfit}. The main RV parameters of the joint fits are reported in Table~\ref{table:rvcomponent}. All models and datasets detect the planetary signal with a coherent semi-amplitude $K$. Models with an activity GP ($\mathcal{M_{\rm 1cp+GP}}$ and $\mathcal{M_{\rm 1ep+GP}}$) produced the highest Bayesian log-evidence (Fig.~\ref{fig:modelselect}) and needed the smallest amount of additional white noise ($\sigma_{\rm SPIRou}$, $\sigma_{\rm IRD}$). We ultimately adopted the results of the SPIRou only joint fit because including the seven IRD RV measurements yields similar or smaller Bayesian log-evidences (Fig.~\ref{fig:modelselect}), with extra white noise term $\sigma_{\rm IRD}$ about three times the level of the planetary signal (see Table~\ref{table:rvcomponent}).

\begin{table}[b!]
\end{table}
\begin{deluxetable}{lcccc}
\tablecaption{RV component of the joint transit-RV fit for different models ($\mathcal{M}$) and datasets}
\tablehead{\colhead{Parameter} &
\colhead{1cp} & \colhead{1cp+GP} & \colhead{1ep} & \colhead{1ep+GP}
}
\startdata
\multicolumn{5}{c}{\textit{SPIRou only}}\\
$K$ (m/s) & 4.2$\pm$0.9 & 3.5$\pm$0.9 & 4.7$\pm$0.9 & 3.6$\pm$0.9\\
$e$ & --- & --- & 0.20$\pm$0.09 & 0.12$^{+0.12}_{-0.08}$\\
$A_{\rm GP}$ (m/s) & --- & 4.5$^{+2.0}_{-1.2}$ & --- & 4.4$^{+1.9}_{-1.2}$\\
$\ell_{\rm GP}$ (days) & --- & 11.3$^{+12.0}_{-6.4}$ & --- & 11.6$^{+12.4}_{-6.4}$\\
$\sigma_{\rm SPIRou}$ (m/s) & 4.9$\pm$0.9 & 2.3$\pm$1.3 & 4.7$^{+0.9}_{-1.0}$ & 2.2$\pm$1.3\\
\hline
\multicolumn{5}{c}{\textit{SPIRou\,+\,IRD}}\\
$K$ (m/s) & 4.1$\pm$0.9 & 3.5$\pm$0.9 & 4.7$\pm$0.9 & 3.6$\pm$0.9\\
$e$ & --- & --- & 0.19$\pm$0.09 & 0.12$^{+0.12}_{-0.08}$\\
$A_{\rm GP}$ (m/s) & --- & 4.7$^{+2.3}_{-1.3}$ & --- & 4.6$^{+2.2}_{-1.2}$\\
$\ell_{\rm GP}$ (days) & --- & 11.1$^{+11.2}_{-6.2}$ & --- & 11.7$^{+11.2}_{-6.2}$\\
$\sigma_{\rm SPIRou}$ (m/s) & 4.9$\pm$0.9 & 2.3$\pm$1.3 & 4.7$^{+0.9}_{-1.0}$ & 2.3$^{+1.2}_{-1.3}$\\
$\sigma_{\rm IRD}$ (m/s) & 13.6$^{+5.4}_{-3.6}$ & 11.5$^{+5.7}_{-4.3}$ & 14.2$^{+5.7}_{-3.8}$ & 11.9$^{+5.5}_{-4.1}$\\
\enddata
\tablecomments{$\mathcal{M_{\rm 1cp}}$: single circular orbit planet\\
$\mathcal{M_{\rm 1cp+GP}}$: single circular orbit planet and activity GP\\
$\mathcal{M_{\rm 1ep}}$: single eccentric orbit planet\\
$\mathcal{M_{\rm 1ep+GP}}$: single eccentric orbit planet and activity GP}
\label{table:rvcomponent}
\end{deluxetable}

\clearpage

\section{Summary of interior parameters}
\label{mcmc:summary}
    We present the summary plot for the interior analysis from MCMC modeling.
    
    \begin{figure}[ht!]
    \centering
        \includegraphics[width=1\linewidth]{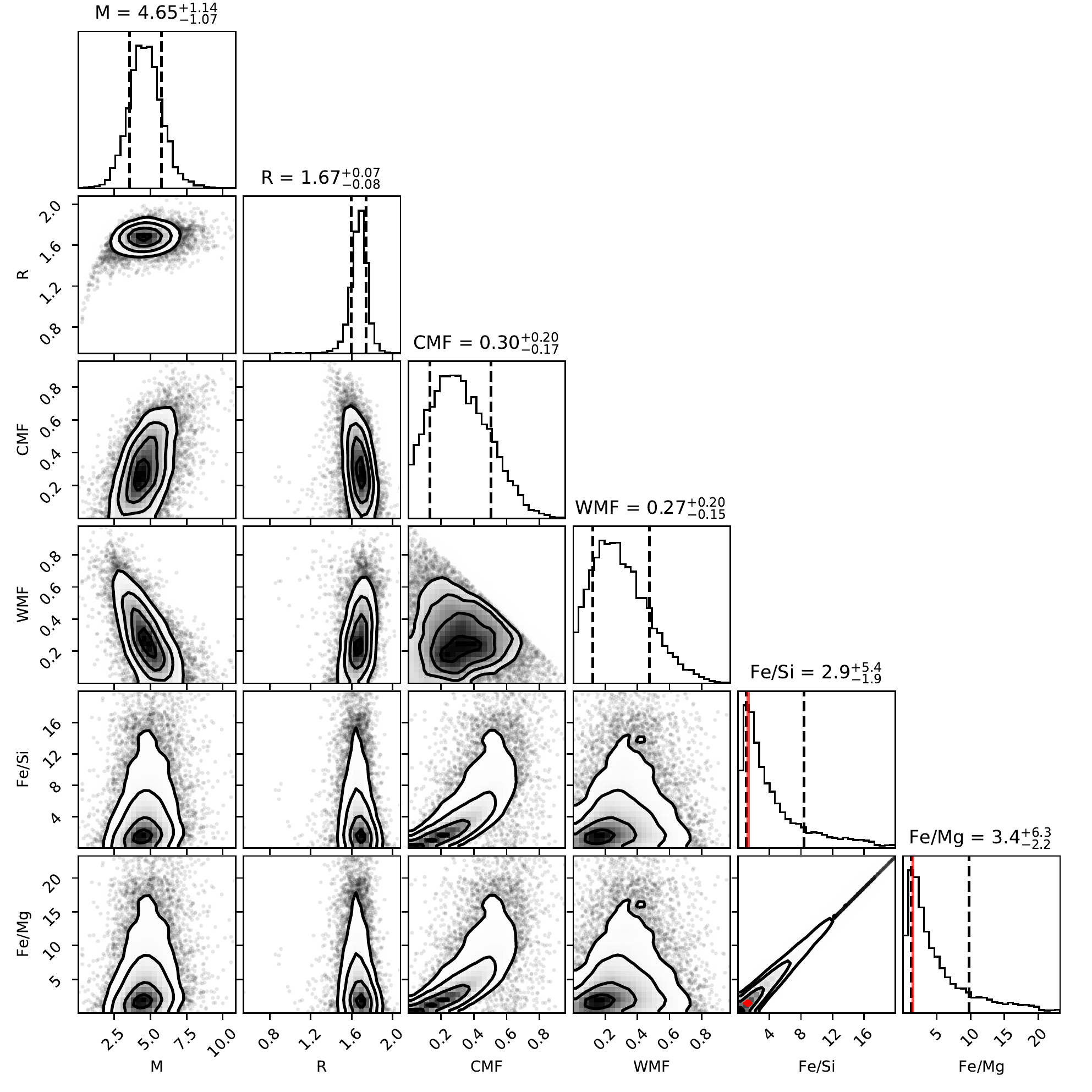}
        \linespread{1}
      \caption{No assumptions, corner plot summary for TOI-1452\,b interior parameters where core mass fraction (CMF) and water mass fraction (WMF) are simulated quantities for a given planetary mass and radius. Chemical ratios Fe/Si and Mg/Si are derived quantities, the red truths in Fe/Si, Fe/Mg space are the mean stellar refractory ratio and the dotted truths represent 16$^{\rm th}$ and 84$^{\rm th}$ percentiles.
      }
      \label{fig:corner}
    \end{figure}
    
    \begin{figure}[ht!]
    \centering
        \includegraphics[width=1\linewidth]{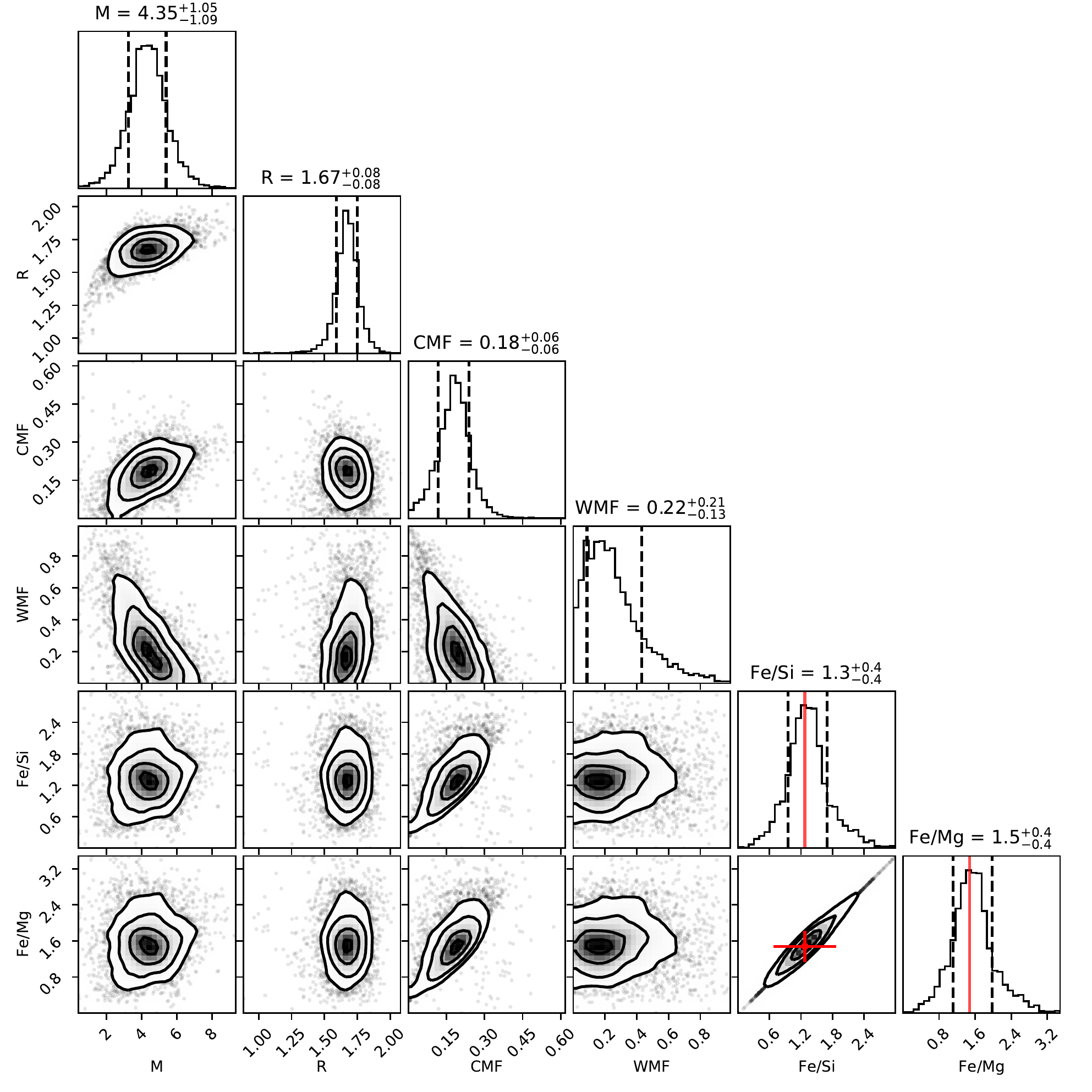}
        \linespread{1}
      \caption{Stellar prior, corner plot summary for TOI-1452\,b interior parameters where we assume that the planet follows stellar refractory ratios for a given planetary mass and radius. Thus, the log-probability function is modified to include the restriction posed by Fe/Mg ratio of the star and can be written as $\mathrm{Fe/Mg} \sim \mathcal{N}\mathrm{(\ {Fe_{*}}/{Mg_{*}},\sigma_{Fe_{*}/Mg_{*}})}$. The red and dotted truths represent stellar ratio and 16$^{\rm th}$ and 84$^{\rm th}$ percentiles of the posterior respectively.
      }
      \label{fig:corner_wp}
    \end{figure}
    
    \begin{figure}[ht!]
    \centering
        \includegraphics[width=1\linewidth]{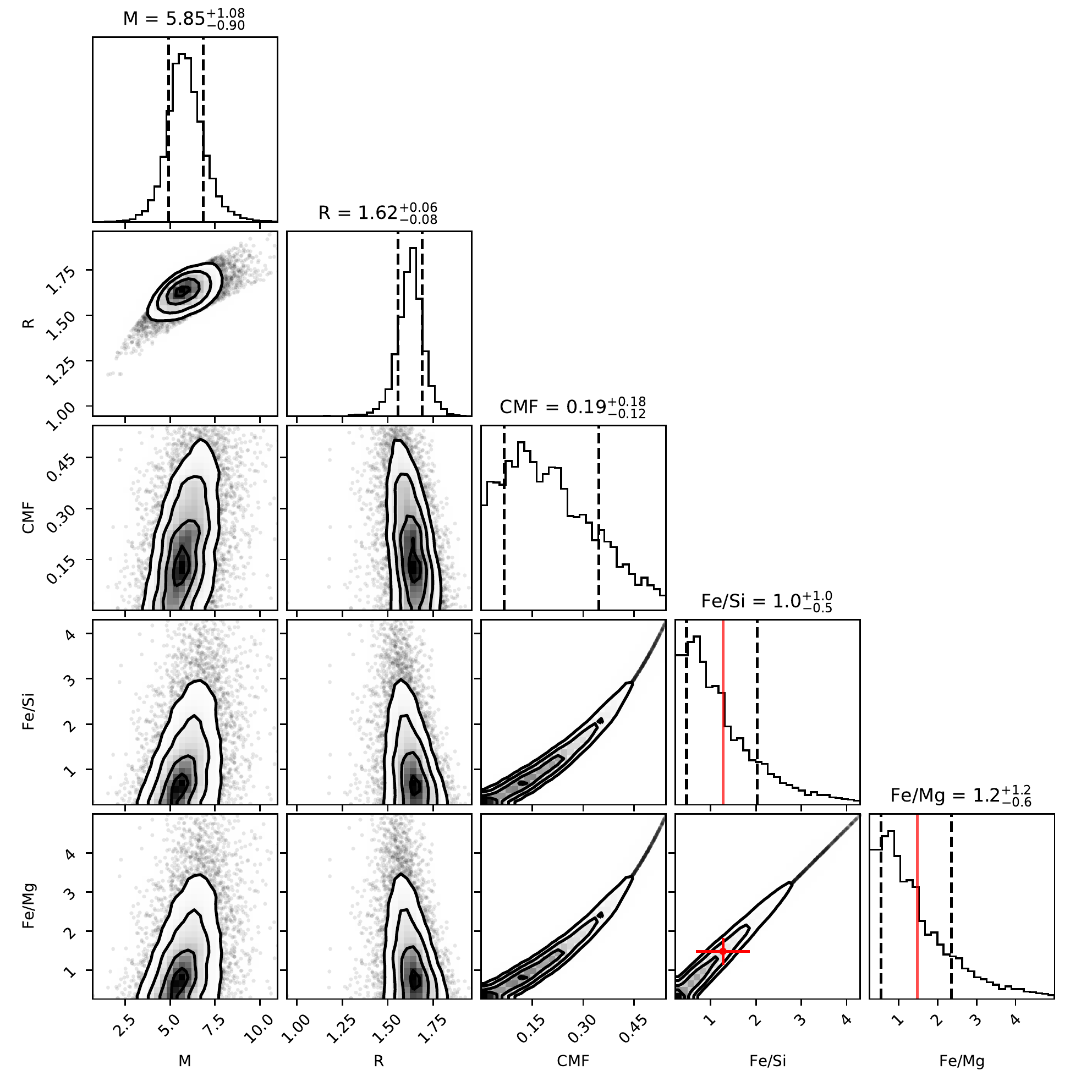}
        \linespread{1}
      \caption{Bare-rock, corner plot summary for TOI-1452\,b interior parameters where we assume only rocky composition is possible (no water) for a given planetary mass and radius. Chemical ratios Fe/Si and Mg/Si are derived quantities, the red and dotted truths represent stellar ratio and 16$^{\rm th}$ and 84$^{\rm th}$ percentiles of the posterior respectively.
      }
      \label{fig:corner_rock}
    \end{figure}
    \vspace{2cm}
    
\clearpage
    
\section{Radial Velocity Measurements}
\setcounter{table}{0}
\renewcommand{\thetable}{D.\arabic{table}}

We present the radial velocity measurements of TOI-1452 from SPIRou and IRD in the online Table \ref{table:spirou_rv}.
\begin{table}[b!]
\end{table}
\begin{deluxetable}{lccc}
%\linespread{0.85}
\tablecaption{SPIRou and IRD RV measurements \label{table:spirou_rv}}
\tablehead{
\colhead{Instrument} & \colhead{BJD - 2\,400\,000} & \colhead{RV (m/s)} & \colhead{$\sigma_{\rm RV}$ (m/s)}
}
\startdata
SPIRou & 59004.995291 & -33983.32 & 8.39 \\
SPIRou & 59005.006067 & -33975.16 & 8.12 \\
SPIRou & 59005.016836 & -33980.21 & 8.25 \\
SPIRou & 59005.027551 & -33995.65 & 8.28 \\
SPIRou & 59009.008115 & -33975.18 & 8.02 \\
SPIRou & 59009.018950 & -33974.66 & 8.20 \\
SPIRou & 59009.029725 & -33975.25 & 7.96 \\
\ldots & \ldots & \ldots & \ldots \\
IRD & 59118.766592 & 15.17 & 4.08 \\
IRD & 59122.769496 & 17.14 & 3.86 \\
IRD & 59122.787278 & 14.54 & 3.77 \\
IRD & 59156.837456 & -14.4 & 4.86 \\
IRD & 59372.001844 & -9.95 & 3.71 \\
IRD & 59373.904952 & 2.00 & 5.70 \\
IRD & 59390.850628 & 12.31 & 4.03 \\
\enddata
\tablecomments{Table \ref{table:spirou_rv} is published in its 
entirety in the machine-readable format. A portion is shown here for guidance 
regarding its form and content.}
\end{deluxetable}

\end{document}